\documentclass{ws-procs9x6}

\begin{document}

\title{Cosmological perturbations}

\author{J. LESGOURGUES}

\address{ITP, EPFL, 1015 Lausanne, Switzerland\\
Theory Division, CERN, 1211 Geneva 23, Switzerland\\
LAPTh (CNRS - Universit\'e de Savoie), BP 110, 74941 Annecy-le-Vieux Cdx, France\\
E-mail: Julien.Lesgourgues@cern.ch\\
www.cern.ch/Julien.Lesgourgues}

\begin{abstract}
We present a self-contained summary of the theory of linear cosmological perturbations. We emphasize the effect of the six parameters of the minimal cosmological model, first, on the spectrum of Cosmic Microwave Background temperature anisotropies, and second, on the linear matter power spectrum. We briefly review at the end the possible impact of a few non-minimal dark matter and dark energy models.
\end{abstract}

\keywords{Cosmological Perturbations; Cosmic Microwave Background; Large Scale Structure of the Universe.}

\bodymatter

\section*{Purposes}

Cosmology is progressing by leaps and bounds thanks to a spectacular amount of observational data. It provides crucial clues for particle physics, and more generally for high energy physics. A key role is played by Cosmic Microwave Background (CMB)  and Large Scale Structure (LSS) data. Their interpretation relies on the {\it theory of cosmological perturbations}.

The aim of this course is to introduce this theory, staying at the linear level, and hopefully with a simple, self-contained and original approach. It is targeted at PhD students in astrophysics and cosmology, as well as researchers in particle physics willing to follow current developments in cosmology. We discuss in particular the effect of the parameters of the minimal cosmological model on the spectrum CMB temperature anisotropies, and on the matter power spectrum. We briefly review at the end the possible impact of non-minimal dark matter and dark energy models.

These notes are based on a cycle of five lectures of 90 minutes. Since this is not much for introducing linear cosmological perturbation theory, the lectures remained at a qualitative level, with more emphasis on physical ideas than on the algebra. These written notes contain roughly the same amount of information as the lectures. After reading these few pages, the reader may wish to improve his/her level of understanding, while sticking to the same structure and same notations. For this purpose, we suggest to read chapters 2, 5 and 6 of~Ref.~\refcite{CUP}: they contain a similar presentation, still simple and qualitative, but more detailed. Optionally, the reader can also grab there a description of neutrino effects, which are completely neglected here.

For a quantitative description of perturbation theory, gauge transformations, CMB physics and linear or non-linear structure formation, the reader should refer to the specialized literature. In the bibliography, we list a very incomplete selection of outstanding references.

Sections \ref{jl:sec2} and \ref{jl:sec3} of these notes are illustrated with several figures that represent the qualitative evolution of cosmological perturbations. It would have been easy to produce them with a Boltzmann code, as we did in Ref.~\refcite{CUP}. We deliberately chose instead to include only scanned hand-drawings. The first reason is that drawings offer the possibility to exaggerate some effects for pedagogical purposes: they are often more illuminating than exact numerical plots. The second reason is that they better render the conviviality of the blackboard presentation at TASI.

%\tableofcontents

\section{Linear cosmological perturbations}\label{jl:sec1}
\subsection{Classification}\label{jl:sec11}

We decompose the metric and stress-energy tensor of the universe into spatial averages and linear perturbations,
\begin{eqnarray}
g_{\mu \nu} (t,\vec{x}) &=& \bar{g}_{\mu \nu} (t) + \delta g_{\mu \nu} (t,\vec{x})~,\\
T_{\mu \nu} (t,\vec{x}) &=& \bar{T}_{\mu \nu} (t) + \delta T_{\mu \nu} (t,\vec{x})~,
\end{eqnarray}
where $ \bar{g}_{\mu \nu}$ stands for the metric of the homogeneous and isotropic Friedmann--Lema\^{\i}tre (FL) model. Being symmetric, the two perturbed tensors contain ten degrees of freedom each, describing different aspects of gravity. Bardeen showed in 1980 that they can be decomposed on the basis of scalars, vectors and tensors under spatial rotations (spatial rotations play a special role because they leave the FL background invariant). These three sectors are decoupled at first order in perturbation theory.

In the vacuum, scalar and vector perturbations vanish, while tensor perturbations can propagate if they have been excited: they account for gravitational waves, the only ``real'' (propagating) gravitational degrees of freedom in General Relativity (GR). In the presence of matter, scalars represent the response of the metric to an irrotational distribution of matter, and generalize Newton's theory of gravitation. Vectors represent the response of the metric to vorticity, and describe phenomena with no equivalent in Newton's theory, called ``gravito-magnetism''.

In minimal cosmological models, the vorticity of the various matter components decays with time, and vectors can be neglected. Tensors may play a small role in CMB anisotropies, that we will mention briefly in Sec.~\ref{jl:sec28}. They can be studied separately, since they decouple from the scalar sector at first order in perturbations. Hence this course will be essentially focused on scalar perturbations. 

The four scalar components of both the metric and stress-energy perturbed tensors are contained in:
\begin{enumerate}
\item the $(00)$ term, 
\item the trace of the $(ij)$ matrix, 
\item the irrotational part of the $(0i)$ vector, 
\item the traceless longitudinal part of the $(ij)$ tensor.
\end{enumerate}
For the perturbed metric  $\delta g_{\mu \nu}$, these components correspond (in the same order) to:
\begin{enumerate}
\item the generalized gravitational potential $\psi$, 
\item the local distortion $\phi$ of the average scale factor $a(t)$: the ``local scale factor'' is given by $(1-\phi)a$, 
\item the potential $b$ such that $\delta g_{0i} = \partial_i b$, 
\item the potential $\mu$ of the metric shear: $\delta g_{ij} = (\partial_i \partial_j - \frac{1}{3} \delta_{ij} \Delta) \mu$~.
\end{enumerate}
For the perturbed stress-energy tensor  $\delta T_{\mu \nu}$, these components (still in the same order) represent:
\begin{enumerate}
\item the energy density perturbations $\delta \rho$ (we will usually refer to the relative perturbations $\delta \equiv \delta \rho / \bar{\rho}$), 
\item the pressure perturbations $\delta p$, 
\item the potential $v$ of the irrotational component of the flux of energy, $\delta T^0_i=\partial_i v$ ($v$ is sometimes called the velocity potential, since in the case of a fluid it is related to the bulk velocity),
\item the potential $s$ of the shear stress or anisotropic stress: $\delta T^i_j = (\partial_i \partial_j - \frac{1}{3} \delta_{ij} \Delta) s$~.
\end{enumerate}
It is equivalent to use as a variable the ``velocity potential'' $v$ or the ``velocity divergence'' $\theta$ defined as
\begin{equation}
\partial_i \delta T^0_i \equiv (\bar{\rho} + \bar{p}) \theta = \Delta v~.
\end{equation}
Similarly, we can use the function $\sigma$ instead of $s$, with the definition
\begin{equation}
(\partial_i \partial_j - \frac{1}{3} \delta_{ij} \Delta)  \delta T^i_j \equiv (\bar{\rho} + \bar{p}) \sigma = \Delta (\Delta s)~.
\end{equation}
The function $\sigma$ is usually called the anisotropic stress, although the true anisotropic stress is the component of $\delta T^i_j$ deriving from the potential $s$. The factors $(\bar{\rho} + \bar{p})$ in the two previous equations are introduced in the definitions in order to obtain simple equations (some authors use alternative notations without these factors or with different ones). To summarize, we see that we can manipulate four degrees of freedom representing the scalar perturbations of matter fields, that can be chosen to be the density fluctuation, pressure perturbation, velocity divergence and anisotropic stress: $\{\delta, \delta p, \theta, \sigma\}$.

\subsection{Gauges}\label{jl:sec12}

In an idealised FL universe, there is only one time slicing compatible with the assumption of homogeneity. Instead, in a perturbed universe, there is an infinity of time slicings compatible with perturbation theory (i.e. such that on each slice, all quantities remain close to their average value). 

The perturbation of any quantity in a given point is the difference between the true and the average quantity in this point. For instance, for the total energy density $\rho$, 
\begin{equation}
\delta \rho (t,\vec{x}) = \rho (t,\vec{x}) - \bar{\rho} (t)~.
\end{equation}
While $\rho (t,\vec{x})$ is a locally, unambiguously defined quantity, $\bar{\rho} (t)$ depends on the choice of equal-time hypersurface going through the point $(t,\vec{x})$. With a different choice, $\rho (t,\vec{x})$ would be compared to the average performed on a different sheet, that would take a different value. Hence $\delta \rho (t,\vec{x})$ also depends on the choice of time slicing.

A gauge is a choice of time slicing. Gauge transformations are induced by coordinate transformations $x_\mu \mapsto x_\mu + \epsilon_\mu$ mapping the points of one time slicing to those of another time slicing. All coordinate transformations do not induce a valid gauge transformation: the condition that perturbations must still be linear after the transformation restricts $\epsilon_\mu$ to be small in every point.

A naive study of the equations of motion of perturbed quantities would be plagued by the freedom to change the gauge without changing physical results: some solutions of the full equations would be ``gauge modes'' with no observable consequences. To deal with this issue, one can adopt one of two point of views:
\begin{itemize}
\item
one can derive gauge-invariant quantities (i.e., non-trivial integro-differential combinations of the metric and stress-energy tensor components left invariant by a gauge transformation), and gauge-invariant equations of motions for these quantities. Note that there are four scalar degrees of freedom in $\delta g_{\mu \nu}$ and two scalar degrees of freedom in the four-vector field $\epsilon_\mu$ inducing gauge transformations: namely, $\epsilon_0$ and the potential $e$ such that $\epsilon_i = \partial_i e$. Hence we can use gauge transformations to cancel two scalar degrees of freedom, and build up two independent gauge-invariant scalars. One way to define them is through the two Bardeen potentials $\Phi_\mathrm{A}$ and $\Phi_H$, defined by Bardeen (1980) as two integro-differential combinations of $\psi$, $\phi$, $b$, $\mu$.
\item
one can fix the gauge, i.e. introduce a condition such that the time slicing is unique. Then the number of independent solutions to the equations will be the same as in the gauge-invariant formalism. In any case, one can show that truly observable quantities are always independent of the gauge. Obtaining them after solving equations in the gauge-invariant formalism or in one particular gauge should not make any difference in practice.
\end{itemize}

A convenient gauge choice for a pedagogical introduction to CMB and matter scalar perturbations is the so-called Newtonian gauge or longitudinal gauge, in which one imposes that non-diagonal scalar perturbations of the metric vanish: $b=\mu=0$. This prescription can be showed to fix a unique time slicing. In this gauge, adopting units such that $c=1$ and using proper time $t$, the line element reads:
\begin{equation}
ds^2 = - (1+2\psi) dt^2 + (1-2\phi) a^2 d\vec{l}^{\,\,2}~,
\end{equation}
where $d\vec{l}^{\,\,2}$ stands for the cartesian measure $(dx^2+dy^2+dz^2)$ for a flat FL model, or for $(1 \pm kr^2)^{-1}dr^2 + r^2(\sin\theta^2 d\theta^2 + d\varphi^2)$ for an open/closed FL model in spherical coordinates. We are still free to redefine time (by definition, a time redefinition leaves the time slicing invariant). Lots of results in cosmological perturbation theory look simpler when using conformal time $\eta$, defined up to a constant by $d \eta = dt/a$. In this course we fix the constant in such way that $\eta \longrightarrow 0$ at the vicinity of the initial singularity, when $\rho \longrightarrow \infty$ (this prescription would not work if we were studying cosmological inflation, but in this course, we are not). Conformal time is convenient because photons traveling in a flat unperturbed FL universe along geodesics crossing the origin of the system of coordinates obey to $dr = d\eta$ (this comes from $a \, dl = dt$, i.e. from $ds=0$ with the restriction $d\theta=d\phi=0$). Hence conformal time is a measure of time based on the comoving distance travelled by a photon\footnote{In a universe with non-zero spatial curvature, this remains true, provided that the comoving distance is defined not like $r$, but like $\chi\equiv\int (1 \pm kr^2)^{-1/2}dr$.}, and the comoving distance to a given object coincides with its ``look-back conformal time''. In this course we will use dots for derivatives with respect to proper time and primes for derivatives with respect to conformal time. The Hubble parameter (or expansion rate parameter) reads
\begin{equation}
H = \frac{\dot{a}}{a} = \frac{a'}{a^2}~,
\end{equation}
the Hubble radius is $R_H=1/H$,
and the condition that a Fourier mode of physical wavelength $\lambda$ crosses the Hubble radius is
\begin{equation}
\lambda= R_H ~~\Leftrightarrow~~ \frac{2\pi}{k} a = \frac{1}{H} ~~\Leftrightarrow~~ k \sim aH = \frac{a'}{a}~.
\end{equation}

One advantage of the Newtonian gauge is that the gauge-invariant Bardeen potentials $\{\Phi_\mathrm{A}, \Phi_H\}$ reduce in this gauge to the metric perturbations $\{ \phi, \psi \}$: the evolution of the Newtonian metric perturbations informs us directly on that of two gauge-invariant quantities. Other interesting properties of $\{ \phi, \psi \}$ appear when writing the Einstein equations in the Newtonian gauge. The full Einstein equations linearized at first order in perturbations feature four equations relating scalar degrees of freedom. One of them, associated to the traceless longitudinal part of $\delta G^i_j = 8 \pi G T^i_j$, gives (assuming a flat FL background)
\begin{equation}
\frac{2}{3}(k/a)^2 (\phi-\psi) = 8 \pi G \sum_x (\bar{\rho}_x + \bar{p}_x) \sigma_x~,
\end{equation}
where the index $x$ runs over all the species contributing to the total stress-energy tensor. This means that when the universe contains only shearless components with $\sigma_x=0$ (as would be the case in the presence of perfect fluids), the two metric perturbations are equal. Next, the Einstein equation $\delta G^0_0 = 8 \pi G T^0_0$ gives (still assuming a flat FL background)
\begin{equation}
2 a^{-2} \left[ k^2 \phi + 3 \frac{a'}{a}\left(\phi'+\frac{a'}{a}\psi\right)\right] = - 8 \pi G \sum_x \bar{\rho}_x \delta_x~.
\end{equation}
The term on the right-hand side involves the total energy perturbation $\delta \rho_\mathrm{tot}=\sum_x \bar{\rho}_x \delta_x$.
In the short scale (more precisely, sub-Hubble) limit, the term containing $k^2$ dominates the other terms in the square brackets, and we recover the Poisson equation
\begin{equation}
- \frac{k^2}{a^2} \phi = 4 \pi G \, \delta \rho_\mathrm{tot}~,
\end{equation}
where the factor $-k^2/a^2$ represents the Fourier transform of the physical Laplace operator in an expanding universe. Note that in the Poisson equation one may have expected to see the generalized gravitational potential $\psi$ instead of $\phi$: however, in the sub-Hubble limit, the shear of individual components is usually either null or negligible, so that $\phi=\psi$.

\subsection{Equations of motion}\label{jl:sec13}

In the minimal cosmological scenario, the universe features several species with spatial fluctuations, described with different equations because of their distinct properties: cold dark matter (CDM) is non-relativistic and collisionless, neutrinos are ultra-relativistic and collisionless at the times of interest, baryons are non-relativistic and smoothly interpolating from a strongly coupled to decoupled regime, and finally photons are ultra-relativistic and interpolating between the same two regimes. 

The equation of conservation of the total stress-energy tensor, $D_\mu T^\mu_\nu=0$ (deriving from Bianchi identities), yields two scalar and two vector equations. The scalar ones are the conservation of energy equation and the Euler equation. 

For an single component experiencing no interaction with other species (other than gravitational), the equation $D_\mu T^\mu_\nu=0$ applies to the individual stress-energy tensor: it gives one continuity and one Euler equation for that component. The evolution of  single component experiencing interactions is also given by the continuity and Euler equation, but with an extra source term accounting for stress-energy injection/leak caused by the interaction.

We have seen that the perturbations of each component $x$ can be described by four variables $\{\delta_x, \delta p_x, \theta_x, \sigma_x \}$. Hence, in general, two equations of motion are not sufficient for closing the system. However:
\begin{itemize}
\item for a perfect fluid, microscopic interactions impose local thermodynamical equilibrium. The pressure is then isotropic\footnote{An anisotropic shear stress $\sigma\neq0$ reflects the fact that in each given point, particles travel with different velocities (due to some intrinsic velocity dispersion and/or a superposition of several flows in phase space), leading to anisotropic pressure. This contradicts the assumption of a perfect fluid, in which local interactions result in a unique bulk velocity (after coarse-graining over microscopic scales), and erase anisotropic pressure.}, with $\sigma_x=0$. In addition, pressure perturbations obey to $\delta p_x = c_\mathrm{a}^2 \delta \rho_x$, where $c_\mathrm{a}$ is the adiabatic sound speed inferred from the equation of state of the fluid. If $\sigma_x$ vanishes and $\delta p_x$ is a function of $\delta \rho_x$, perturbations in the fluid are described by only two independent functions $\delta_x\equiv \delta \rho_x/\bar{\rho}_x$ and $\theta_x$. If the collision term also vanishes or is specified, the two equations of motion (continuity and Euler) are sufficient for closing the system and computing the evolution of perturbations.
\item for a decoupled or weakly interacting species, there are no such simplifications concerning the anisotropic stress and pressure perturbation. Hence, in general, the two equations inferred from stress-energy conservation are not sufficient. For such species, one has to use the more general Boltzmann equation, giving the evolution of each phase-space distribution:
\begin{equation}
\frac{d}{d \eta} f_x = \sum_y C_{xy}[f_x,f_y]~,
\end{equation}
where the sum holds over the species $y$ coupled with $x$. Each phase-space distribution can be decomposed into a background and perturbation part:
\begin{equation}
f_x(\eta,\vec{x},\vec{p}) = \bar{f}_x(\eta,|\vec{p}|) + \delta f_x(\eta,\vec{x},\vec{p})~,
\end{equation}
where $\vec{p}$ stands for momentum, $|\vec{p}|$ for its modulus, and the background part does not depend on the direction of $\vec{p}$ by assumption of isotropy.
\item fully decoupled CDM is a particular case of a collisionless species with negligible velocity dispersion (the word ``cold'' refers precisely to this last assumption). Hence, it behaves in the same way as a pressureless perfect fluid, although in reality it has no interactions and should not be called a fluid. Since the velocity dispersion is negligible, in a given point, all particles share the same velocity, imposed by gravitational flows (while for  non-cold collisionless species, the velocity would get two contributions, one from gravitational flows, and one from the phase-space distribution function). Hence the anisotropic pressure vanishes (see the previous footnote). The pressure perturbation $\delta p_x$ is also related to the velocity dispersion in a given point, anf can be neglected with respect to the density perturbation in the CDM case. Hence CDM is formally equivalent to a perfect fluid with no anisotropic stress and no pressure perturbation (or in other words, with a sound speed $c_\mathrm{s}=0$). In that case, the two equations of motion inferred from $D_\mu T^\mu_\nu=0$ are sufficient, like for a fluid.
\end{itemize}
Gravitational interactions between species are accounted by the presence of metric perturbations in each equation of motion (more specifically, terms in $k^2 \psi$ accounting for gravitational forces, and terms in $\phi'$ accounting for dilation effects, i.e. for local distortions of the scale factor with respect to $a$). Hence, in order to close the full system of equations, we still need two independent relations, to be chosen among the four scalar Einstein equations: they provide the value of $k^2 \psi$ and $\phi'$ at each time, as a function of all matter fields.

\subsection{Initial conditions}\label{jl:sec14}

We wish to study the evolution of matter perturbations, starting from some early time at which all Fourier modes of interest (those which are observable in the CMB spectrum and in the matter power spectrum on linear or mildly non-scales) are still outside the Hubble radius. Indeed, super-Hubble modes experience a trivial evolution, unaffected by small scale interactions (Thomson or Coulomb scattering, usual gravitational force $\vec{\nabla} \phi$, etc.) Hence, the perturbations evaluated at some arbitrary time but on super-Hubble scales reflect directly the mechanism responsible for the formation of perturbations in the very early universe. In the standard cosmological model, these initial conditions can be inferred from inflation.

Typically, a good time for setting initial conditions is when the redshift $z=a_0/a-1$ is of the order of $10^5$: at this time, all comoving scales that are observable in the CMB and linear matter power spectrum still verify $k \ll aH$.

It is crucial to understand that, as long as the background cosmology is assumed to be of the FL type, the perturbed stress-energy momentum tensor $\delta T_\mu^\nu$ must be diagonal on super-Hubble scales. Indeed, the background tensor $\bar{T}_\mu^\nu$ is diagonal, and of the form: ${\rm diag}(-\bar{\rho},\bar{p},\bar{p},\bar{p})$. This can be showed to be the most general assumption compatible with homogeneity and isotropy. Let us assume that we Taylor-expand  $\delta T_\mu^\nu$ in powers of the variable $(k\eta)$. For any power-law scale factor, $aH$ is given by $1/\eta$ times a factor of order one. Hence the limit $k\eta \ll 1$ represents precisely the super-Hubble limit. In the Taylor expansion, the zero mode should share the same properties as the background solution, and be diagonal. Higher order terms account for contributions to $\delta T_\mu^\nu$ growing with time, and possibly becoming important around the time of Hubble crossing.

The total scalar perturbations $\delta \rho$ and $\delta p$ are the only one preserving the diagonal form of $\delta T_\mu^\nu$: we conclude that they are the only one that do not vanish at order zero in the $(k \eta)$ expansion. Using stress-energy conservation equations, one can show that the part of $\delta T_\mu^\nu$ associated with the velocity divergence is of order one in $(k \eta)$, while the part associated to the total anisotropic stress is of order two.

Suppose that the universe contains initially $N$ uncoupled perfect fluids\footnote{The discussion presented in this section could be generalized to $N$ coupled species, not all of them being perfect fluids: the conclusions would not change qualitatively, and we restrict here to $N$ uncoupled fluid for simplicity.},  with $N$ known sound speeds $c_x^2=\delta p_x/\delta \rho_x$. There are $2N$ independent initial conditions, corresponding to possible initial values of each $\delta_x$ and each $\delta_x'$. Importantly, in the $2N$-dimensional basis of IC's, one basis vector is very special, as we shall see below.

Before studying perturbations, one should have solved for the background evolution: all background quantities should be known, including for instance the density $\bar{\rho}_x(\eta)$ and pressure $\bar{p}_x(\eta)$ of each species $x$. Now, let us assume that the real universe is perturbed initially by a single degree of freedom (one may say, by a single initial time shifting function). This is the case in single-field inflationary cosmology: during inflation, there is a single clock (the inflaton), and perturbations arise from a single time shifting function (the inflaton perturbation).

As long as we are dealing with super-Hubble modes, we can neglect microscopic interactions and say that the evolution in each point (in fact, in each Hubble patch) is still given by homogeneous cosmology, taking this shift function $\delta \eta(\vec{x})$ into account:
\begin{eqnarray}
\forall x, \qquad \rho_x(\eta,\vec{x}) &=& \bar{\rho}_x\left(\eta+\delta \eta(\vec{x})\right) = \bar{\rho}_x(\eta) + \bar{\rho}_x'(\eta) \, \delta \eta(\vec{x}) \nonumber \\
p_x(\eta,\vec{x}) &=& \bar{p}_x\left(\eta+\delta \eta(\vec{x})\right) = \bar{p}_x(\eta) + \bar{p}_x'(\eta) \, \delta \eta(\vec{x})
\label{jl:ad_ic}
\end{eqnarray}
where in the last equalities, terms of order two or higher in $\delta \eta$ have been neglected. The above ansatz restricts a lot the choice of possible initial conditions. Indeed, for each uncoupled species, the energy conservation law for the background gives
\begin{equation}
\bar{\rho}_x' = -3 \frac{a'}{a} (\bar{\rho}_x + \bar{p}_x)~.
\end{equation}
Combining this equation with the above ansatz, we get
\begin{equation}
\forall x, \qquad \frac{\delta \rho_x}{\bar{\rho}_x + \bar{p}_x} = - 3 \frac{a'}{a} \frac{\delta \rho_x}{\bar{\rho}_x'} = -3 \frac{a'}{a}  \, \delta \eta(\vec{x})~.
\end{equation}
Since the last term is independent of the index $x$, we see that for each pair of fluids,
\begin{equation}
\forall x,y, \qquad \frac{\delta \rho_x}{\bar{\rho}_x + \bar{p}_x} = \frac{\delta \rho_y}{\bar{\rho}_y + \bar{p}_y}~.
\label{jl:ad_ic2}
\end{equation}
Note that this famous relation is a consequence of Eqs.~(\ref{jl:ad_ic}). It shows that in presence of such initial conditions, everything is fixed up to a single function of $\vec{x}$. Let us take the example of a universe containing only photons, baryons, cold dark matter and neutrinos. We can use the fact that for non-relativistic species $\bar{p}_x \ll \bar{\rho}_x$, while for ultra-relativistic ones  $\bar{p}_x =  \bar{\rho}_{x}/3$. Hence, if one function is known --- for instance, $\delta_\gamma(\vec{x})$ at initial time --- the others can be derived from
\begin{equation}
\delta_\mathrm{b} = \delta_\mathrm{cdm} = \frac{3}{4} \delta_\nu = \frac{3}{4} \delta_\gamma~.
\label{jl:ad_ic4}
\end{equation}
It is a simple exercise to prove that the ansatz of Eqs.~(\ref{jl:ad_ic}) implies additional important relations:
\begin{itemize}
\item
first,
\begin{equation}
\forall x, \qquad \delta p_x(\eta,\vec{x}) = \frac{{\bar{p}_x}'(\eta)}{{\bar{\rho}_x}'(\eta)} \, \delta \rho_x(\eta,\vec{x})~,
\label{jl:ad_ic3bis}
\end{equation}
i.e., as long as we are dealing with super-Hubble modes and Eqs.~(\ref{jl:ad_ic}) are satisfied, each component features an adiabatic sound speed independent of $x$, obeying to $c_x^2(\eta) \equiv {{\bar{p}_x}'(\eta)}/{{\bar{\rho}_x}'(\eta)}$,
\item
second, 
\begin{equation}
\delta p_{\rm tot}(\eta,\vec{x}) =  c_\mathrm{s}^2(\eta)  \, \delta \rho_{\rm tot}(\eta,\vec{x})~,
\label{jl:ad_ic3}
\end{equation}
where the squared total sound speed $c_\mathrm{s}^2$ can be easily expressed as a weighted average over the squared adiabatic sound speed of individual components. Hence, the total matter content resulting from the sum of all components also features an adiabatic sound speed. 
\end{itemize}
For the reason expressed in the last item, such initial conditions are usually called {\it adiabatic initial conditions}. The word adiabatic refers to thermodynamics, but it is actually more helpful to think of adiabatic initial conditions as resulting from the assumption of a single inhomogeneous degree of freedom in the early universe: then Eqs.~(\ref{jl:ad_ic}) stands for the true definition of adiabatic initial conditions, and Eqs.~(\ref{jl:ad_ic2}, \ref{jl:ad_ic4}, \ref{jl:ad_ic3bis}, \ref{jl:ad_ic3}) for its consequences.

More general initial conditions not obeying to Eqs.~(\ref{jl:ad_ic}) can be expanded on different bases. A famous basis is formed by (i) the adiabatic mode; (ii) $(N-1)$ non-decaying isocurvature modes, getting their name from the property that for each of them, the total density (and spatial curvature) perturbations vanish asymptotically in the super-Hubble limit, while two species have opposite density perturbations compensating each other; (iii) $N$ decaying modes that are irrelevant for most purposes. 

It is clear from the previous discussion that non-adiabatic initial conditions should only be considered when assuming that primordial perturbations are generated by more than one degree of freedom (for instance, two inflaton fields, or one inflaton and one axion, etc.)
The assumption of several degrees of freedom is necessary but not sufficient. The primordial universe may contain a mixture of adiabatic and isocurvature modes until a time at which all species are brought in thermal equilibrium. At that time, if we further assume that all chemical potentials vanish, the perturbations of each species can be inferred from those of temperature, $T(\eta,\vec{x})=\bar{T}(\eta) + \delta T(\eta,\vec{x})$. Then, temperature plays precisely the role of a single time-shifting function. Any non-adiabatic initial condition is washed out and becomes irrelevant. Hence, isocurvature modes can be observable --- for instance in the CMB --- only under additional assumptions. For instance, one species carrying isocurvature perturbations might remain out of equilibrium at all times, or might feature a chemical potential with spatial fluctuations. There exist a few non-minimal, but still reasonable scenarii featuring isocurvature perturbations (with axions, curvatons etc.). Ultimately, the presence of isocurvature modes is to be tested with observations. Current CMB observations put strong limits on the amplitude of such modes, and prefer purely adiabatic initial conditions. We will restrict to this case in the rest of this course.

For adiabatic initial conditions, we found the relation~(\ref{jl:ad_ic2}) holding between density fluctuations. Also, if we do not consider the case of species with a non-zero anisotropic stress, we can assume that $\phi=\psi$ at initial time. By plugging Eqs.~(\ref{jl:ad_ic4}), (\ref{jl:ad_ic3})  and $\phi=\psi$ into the Einstein equations, one is led to a second-order differential equation for $\psi$ only. During the radiation dominated era, one can show that this equation has two solutions, one constant in time, and one decaying. The (00) Einstein equation then shows that the constant solution is related to density fluctuations through
\begin{equation}
-2 \psi = -2 \phi = \delta_{\rm tot} \simeq \delta_\gamma = {\rm constant}~.
\label{jl:ad_ic5}
\end{equation}
Hence, in the Newtonian gauge and for adiabatic initial conditions, super-Hubble metric fluctuations and density fluctuations are static. In fact, on super-Hubble scales, there can only be some time evolution when the universe changes of total equation of state (or equivalently, of expansion law). This is the case at the time of equality between radiation and matter. During matter domination and on super-Hubble scales, they freeze out again. Then, the relation $-2 \psi = -2 \phi = \delta_{\rm tot}$ and  Eq.~(\ref{jl:ad_ic4}) are still satisfied, but now $\delta_{\rm tot}\simeq \delta_\mathrm{b}=\delta_\mathrm{cdm}=\frac{3}{4} \delta_\gamma$. This detail will become important  when studying the Sachs-Wolfe effect in section~\ref{jl:sec23}.

\subsection{Power spectra and transfer functions}\label{jl:sec15}

The theory of cosmological perturbations is a stochastic theory: the fluctuations of a given quantity in a given point, $A(\eta,\vec{x})$, obey to a distribution of probability. As long as we stick to linear perturbation theory, there is a ``linear transport of probability'' from one time to another time. Let the probability of $A$ in a given point at time $\eta_1$ be ${\cal P}_1(A)$. In the same point and at $\eta_2$, the linear evolution would have transformed each value $A$ into $\alpha A$, where $\alpha$ is the linear growth (or decrease) factor of $A$ between $\eta_1$ and $\eta_2$. So the probability of $A$ at time $\eta_2$ is given by ${\cal P}_2(A) = {\cal P}_1(A/\alpha)$. This ``linear transport of probability'' implies that the linear evolution respects the shape of the probability distribution, and rescales all its statistical moments of order $n$ by $\alpha^n$. In particular, if the initial probability is Gaussian, the statistics will remain Gaussian at any later time, and the evolution of the system can be formulated as an evolution of the root mean square of all quantities. In summary, as long as we assume linear perturbations with  Gaussian initial conditions, our goal is to solve for the evolution of the root mean squares of the fluctuations.

The equal-time 2-point correlation function of any quantity $A$ in real space is given by
\begin{equation}
\langle A(\eta,\vec{x}) A(\eta,\vec{x}') \rangle \equiv \xi(\eta,\vec{x},\vec{x}') \stackrel{SHI}{=} \xi(\eta,|\vec{x}' - \vec{x}|)~,
\end{equation}
where the last equality holds as a consequence of Statistical Homogeneity and Isotropy (SHI, assumed to hold in a perturbed FL universe). Indeed, the correlation function should be invariant under spatial translations and rotations.

We can go to Fourier space and use the same letter to denote the Fourier transform of $A$. If $A$ is real, $A(\eta,-\vec{k})=A^*(\eta,\vec{k})$. It is easy to show that the previous equation (using the assumption of SHI) implies that the two-point correlation function in Fourier space reads
\begin{equation}
\langle A(\eta,\vec{k}) A^*(\eta,\vec{k}') \rangle \stackrel{SHI}{=} \delta_D(\vec{k}' - \vec{k}) P_\mathrm{A}(k)~,
\end{equation}
where $\delta_D$ is the Dirac distribution.
Here, statistical homogeneity is responsible for the fact that the two-point correlation vanishes for $\vec{k}\neq\vec{k'}$, and statistical isotropy for the fact that $P_\mathrm{A}$ only depends on the modulus $k=|\vec{k}|$. The function $P_\mathrm{A}(k)$ is usually called the power spectrum of $A$. Some authors prefer to refer to the ``dimensionless power spectrum'', defined as
\begin{equation}
{\cal P}_\mathrm{A}(k) \equiv \frac{k^3}{2 \pi^2} P_\mathrm{A}(k)~.
\end{equation}
The reason is that typical expressions for the average of various quantitites in real space consist in the convolution of the power spectrum with some window function $f(k)$, like in
\begin{equation}
\int \frac{d^3 \vec{k}}{(2 \pi)^3} P_\mathrm{A}(k) f(k) = \int \frac{4 \pi k^2 dk}{(2 \pi)^3} P_\mathrm{A}(k) f(k) = \int d \! \log \! k \,\, {\cal P}_\mathrm{A}(k) f(k)~.
\end{equation}
Hence the dimensionless spectrum ${\cal P}_\mathrm{A}(k)$ stands for the weight of each logarithmic interval in the integral. The term ``scale-invariant spectrum'' refers to ${\cal P}_\mathrm{A}(k)$ being independent of $k$, i.e. $P_\mathrm{A}(k) \propto k^{-3}$.

We know that for adiabatic initial conditions, all perturbations are related to each other through Eqs.~(\ref{jl:ad_ic4}, \ref{jl:ad_ic5}). Hence, with Gaussian adiabatic initial conditions, specifying the primordial power spectrum of one quantity is sufficient for knowing everything about the system. For instance, if we assume that the power spectrum of the metric perturbation $\psi$ is a given function $P_\psi(k)$, then we infer from Eqs.~(\ref{jl:ad_ic4}, \ref{jl:ad_ic5}) that the photon and baryon primordial spectra are given by 
$P_\gamma(k)=4 P_\psi(k)$ and $P_\mathrm{b}(k)=\frac{9}{16} P_\gamma(k) = \frac{9}{4} P_\psi(k)$. 

By convention, the primordial spectrum is usually given for the variable ${\cal R}$, which represents the spatial curvature perturbation on one initial comoving hypersurface (i.e. an hypersurface orthogonal to the energy flux of the total cosmic fluid in each point). The advantage of using this quantity is that it is conserved on super-Hubble scales for adiabatic initial conditions (while $\phi$ and $\psi$ get rescaled when the equation of state of the universe changes, e.g. at radiation-matter equality). In the Newtonian gauge, ${\cal R}$ relates to $\psi$ and to the total density perturbation through
\begin{equation}
{\cal R} = \psi - \frac{1}{3} \frac{\delta \rho_{\rm tot}}{\bar{\rho}_{\rm tot}+\bar{p}_{\rm tot}}~.
\end{equation}

The power spectrum of a given quantity at some arbitrary time can be decomposed into two parts, one accounting for initial conditions, and one accounting for linear evolution with time:
\begin{equation}
\langle A(\eta,\vec{k}) A^*(\eta,\vec{k}') \rangle = \delta_D(\vec{k}' - \vec{k}) \left[\frac{A(\eta,\vec{k})}{{\cal R}(\vec{k})}\right]^2 P_{\cal R}(k) ~.
\end{equation}
In the above equation, there is no time argument for ${\cal R}(\vec{k})$ since this quantity is conserved on super-Hubble scales. We just assume that ${\cal R}$ is evaluated at a time such that $k\ll aH$ for all modes of interest. In a FL universe, the equations of motion of all perturbations respect isotropy, and do not depend on the direction of the wavector $\vec{k}$. Hence the ratio between brackets in the last equation is a function of $k$, not $\vec{k}$. This function accounts for the linear evolution, independently of initial conditions. It is called the ``transfer function'' of $A$. In this course, we will denote transfer functions with the same letter as the perturbations themselves, but with the modulus of $k$ as an argument, i.e.
\begin{equation}
A(\eta,k) \equiv \frac{A(\eta,\vec{k})}{{\cal R}(\vec{k})}~.
\end{equation}
In summary of this section, solving for cosmological perturbations (in a model with adiabatic Gaussian IC's) amounts in
\begin{itemize}
\item postulating a primordial spectrum, or calculating it within the framework of a model, for instance of inflation;
\item solving the equations of motion of all perturbations, with quantities normalized initially to ${\cal R}(\vec{k})=1$, in such a way to obtain all transfer functions $A(\eta,k)$ and to infer the evolution of the various root mean squares.
\end{itemize}

\section{CMB temperature anisotropies}\label{jl:sec2}

From now on, we will assume for simplicity that the universe is flat.

\subsection{Photon scattering rate}\label{jl:sec21}

CMB physics consists is the study of electron, baryon and photon perturbations on cosmological scales, taking into account their gravitational coupling with collisionless species such as decoupled neutrinos and CDM. Electrons and baryons carry opposite electric charges and are coupled to each other through very efficient Coulomb scattering processes. Electrons and photons are coupled through Thomson scattering, which is the limit of Compton scattering when electrons are non-relativistic and photons carry a smaller energy than the rest mass of the electron. Then, the scattering process results mainly in a deflection of the photon, with a negligible transfer of energy between the two particles. The leading interaction between baryons and photons is the gravitational one. 

In units such that $c=1$, the Thomson scattering rate (with respect to conformal time) is given by $\Gamma = \sigma_T a n_e x_e$, where $\sigma_T$ is the Thomson scattering rate, $a$ is the scale factor, $n_e$ is the total electron number density (scaling like $a^{-3}$ due to dilution), and $x_e$ is the ionized electron fraction. The product $a n_e$ scales like $a^{-2}$. The ionized fraction is close to one at high energy. Then, at the time of recombination between electrons and nuclei (around $z\sim1080$), which falls at the beginning of matter domination, $n_e$ drops abruptly to very small values. This causes Thomson scattering to become suddenly very inefficient, and photons to decouple from electrons.

Hence photon decoupling is the story of Thomson scattering becoming inefficient, while Coulomb scattering remains very strong. For that reason, one can describe electrons and baryons as a single tightly-coupled fluid. People often refer only to baryons for simplicity. At early time, the full system of (electrons)-baryons-photons is also tightly coupled, but later on, it splits progressively into two collisonless species, (electrons)-baryons on the one hand, photons on the other hand.

The thermodynamical description of recombination is very technical, due to the different energy level of atoms (in particular, of hydrogen). In order to understand CMB anisotropies, we only need to describe the main results of recombination studies at a very qualitative level.
\begin{itemize}
\item The free electron fraction $x_e(\eta)$ starts from one at high redshift. It decreases sharply at the recombination time (corresponding to $z \sim 1080$ or $T\sim 0.3$~eV), and freezes at a very small value (due to departure from thermal equilibrium). Around $z\sim 10$, star formation causes a reionization of the universe, and $x_e$ goes up again to one.  
\item The Thomson scattering rate $\Gamma = \sigma_T a n_e x_e$ evolves like $a^{-2} x_e$. Before recombination, $\Gamma \gg \frac{a'}{a}$, and the universe is opaque. The sudden drop of $x_e$ at recombination renders the universe transparent: $\Gamma \ll \frac{a'}{a}$. Due to the dilution factor coming from $n_e$,  $\Gamma$ remains much smaller than $\frac{a'}{a}$ even during reionization, and the universe keeps being transparent (this is why despite of reionization, most photons emitted at recombination do not interact anymore, and allow us to observe anisotropies on the last scattering surface).
\item The optical depth $\tau(\eta)\equiv\int_\eta^{\eta_0} d\eta \Gamma(\eta)$ represents the opacity of the universe at a given time, when seen from today (when $\eta=\eta_0$). It tends to infinity when $\eta \longrightarrow 0$, falls below one at recombination and stabilize at a value of the order of 0.1 between  recombination and reionization. After reionization it decreases smoothly and reaches zero today by definition. 
\item The visibility function $g(\eta)\equiv-\tau' e^{-\tau}$ gives the probability that a CMB photon seen today experienced its last scattering at time $\eta$. It starts from negligible values at high redshift (suppressed by the $e^{-\tau}$ factor). It has a narrow spike around the time of recombination, and then it falls again to negligible values due to the smallness of $\tau'$ between recombination and reionization. It develops a second smaller and wider spike around reionization. This function shows that most CMB photons did not interact between the last scattering surface and today, while a minority rescattered at reionization. The width of the recombination spike gives an indication on the thickness of the last scattering ``surface''.
\item The diffusion length $\lambda_\mathrm{d}(\eta)$ is an important quantity for understanding the damping of temperature anisotropies on small scales. At any given time, the comoving mean free-path (mfp) of photons is given by $r_{\rm mfp} = 1/\Gamma(\eta)$ (still in units where $c=1$). If the photons experience a random walk analogous to Brownian motion in a gas, the comoving distance over which they travel between time $\eta_\mathrm{ini}$ and $\eta$ can be approximated by
\begin{equation}
r_\mathrm{d} \sim \left[ \int_{\eta_\mathrm{ini}}^{\eta_0} d\eta \,\, \Gamma \,\, r_{\rm mfp}^2 \right]^{1/2} =  \left[ \int_{\eta_\mathrm{ini}}^{\eta_0} d\eta\,\, \Gamma^{-1} \right]^{1/2}~.
\label{jl:diffusion_scale}
\end{equation}
This physical diffusion length of photons is given in this approximation by $\lambda_\mathrm{d} \simeq a r_\mathrm{d}$. It grows quickly from very small to very large scales (comparable to the Hubble scale) around the time of recombination.
\end{itemize}

\subsection{Boltzmann equation}\label{jl:sec22}

Since photons decouple from baryons near the recombination time, we cannot describe them with fluid equations, and need to solve the Boltzmann equation 
\begin{equation}
\frac{d}{d\eta} f = C\left[f, f_e\right]
\end{equation}
at order one in perturbations. The right-hand side stands for the photon-electron coupling due to Thomson scattering. As explained in the previous section, electrons and baryons are so tightly coupled that it makes no difference to think of this term as a photon-electron or photon-baryon coupling term. Solving this equation is involved because the photon phase-space distribution $f$ involves many arguments, $f(\eta,\vec{x},\vec{p})$. Fortunately one can reduce the dimensionality of the problem. First, we notice that as long as photons are in thermal equilibrium with electrons (and hence with baryons), they are entirely described in any point by the local value of the equilibrium temperature $T(\eta,\vec{x})$. The phase space distribution is then of the Bose-Einstein form:
\begin{equation}
f(\eta,\vec{x},\vec{p}) = \frac{1}{e^{\frac{p}{T(\eta,\vec{x})}}-1}~.
\end{equation}
It can be expanded into a background part and a first-order perturbation, $f=\bar{f}+\delta f$, with:
\begin{equation}
\bar{f}(\eta,p) = \frac{1}{e^{\frac{p}{\bar{T}(\eta)}}-1}~, \qquad
\delta f(\eta,\vec{x},\vec{p}) =  \frac{d \bar{f}}{d \log p} \,\, \frac{\delta T(\eta,\vec{x})}{\bar{T}(\eta)}~.
\label{jl:f}
\end{equation}
We see that in the tightly-coupled regime we could replace the variable $f(\eta,\vec{x},\vec{p})$ by the lower-dimensional variable $\Theta(\eta,\vec{x}) \equiv [{\delta T(\eta,\vec{x})}/{\bar{T}(\eta)}]$. The Boltzmann equation leads to an equation of motion for $\Theta(\eta,\vec{x})$. 

At later times, when photons decouple, the shape of the Bose-Eisntein distribution of photons is preserved. This can be infered from the geodesic equation. This equation tells how the individual momentum $p$ of photons evolve when they are decoupled and they travel in the perturbed universe. In the Newtonian gauge, it reads
\begin{equation} 
\frac{d (a\, p)}{d \eta} = - a \, p\, \phi' - a \, \epsilon \, \hat{n}
\cdot \vec{\nabla}\psi~.
\end{equation}
The left-hand side represents the time evolution of the product $(ap)$ for a photon of momentum $p$ traveling over a geodesic. In a perfectly homogeneous universe, the photon would only experience the average cosmological redshifting, $p \propto a^{-1}$, and the product $(ap)$ would be conserved. Due to presence of perturbations in the universe, photons experience gravitational interactions and the product $(ap)$ varies. The first term $-a \, p \, \phi'$ accounts for dilation, i.e. for the fact that locally, the expansion of the universe is a bit more advanced or delayed than the average (remember that $a(1+\phi)$ can be seen as the ``local scale factor''). This means that the redshifting of the photon is also a bit more advanced or delayed locally. The second term accounts for the gravitational blueshifting of photons falling in gravitational potential wells (or redshifting  of those leaving potential wells).  The energy $\epsilon\equiv\sqrt{p^2+m^2}$ can be simply replaced by $p$ for massless photons. In that case we can rewrite the geodesic equation as
\begin{equation} 
\frac{d \ln (a\,p)}{d \eta} = - \phi' - \hat{n} 
\cdot \vec{\nabla}\psi~.
\end{equation}
The fact that $p$ disappeared from the right-hand side is crucial: it shows that photons of different momenta traveling in the perturbed universe along a given geodesic all experience the same relative momentum variation. The consequence is that there can be no distortions of the Bose-Einstein shape of $f$. However, photons traveling along different geodesics and in different directions experience different redshifting. Hence the distribution function acquires a dependence on one extra argument, the direction $\hat{n}$ of propagation ($\hat{n} \equiv \vec{p}/p$):
\begin{equation}
f(\eta,\vec{x},\vec{p}) = \frac{1}{e^{\frac{p}{T(\eta,\vec{x},\hat{n})}}-1}~.
\end{equation}
We can perform the same decomposition in terms of background and perturbations as in Eq.~(\ref{jl:f}). The difference is that $\Theta=\frac{\delta T}{\bar{T}}$ is now a function of $(\eta, \vec{x},\hat{n})$. The Boltzmann equation can now be used to derive an equation of motion for $\Theta(\eta, \vec{x},\hat{n})$. 

If we work in Fourier space, we can derive the equation of motion for the function $\Theta(\eta, \vec{k},\hat{n})$. Because of the statistical isotropy of the FL universe, this equation does not depend explicitly on  $\vec{k}$ nor $\hat{n}$, since there is no preferred direction: it depends only on the direction of propagation relatively to the considered wavenumber , i.e. on the product $(\vec{k}\cdot\hat{n})$. Hence the equation of motion can be written in terms of $k$ and of the angle $\theta$ such that  $(\vec{k}\cdot\hat{n})=k\cos\theta$. The initial conditions for $\Theta$ do depend on the wavevector $\vec{k}$ (since each mode gets random initial conditions), but they also depend only on $\theta$ rather than $\hat{n}$, for a reason that will become clear in the paragraphs below. Hence we can entirely eliminate $\hat{n}$ from the problem, and solve the equation of motion for $\Theta(\eta,\vec{k},\theta)$.

Finally, we can expand the temperature anisotropy with respect to $\theta$ using a Legendre transformation:
\begin{equation}
\Theta(\eta,\vec{k},\theta) = \sum_l (-i)^l (2l+1) \Theta_l(\eta,\vec{k}) P_l(\cos\theta)~.
\label{jl:legendre}
\end{equation}
Here the $\Theta_l$'s are the temperature anisotropy multipoles, and $P_l$ the Legendre polynomials. It can be shown that the monopole $\Theta_0$ is related to the photon density fluctuation $\delta_\gamma$ in a given point, the dipole $\Theta_1$ to its velocity divergence $\theta_\gamma$, and the quadrupole $\Theta_2$ to its anisotropic stress $\sigma_\gamma$. The Boltzmann equation can be written as an infinite hierarchy of equations of motion for the coupled multipoles $\Theta_l$.

Actually, the equation of motion for $\Theta$ takes a striking form when written in real space, before Fourier and Legendre expansions:
\begin{equation}
\Theta' + \hat{n} \cdot \vec{\nabla} \Theta - \phi' + \hat{n} \cdot \vec{\nabla} \psi = - \Gamma \, (\Theta - \Theta_0 - \hat{n} \cdot \vec{v}_e)~.
\label{jl:boltz1}
\end{equation}
We recall that $\Gamma$ is  the conformal Thomson scattering rate, $\Theta_0$ the temperature monopole (i.e. the average of $\Theta(\eta,\vec{x},\hat{n})$ over all directions $\hat{n}$), and $\vec{v}_e$ the bulk velocity of electrons, equal to that of baryons due to tight Coulomb interactions. We recall that the variable $\theta_\mathrm{b}$ is defined as the divergence of $\vec{v}_\mathrm{b}=\vec{v}_e$. 

This equation is illuminating, since it shows that at early times, when photons, electrons and baryons form a tightly-coupled fluid, the fact that $\Gamma$ is huge forces $\Theta$ to evolve in such way that the parenthesis on the right-hand side vanishes. When this is the case, $\Theta$ can only have two non-zero component: a monopole $\Theta_0$, and a dipole equal to $\hat{n} \cdot \vec{v}_\mathrm{b}$. The condition on the dipole can be written equivalently as $\theta_\gamma=\theta_\mathrm{b}$. 

This should remind us of the discussion of Sec.~\ref{jl:sec13}, when we noticed that a perfect fluid can be described in terms of $\delta_x$, $c_\mathrm{s}^2$ and $\theta_x$ only, with a vanishing anisotropic stress $\sigma_x$. Here, we see concretely how this conclusion emerges from the Boltzmann equation in the tightly-coupled regime: $\Theta_2$ (related to $\sigma_\gamma$) and all higher multipoles vanish; the photon perturbations can be described in terms of only two independent variables $\Theta_0$ and $\Theta_1$ (or $\delta_\gamma$ and $\theta_\gamma$), like in a perfect fluid.

The fact that in the tightly-coupled limit the system is driven towards $\theta_\gamma=\theta_\mathrm{b}$ simply shows that the interaction imposes a common bulk velocity to photons, baryons and electrons, as it should be the case in any tightly-coupled fluid (note that we are referring to bulk velocities, not to individual velocities of particles, which are still equal to $c$ for interacting photons).

In the tightly-coupled regime, the dipole component of $\Theta(\eta,\vec{x},\hat{n})$ is given by $\hat{n} \cdot \vec{v}_\mathrm{b}$, with $\theta_\mathrm{b} \equiv \vec{\nabla} \cdot \vec{v}_\mathrm{b}$. Hence in Fourier space this component reads $i (\hat{n} \cdot \vec{k}) k^{-2}  \theta_\mathrm{b}=i(\cos\theta) k^{-1} \theta_\mathrm{b}$. This justifies the fact that initial conditions for $\Theta$ in Fourier space depend on $\theta$, but not on the two degrees of freedom of $\hat{n}$. As mentioned above, this property is preserved by the isotropic equations of motion, so that at all times we can study $\Theta$ as a function of the arguments $(\eta,\vec{k},\theta)$ instead of $(\eta,\vec{k},\hat{n})$.

\subsection{Temperature anisotropy in a given direction}\label{jl:sec23}

The map of temperature anisotropies that we observe today ($\eta=\eta_0$) in our location of the universe ($\vec{x}=\vec{o}$ with a proper choice of origin) when looking in a direction $\hat{n}$ is represented mathematically by
\begin{equation}
\frac{\delta T}{\bar{T}}(\hat{n}) = \Theta(\eta, \vec{o},-\hat{n})
\end{equation}
(since in a direction $\hat{n}$, we see photons traveling towards $-\hat{n}$). Our scope is now to relate this quantity to perturbations on the point of the last scattering surface seen in the same direction $\hat{n}$. This can be done by integrating the Boltzmann equation along the corresponding line-of-sight.

A good starting point consists in computing the total derivative of the product $e^{\tau}(\Theta + \psi)$ along the trajectory of photons between the last scattering surface and the observer. The reason for choosing this product rather than just $\Theta$ will become clear in a few lines: the derivative of this term will be easy to simplify, using the Boltzmann equation.

The total derivative of an arbitrary function ${\cal F}$ of $(\eta,\vec{x},\hat{n})$ along the trajectory of photons going in a direction $\hat{n}$ reads
\begin{equation}
\frac{d}{d\eta} {\cal F}(\eta,\vec{x},\hat{n}) = {\cal F}' + \frac{dx_i}{d \eta} \frac{\partial {\cal F}}{\partial x_i} + \frac{dn_i}{d \eta} \frac{\partial {\cal F}}{\partial n_i}~.
\end{equation}
If ${\cal F}$ is of order one in perturbations, the first two terms on the right-hand side are also of order one. Instead the last term is of order two, since $\frac{dn_i}{d \eta}$ is of order one (this is clear from the fact that in an unperturbed universe, photons would travel in straight line with $\frac{dn_i}{d \eta}=0$). Hence we can drop this term in first-order perturbation theory, and for $\frac{dx_i}{d \eta}$ we only need to keep the zero-th order contribution, that can be computed assuming a homogeneous universe:
\begin{equation}
\frac{dx_i}{d \eta} = \hat{n}~.
\end{equation}
The previous relation is just telling that photons are traveling in the direction $\hat{n}$ and at the velocity of light: hence, in units where $c=1$, $d \vec{x}^2 = d \eta^2$. In summary, the total derivative of ${\cal F}$ is given at first order by
\begin{equation}
\frac{d}{d\eta} {\cal F}(\eta,\vec{x},\hat{n}) = {\cal F}' + \hat{n} \cdot \vec{\nabla} {\cal F}~.
\end{equation}
Let us now replace the generic function ${\cal F}$ by 
\begin{equation}
{\cal F}(\eta,\vec{x},\hat{n}) = e^{-\tau(\eta)}\left(\Theta(\eta,\vec{x},\hat{n}) + \psi(\eta,\vec{x})\right)~.
\end{equation}
The total derivative of this function reads
\begin{equation}
\frac{d}{d \eta} \left[ e^{-\tau} (\Theta+\psi)\right] = e^{-\tau} \left(\Theta' + \psi' + \hat{n} \cdot \vec{\nabla} (\Theta+\psi)\right) - \tau' e^{-\tau} (\Theta+\psi)~.
\end{equation}
We now use the linearized  Boltzmann equation~(\ref{jl:boltz1}) and the fact that $\tau'=-\Gamma$ to write the result as
\begin{equation}
\frac{d}{d \eta} \left[ e^{-\tau} (\Theta+\psi)\right] = - e^{-\tau} \tau' (\Theta_0 + \psi + \hat{n} \cdot \vec{v}_\mathrm{b}) + e^{-\tau} (\phi'+\psi')~.
\end{equation}
Finally, using the definition of the visibility function $g$ given in Sec.~\ref{jl:sec21}, we get
\begin{equation}
\frac{d}{d \eta} \left[ e^{-\tau} (\Theta+\psi)\right] = g \left(\Theta_0 + \psi + \hat{n} \cdot \vec{v}_\mathrm{b}\right) + e^{-\tau} (\phi'+\psi')~.
\end{equation}
We can integrate this relation along the line of sight, i.e. along a straight line seen by the observer in a given direction $-\hat{n}$ (since the photons go in the direction $\hat{n}$), starting from an early time {\it before} recombination (such that $e^{-\tau(\eta_\mathrm{ini})}\simeq 0$) until the present time at which photons reach the observer (with by definition $e^{-\tau(\eta_0)}=1$). The result reads
\begin{equation}
(\Theta+\psi)|_\mathrm{obs} = \int_{\eta_\mathrm{ini}}^{\eta_0} d \eta \left[
g \left(\Theta _0 + \psi + \hat{n} \cdot \vec{v}_\mathrm{b}\right) + e^{-\tau} (\phi'+\psi')\right]~,
\label{jl:los_nid}
\end{equation}
where the notation $|_\mathrm{obs}$ means ``evaluated at the observer location, along this line of sight'', i.e. at the coordinate $(\eta_0, \vec{o}, \hat{n})$.

We can gain further intuition from this equation if we use the instantaneous decoupling approximation, in which all photons are assumed to decouple precisely at the time $\eta_\mathrm{dec}$. In this limit, we can replace the visibility function $g$ by the Dirac function $\delta_D(\eta-\eta_\mathrm{dec})$, and $e^{-\tau}$ by the Heaviside function $H(\eta-\eta_\mathrm{dec})$. Note that the approximation $g(\eta)=\delta_D(\eta-\eta_\mathrm{dec})$ is correctly normalized, since the definition of $g$ implies $\int d\eta \, g(\eta) = 1$. In  the instantaneous decoupling limit, Eq.~(\ref{jl:los_nid}) reads:
\begin{equation}
(\Theta+\psi)|_\mathrm{obs} = \left( \Theta _0 + \psi + \hat{n} \cdot \vec{v}_\mathrm{b})\right)|_\mathrm{dec} + \int_{\eta_\mathrm{dec}}^{\eta_0} \!\!\! d\eta \, (\phi'+\psi')~,
\end{equation}
where the notation $|_\mathrm{dec}$ means ``evaluated on the last scattering surface, along this line of sight'', i.e. at the coordinate $(\eta_\mathrm{dec}, -r_\mathrm{dec}\hat{n}, \hat{n})$ (here $r_\mathrm{dec}$ is the comoving radius of the last scattering surface). Let us now give the interpretation of each term in this crucial equation.

First, $\Theta|_\mathrm{obs}$ is the temperature anisotropy measured by the observer in the direction $-\hat{n}$, while $\Theta _0|_\mathrm{dec}$ is the temperature anisotropy in the point of the last scattering surface seen in the same direction. If only these two terms where present, this relation would simply tell us that that the temperature anisotropy seen today in a given direction is equal to the intrinsic anisotropy in the point where the observed photons last scattered.

Second, the term $\hat{n} \cdot \vec{v}_\mathrm{b}|_\mathrm{dec}$ stands for the correction to this temperature coming from by the usual Doppler. Indeed, this correction is caused by the velocity of the baryon-photon fluid (that we assumed to be tightly coupled until $\eta_\mathrm{dec}$) projected along the line of sight.

Next, we expect a correction from gravitational effects. The redshifting and blueshifting of the photons traveling along gravitational potential fluctuations should affect the observed temperature anisotropy relatively to the intrinsic one. It turns out that if the gravitational potential was constant in time (but, of course, not in space), this effect would be conservative, and would only depend on $\psi|_\mathrm{obs}-\psi|_\mathrm{dec}$. This explains the second term on the left-hand and right-hand sides. But if $\psi$ varies in time, the effect is not conservative anymore: intuitively, the amount of blueshifting and redshifting experienced by photons traveling across a potential well do not compensate each other if the gravitational well gets deeper between the time at which the photon enters and leaves the well. This explains the addition term $\int d\eta \, \psi$. A similar effect is caused by dilation effects along the line-of-sight, and contributes like $\int d\eta \, \phi$.

Finally, we can drop the second term on the left-hand side, because this term represents only a tiny {\it isotropic} correction to the observed anisotropies. It is impossible to measure it with the CMB map only, because it is formally equivalent to a redefinition of the average temperature $\bar{T}$, but only by a tiny amount of the order of $10^{-5} \bar{T}$.

Let us write once more our result, dropping this unobservable correction, and grouping the terms in a particular way:
\begin{equation}
\Theta|_\mathrm{obs} = \mathop{\underbrace{(\Theta _0 + \psi) |_\mathrm{dec}}}_\mathrm{SW} + \mathop{\underbrace{\hat{n} \cdot \vec{v}_\mathrm{b} |_\mathrm{dec}}}_\mathrm{Doppler} + \mathop{\underbrace{\int_{\eta_\mathrm{dec}}^{\eta_0} \!\!\! d\eta \, (\phi'+\psi')}}_\mathrm{ISW}~.
\label{jl:realspace}
\end{equation}
The first term is conventionally called the Sachs-Wolfe (SW) term, and includes the intrinsic temperature term $\Theta_0$ and the ``gravitational Doppler shift'' term $\psi$ at one point on the last scattering surface. The second term is the conventional Doppler term. The last term is called the Integrated Sachs-Wolfe (ISW) term and contains all non-conservative gravitational effects occurring in a universe with non-static metric fluctuations.

We can gain further insight on the Sachs-Wolfe term. We will try to find a simpler expression for $ (\Theta _0 + \psi) |_\mathrm{dec}$, that applies at least for describing large angular patterns on CMB maps, i.e. maps smoothed over small scales. For instance, this expression would describe very well the map of the COBE satellite, which had limited angular resolution. In more precise terms, we wish to calculate the contribution to the Sachs-Wolfe term of large wavelengths, which are bigger than the Hubble radius at the time of recombination.

Let us first focus on the term $\Theta_0 |_\mathrm{dec}$. We have seen that on super-Hubble scales, temperature anisotropies only have a monopole and a dipole component, related respectively to $\delta_\gamma$ and $\theta_\gamma$. We can be more precise now. We know from thermodynamics that the local value of the photon density is proportional to the temperature to the power four. Taking the derivative of $\log \rho_\gamma = \log T^4$, we get $\delta_\gamma = 4 \, \delta T/\bar{T}$, where on the right-hand side the temperature anisotropy is averaged over all directions $\hat{n}$: hence $\delta_\gamma = 4 \Theta_0$. 

Let us now focus on the term $\psi |_\mathrm{dec}$. We have seen in Sec.~\ref{jl:sec14} that on super-Hubble scales and for adiabatic initial conditions, $\delta_\gamma = \frac{4}{3} \delta_\mathrm{b}$. Also, we know that decoupling takes place at the beginning of matter domination, and we mentioned at the very end of section~\ref{jl:sec14} that for super-Hubble scales and during matter domination, one has $-2 \phi= -2 \psi= \delta_\mathrm{tot} = \delta_\mathrm{b} = \delta_\mathrm{cdm}$. Hence, $-2 \psi = \frac{3}{4} \delta_\gamma$. Putting all these equalities together, we conclude that on the last scattering surface,
\begin{equation}
\Theta _0 + \psi  = \frac{1}{4} \delta_\gamma + \psi = (- \frac{1}{4} \,2\, \frac{4}{3} + 1) \,\psi = (-\frac{2}{3}+1)\, \psi = \frac{1}{3} \psi~.
\end{equation} 

Moreover, on super-Hubble scales, we can neglect the Doppler term, which can be shown to be important only on sub-Hubble scales. We will see later that the integrated Sachs-Wolfe term plays a role on large scales, but only a small role, because $\phi$ and $\psi$ are static during most of the evolution after decoupling. Hence we get an approximation for CMB anisotropies smoothed over small scales (that was first derived by Sachs and Wolfe in 1967):
\begin{equation}
\Theta|_\mathrm{obs, \, large \, scales} \simeq  \frac{1}{3} \psi |_\mathrm{dec} = - \frac{1}{8} \delta_\gamma |_\mathrm{dec}~.
\label{jl:SW}
\end{equation}
In this calculation, we have seen that the term $\psi$ wins over the term $\Theta_0$, leading to a minus sign in front of $\delta_\gamma$ in the above relation. This means that an over-density on the last scattering surface ($\delta_\gamma>0$), corresponding to a potential well ($\psi<0$), leads to a cold spot in the observed map ($\Theta<0$). Conversely, a hot spot corresponds to an under-density. Hence, due to the ``gravitational Doppler shift'' effect accounted by the term $\psi$ (often called the Sachs-Wolfe effect), the patterns that we observe on CMB map are inverted with respect to intrinsic fluctuations on the last scattering surface.

Equation~(\ref{jl:realspace}) (and its large-scale approximation~(\ref{jl:SW})) are important for pedagogical purposes, but they have no practical application: indeed, what we wish to calculate and to compare to observations is a theoretical prediction for the statistical properties of CMB anisotropies. Hence, we need to compute at least the CMB two-point correlation function.

\subsection{Spectrum of temperature anisotropies}\label{jl:sec24}

{\bf \it Definition.} The map of CMB temperature anisotropies can be expanded in spherical harmonics:
\begin{equation}
\frac{\delta T}{\bar{T}}(\hat{n}) = \Theta(\eta_0, \vec{o},-\hat{n}) = \sum_{lm} a_{lm} Y_{lm} (\hat{n})~.
\end{equation}
Using the Legendre expansion of $\Theta$ introduced in Eq.~(\ref{jl:legendre}), and some basic relations between Legendre polynomials and spherical harmonics, it is easy to  express $a_{lm}$ as a function of $\Theta_l$:
\begin{equation}
a_{lm} = (-i)^l \int\frac{d^3\vec{k}}{2\pi^2} Y_{lm}(\hat{k})  \Theta_l(\eta_0,\vec{k})~,
\end{equation}
where we recall that hats denote unit vectors: $\hat{k} \equiv \vec{k}/k$. In linear perturbation theory and assuming Gaussian initial conditions, both $\Theta_l$ and $a_{lm}$ are Gaussian random variables. Using the orthogonality relation of spherical harmonics and the definitions given in Sec.~\ref{jl:sec15}, we can infer the two-point correlation function of the $a_{lm}$'s as a function of the power spectrum of $\Theta_l$, or even better, of the primordial curvature power spectrum:
\begin{equation}
\langle a_{lm} a_{l'm'}^* \rangle = \delta_{ll'}^K \delta^K_{mm'} \left[ \frac{1}{2 \pi^2} \int \frac{dk}{k} \Theta_l^2(\eta_0,k) {\cal P}_{\cal R} (k) \right]~.
\label{jl:alm_spec}
\end{equation}
Here $\delta^K_{ll'}$ represents the Kronecker symbol. The fact that $\langle a_{lm} a_{l'm'}^* \rangle$ vanishes for $l\neq l'$ or $m\neq m'$ comes out of the algebra, but physically, it is a consequence of the homogeneity of the universe --- just like the fact that all power spectra in Fourier space are proportional to $\delta_D(\vec{k}'-\vec{k})$. Similarily, the fact that the quantity between brackets is a function of $l$ but not of $m$ is a consequence of isotropy --- like the fact that in Fourier space, power spectra are functions of $k$ but not $\vec{k}$.

The quantity between  brackets is usually denoted $C_l$, and is called the power spectrum of temperature anisotropies in harmonic space, or the temperature harmonic power spectrum: 
\begin{equation}
C_l \equiv \frac{1}{2 \pi^2} \int \frac{dk}{k} \Theta_l^2(\eta_0,k) {\cal P}_{\cal R} (k)~.
\label{jl:cl_def}
\end{equation}
In a universe with linear and Gaussian perturbations, the $C_l$'s encode all the information concerning the cosmological model describing our universe that is contained in the CMB temperature map.

It is worth coming back on the meaning of the averaging symbols $\langle ... \rangle$ in Eq.~(\ref{jl:alm_spec}). Since the theory of cosmological perturbation is stochastic, the $a_{lm}$'s should be seen as random numbers, and the average is meant over many realizations of the theory. In a sense, ``a given realization'' means ``a given universe'', and the average holds over many universes, all obeying to the same cosmological model, which is encoded in the spectrum $C_l$.

\begin{figure}
\begin{center}
\includegraphics[width=6cm]{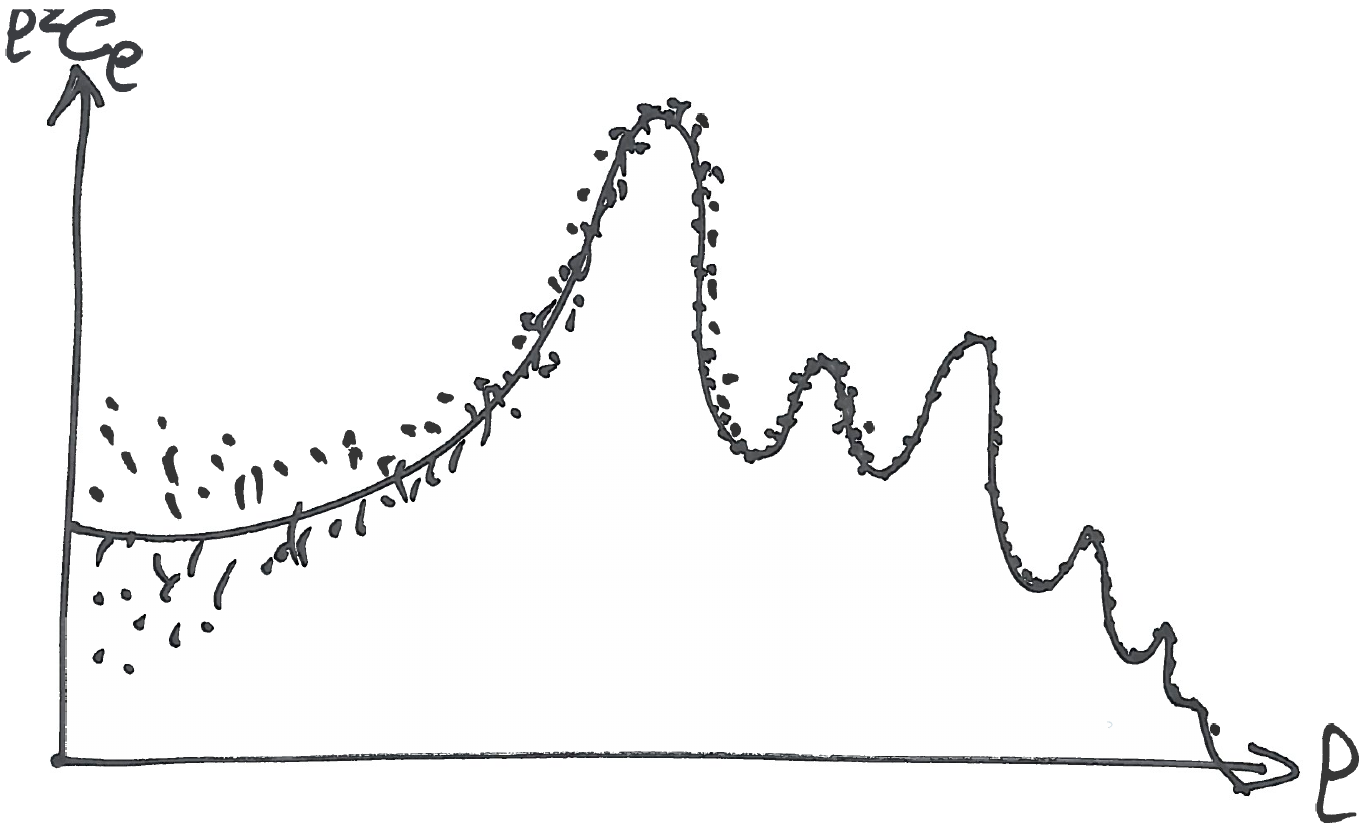}
\end{center}
\caption{Observed values of the temperature harmonic spectrum, $C_l^{obs}$, are expected to be scattered around the true underlying $C_l$'s. The scattering, called cosmic variance, decreases with increasing $l$.}
\label{jl:fig24a}
\end{figure}

However, we observe CMB anisotropies in our universe, i.e. in only one realization. CMB maps allow us to measure a definite value of each $a_{lm}$. Hence the squares $|a_{lm}|^2$ are not expected to be equal to $C_l$, even if we postulated the right model: there should be some scattering around $C_l$. This scattering limits the possibility to find the best theory matching  the observations. However we can reduce considerably the scattering by noticing that for any fixed $l$, the statistical distribution of $|a_{lm}|^2$ is independent of $m$ (as a consequence of isotropy, and as expressed by Eq.~(\ref{jl:cl_def})). Hence, for an ideal full-sky CMB experiment, the best estimator of the underlying $C_l$'s is the average between all observed coefficients $|a_{lm}|^2$ with fixed $l$,
\begin{equation}
C_l^{obs} \equiv \frac{1}{2l+1} \sum_{-l \leq m \leq l} | a_{lm}^{obs} |^2~.
\label{jl:cl_obs}
\end{equation}
In a typical universe, this quantity should be closer to the underlying $C_l$ than a single $|a_{lm}^{obs}|^2$. The way to see this mathematically is to consider again the theoretical (stochastic) $a_{lm}$'s, and to define
\begin{equation}
\hat{C}_l \equiv \frac{1}{2l+1} \sum_{-l \leq m \leq l} | a_{lm} |^2~.
\end{equation}
By performing averages in the same sense as in equation~(\ref{jl:alm_spec}), one can easily show that for Gaussian $a_{lm}$'s,
\begin{equation}
\langle \hat{C}_l \rangle = C_l
\qquad \mathrm{and} \qquad
\langle (\hat{C}_l-C_l)^2 \rangle = \frac{2}{2l+1} C_l^2~.
\end{equation}
The first equality shows that the observed $C_l$'s defined in Eq.~(\ref{jl:cl_obs}) are unbiased estimators of the true underlying $C_l$'s. The second equality gives the typical scattering between the theory and the observations. Since for larger $l$ we perform an average over more values of $m$, the relative scattering decreases. This is the case with true data points, which are distributed qualitatively like in Fig.~\ref{jl:fig24a}. This scattering is called cosmic variance. It can be seen as a theoretical error: because of cosmic variance, we cannot reconstruct the underlying model with infinite precision, even if we have infinitely precise observations. Cosmic variance is large for small $l$'s, meaning that the shape of the true underlying $C_l$'s will always be poorly known at low $l$.

\vspace{0.2cm}

\noindent {\bf \it Line-of-sight integral in Fourier space.} According to equation~(\ref{jl:cl_def}), for a given primordial spectrum, the shape of the CMB spectrum $C_l$ depends on the square of the transfer function $\Theta_l(\eta_0,k)$. We would like to understand this shape at least qualitatively. In real space, we did learn a lot on the behavior of $\Theta(\eta,\vec{x}, \hat{n})$ by using the line-of-sight integral approach presented in section~\ref{jl:sec23}. A similar approach can be worked out in Fourier and harmonic space, i.e. for the variable $\Theta_l(\eta, k)$. We do not present here the intermediate steps, which can be found in Ref.~\refcite{Seljak:1996is}. The final result shows many similarity with its real space counterpart, Eq.~(\ref{jl:los_nid}). It can be decomposed into:
\begin{eqnarray}
\Theta_l (\eta_0,k) &=& \int_{\eta_\mathrm{ini}}^{\eta_0} d \eta \,\, S_T(\eta,k) \,\, j_l(k(\eta_0-\eta))~,\nonumber \\
S_T(\eta,k) &\equiv& \mathop{\underbrace{g\left(\Theta_0 + \psi\right)}}_\mathrm{SW} 
+ \mathop{\underbrace{\left(g \, k^{-2} \theta_\mathrm{b}\right)'}}_\mathrm{Doppler}  
+ \mathop{\underbrace{e^{-\tau} (\phi'+\psi')}}_\mathrm{ISW}~.
\label{jl:los_fou}
\end{eqnarray}
We see that $\Theta_l(\eta_0,k)$ is given by the convolution of spherical Bessel functions $j_l(x)$ with a function $S_T(\eta,k)$, called the temperature source function, which contains the usual three terms: Sachs-Wolfe, Doppler, and Integrated Sachs-Wolfe. Like in the previous section, we can use the instantaneous decoupling approximation, integrate the Doppler term by part, and write $\Theta_l(\eta_0,k)$ as:
\begin{eqnarray}
\Theta_l (\eta_0,k) &\simeq&\left[\Theta_0(\eta_\mathrm{dec},k) + \psi(\eta_\mathrm{dec},k) \right] j_l(k(\eta_0-\eta_\mathrm{dec}))\nonumber \\
&& + k^{-1} {\theta_\mathrm{b}}(\eta_\mathrm{dec},k)\,\, j_l' (k(\eta_0-\eta_\mathrm{dec})) 
\nonumber \\
&&
+ \int_{\eta_\mathrm{dec}}^{\eta_0} d \eta \,\, \left[\phi'(\eta,k)+\psi'(\eta,k)\right]  j_l(k(\eta_0-\eta))
\label{jl:theta_ida}
\end{eqnarray}
(note that in the second line, i.e. in the Doppler term, the prime stands for the derivative of the function $j_l(x)$ with respect to its argument, not with respect to conformal time). This approximate result can be plugged into Eq.~(\ref{jl:cl_def}) to obtain the final spectrum $C_l$. We see that each $C_l$ can be decomposed into six terms: the power spectrum $C_l^\mathrm{SW}$ of the SW term, coming from the first line of Eq.~(\ref{jl:theta_ida}) squared, that of the Doppler term, coming from the second line of Eq.(\ref{jl:theta_ida}) squared, that of the Integrated Sachs-Wolfe term, coming from the third line of Eq.~(\ref{jl:theta_ida}) squared, and finally the three cross-spectra involving each pair of terms.

\begin{figure}
\begin{center}
\includegraphics[width=9cm]{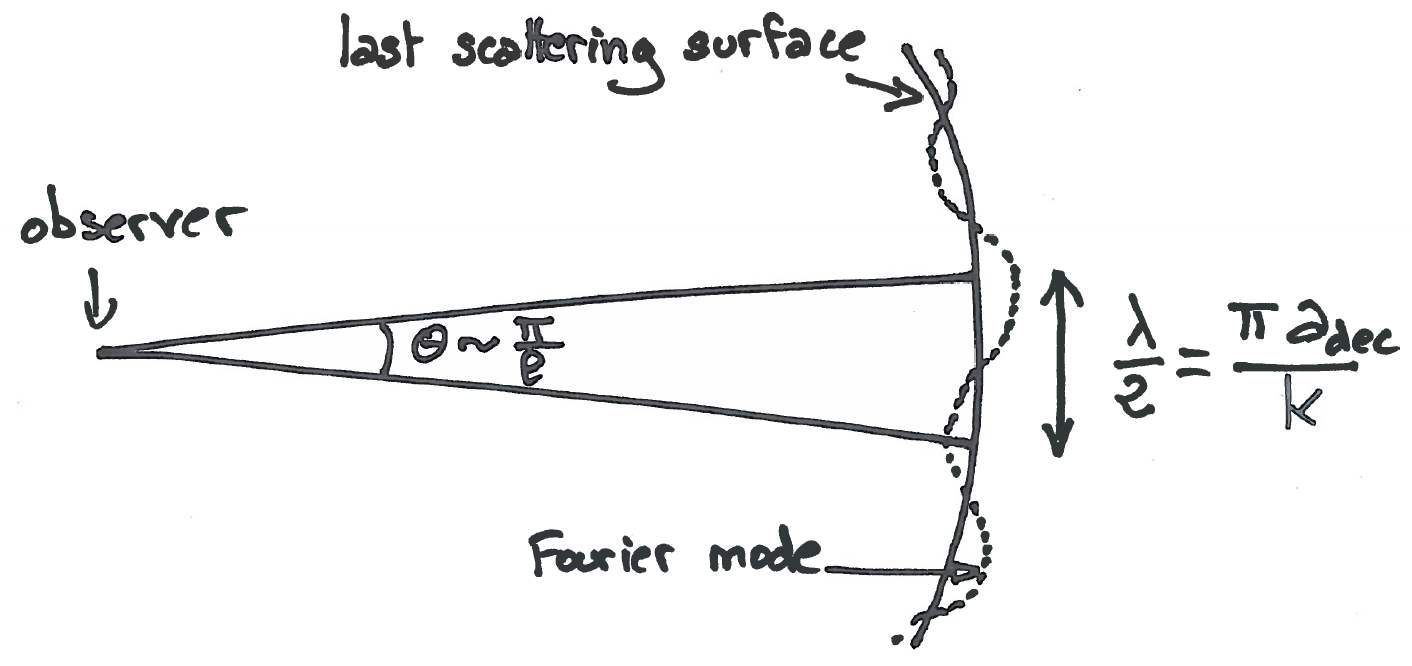}
\end{center}
\caption{A multipole $l$ refers to structures seen on the last scattering surface under an angle $\theta=\pi/l$. These structures are seeded by Fourier modes with a half wavelength $\frac{\lambda}{2}=\frac{\pi a_\mathrm{dec}}{k}$.}
\label{jl:fig24b}
\end{figure}

For large values of $l$, the spherical Bessel functions $j_l(x)$ and $j_l'(x)$ are very peaked near $x \simeq l$. Hence, for the Sachs-Wolfe and Doppler contributions to the spectrum $C_l$, the integral over $k$ in Eq.~(\ref{jl:cl_def}) will pick up mainly modes with $k(\eta_0-\eta_\mathrm{dec}) \simeq l$. This shows that the SW contribution to $C_l$ is given by the product of the primordial spectrum with the squared transfer function $(\Theta_0 + \psi)$ at a given value of $\eta$ and $k$:
\begin{equation}
C_l^\mathrm{SW} \sim \left[\Theta_0(\eta_\mathrm{dec},k) + \psi(\eta_\mathrm{dec},k) \right]^2 {\cal P}_{\cal R} (k)~,
\qquad
k \simeq \frac{l}{(\eta_0-\eta_\mathrm{dec})}
\label{jl:clsw}
\end{equation}
(for simplicity, we did not write numerical factors and powers of $l$ or $k$ in front of this expression). In other words, $C_l^\mathrm{SW}$ depends on the power spectrum of the perturbation $(\Theta_0 + \psi)$, evaluated at the time of decoupling, and for wavenumbers in the vicinity of $k=\frac{l}{(\eta_0-\eta_\mathrm{dec})}$,
\begin{equation}
C_l^\mathrm{SW} \sim \langle \left| \Theta_0 + \psi \right|^2 \rangle_{(\eta,k) \simeq \left(\eta_\mathrm{dec}, \,\, {l}/{(\eta_0-\eta_\mathrm{dec})}\right)}~.
\label{jl:clsw2}
\end{equation}
We reached this result with mathematical arguments, but it has a very simple geometrical interpretation, illustrated in Fig.~\ref{jl:fig24b}. The spectrum $C_l$ encodes the correlation between structures on CMB maps seen under an angle $\theta=\pi/l$. This angle subtends a given physical scale on the last scattering surface, namely $\theta \times d_\mathrm{a}(z_\mathrm{dec})$, where $d_\mathrm{a}(z)$ is the angular diameter distance to objects of redshift $z$. Since the Sachs-Wolfe term $(\Theta_0 + \psi)$ contributes to the temperature map only at the time $\eta \simeq \eta_\mathrm{dec}$, the spectrum $C_l^\mathrm{SW}$ should depend on the power spectrum of the Sachs-Wolfe term $\langle |\Theta_0 + \psi|^2 \rangle$ at that time, and for a wavenumber such that
\begin{equation}
\frac{\lambda}{2} = \frac{\pi a(\eta_\mathrm{dec})}{k} = \theta \, d_\mathrm{a}(z_\mathrm{dec})~.
\label{jl:dadec0}
\end{equation}
The reason for which $\lambda$ has been divided by two is that for spherical harmonics, $\theta=\pi/l$ is the angle between a maximum and a minimum, while for a Fourier mode the distance between a maximum and a minimum is one half of the wavelength, $\frac{\lambda}{2}=\frac{\pi a}{k}$. Eq.~(\ref{jl:dadec0}) leads to
\begin{equation}
\frac{a(\eta_\mathrm{dec})}{k} = \frac{d_\mathrm{a}(z_\mathrm{dec})}{l}~.
\label{jl:dadec}
\end{equation}
We recall that in a flat FL universe the angular diameter distance is given by
\begin{equation}
d_\mathrm{a}(z) = a(t(z)) \int_{t(z)}^{t_0} \frac{dt}{a}~,
\end{equation}
$t(z)$ being the proper time at which an object seen today with a redshift $z$ emitted light. The conformal time $\eta(z)$ is defined similarly. In terms of conformal time,
\begin{equation}
d_\mathrm{a}(z) = a(\eta(z)) \int_{\eta(z)}^{\eta_0} d \eta = a(\eta(z)) \,\, [\eta_0-\eta(z)]~,
\end{equation}
and for the case of a point located on the last scattering surface,
\begin{equation}
d_\mathrm{a}(z_\mathrm{dec}) = a(\eta_\mathrm{dec})  \,\, (\eta_0-\eta_\mathrm{dec})~.
\end{equation}
Hence, Eq.~(\ref{jl:dadec}) can be written as
\begin{equation}
\frac{1}{k} = \frac{(\eta_0-\eta_\mathrm{dec})}{l}~,
\label{jl:cor_kl}
\end{equation}
which is the same relation between $k$ and $l$ as in Eqs.~(\ref{jl:clsw}, \ref{jl:clsw2}). In those equations, we implicitly performed a small-angle approximation. For large angles (small $l$'s), it is inaccurate to say that a given angle/mutipole corresponds to a single Fourier mode on the last-scattering surface, and it is important to keep the spherical Bessel function of Eq.~(\ref{jl:los_fou}) and the integral over $k$ of Eq.(\ref{jl:cl_def}). In summary, Eqs.~(\ref{jl:clsw}, \ref{jl:clsw2}) represent the instantaneous decoupling {\it and} small-angle limit of the true power spectrum $C_l^\mathrm{SW}$.

Using the same two limits, a similar discussion can be carried for the Doppler and Integrated Sachs-Wolfe power spectra. The Doppler term depends on the power spectrum of the baryon velocity divergence evaluated roughly at the same time and scale,
\begin{equation}
C_l^\mathrm{Doppler} \sim \langle \left| \theta_\mathrm{b}\right|^2 \rangle_{(\eta,k) \simeq \left(\eta_\mathrm{dec}, \,\, {l}/{(\eta_0-\eta_\mathrm{dec})}\right)}~,
\label{jl:cld}
\end{equation}
while the ISW term can be written approximately in terms of the integral 
\begin{equation}
C_l^\mathrm{ISW} \sim \int_{\eta_\mathrm{dec}}^{\eta_0} d\eta\, (\eta_0-\eta) \,\, \langle \left| \phi' + \psi' \right|^2 \rangle_{(\eta,k) \simeq \left(\eta, \,\, {l}/{(\eta_0-\eta)}\right)}~.
\label{jl:clisw}
\end{equation}
Again, for simplicity, we did not write numerical factors and powers of $l$ and $k$ in front of these expressions.
 
\vspace{0.2cm}
 
In  the next sections, we will infer the shape of the full spectrum $C_l$ from that of the three power spectra appearing in Eqs.~(\ref{jl:clsw2}, \ref{jl:cld}, \ref{jl:clisw}):
\begin{center}
\begin{tabular}{llcl}
SW&:&$\langle \left| \Theta_0 + \psi \right|^2 \rangle$&at ${(\eta,k) \simeq \left(\eta_\mathrm{dec}, \,\, {l}/{(\eta_0-\eta_\mathrm{dec})}\right)}~,$ \\
Doppler&:&$\langle \left| \theta_\mathrm{b}\right|^2 \rangle$& at ${(\eta,k) \simeq \left(\eta_\mathrm{dec}, \,\, {l}/{(\eta_0-\eta_\mathrm{dec})}\right)}~,$ \\
ISW&:&$\langle \left| \phi' + \psi' \right|^2 \rangle$&for all ${(\eta,k) \simeq \left(\eta, \,\, {l}/{(\eta_0-\eta)}\right)}~.$\\
\end{tabular}
\end{center}
Before entering into details, we can make a guess. Even if the primordial spectrum ${\cal P}_{\cal R}(k)$ is smooth, the temperature spectrum $C_l$ should contain some structure. 
\begin{figure}
\begin{center}
\includegraphics[width=7cm]{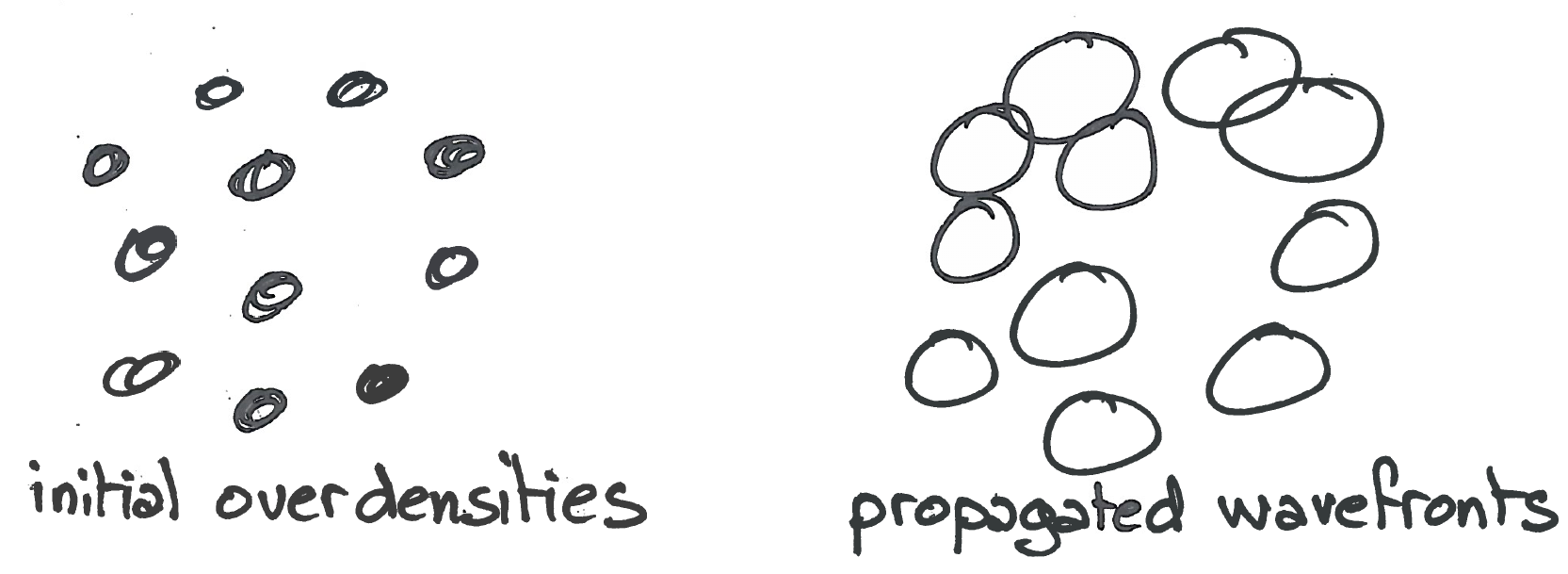}
\end{center}
\caption{Initial over-densities in the early universe propagate in the form of wavefronts. The distance travelled by any wavefront at a given time is given by the sound horizon in the photon-baryon fluid. Here, we represent initial over-densities as spherical patterns at a given scale, while in the real universe primordial over-densities result from a superposition of structures on all scales.}
\label{jl:fig24c}
\end{figure}
Indeed, as illustrated by Fig.~\ref{jl:fig24c}, any primordial over-density is expected to propagate. In Fig.~\ref{jl:fig24c}, we represented naively the primordial over-densities with little dots, giving rise to spherical wavefronts at later time. In the real universe, wavefront patterns are less visible, because primordial over-densities result from a superposition of structures on all scales. However, there is always a characteristic scale in this problem: namely, the distance by which a wavefront travels between some time in the primordial universe and the time of photon decoupling. This distance, called the sound horizon at decoupling $d_\mathrm{s}(\eta_\mathrm{dec})$, obeys to
\begin{equation}
d_\mathrm{s} \equiv a \int^{t}_{t_\mathrm{ini}} \frac{c_\mathrm{s} dt}{a} = a \int^{\eta}_{\eta_\mathrm{ini}} c_\mathrm{s} d\eta~, 
\end{equation}
where $c_\mathrm{s}$ is the sound speed in the photon-baryon fluid (in units of the speed of light). Two points on the last scattering surface separated by this distance should be partially correlated, since density waves have propagated from one point to the other. Hence, in angular space, the two-point correlation function of CMB anisotropies should exhibit a characteristic feature for angular scales corresponding to the sound horizon at decoupling, $\theta \sim d_\mathrm{s}(\eta_\mathrm{dec})/d_\mathrm{a}(z_\mathrm{dec})$. Similarly, the harmonic power spectrum $C_l$ should exhibit a feature at the corresponding scale, $l\sim \pi/\theta \sim \pi \,\, d_\mathrm{a}(z_\mathrm{dec}) / d_\mathrm{s}(\eta_\mathrm{dec})$, and also for all the harmonics of this scale. We will get a confirmation of this in the next section.

\subsection{Acoustic oscillations}\label{jl:sec25}

As long as electrons, baryons and photons are tightly coupled, they form an effective single fluid in which density waves propagate at the sound speed
\begin{equation}
c_\mathrm{s}^2 = \frac{\delta p_\gamma + \delta p_\mathrm{b}}{\delta \rho_\gamma + \delta \rho_\mathrm{b}}
\end{equation}
(the density and pressure of electrons is always negligible with respect to that of photons). The density fluctuation $\delta_x$ of each species $x=\gamma, b$ can be inferred from the local value of the equilibrium temperature. The fact that $\rho_\mathrm{b} \propto T^3$ and $\rho_\gamma \propto T^4$ implies $\delta_\gamma = \frac{4}{3} \delta_\mathrm{b}$, and tight coupling imposes $\theta_\gamma=\theta_\mathrm{b}$, as we already saw in Sec.~\ref{jl:sec22}. We can simplify the expression of the sound speed, using also the fact that $| \delta p_\mathrm{b} | \ll | \delta p_\gamma | $. The result reads
\begin{equation}
c_\mathrm{s}^2 = \frac{1}{3(1+R)}~, \quad R \equiv \frac{4 \bar{\rho}_\mathrm{b}}{3 \bar{\rho}_\gamma} \propto a~.
\end{equation}
It is possible to derive a simple equation of motion for the photon temperature fluctuation $\Theta_0(\eta,\vec{x})$ in the tightly-coupled regime:
\begin{equation}
\Theta_0''+\frac{R'}{1+R} \Theta_0' + k^2 c_\mathrm{s}^2 \Theta_0 = - \frac{k^2}{3} \psi + \frac{R'}{1+R} \phi' + \phi''~.
\label{jl:osc_eq}
\end{equation}
This equation follows from the combination of the continuity and Euler equations for photons and baryons. Given that $R$ is proportional to the scale factor, we could replace $R'$ by $(a'/a)R=aHR$. The second term on the left-hand side is a damping term, increasing with the contribution of baryons to the total energy of the fluid. The third term accounts for pressure forces in the effective fluid. The first term on the right-hand side accounts for the gravitational force, and the last two terms for dilation effects.

This equation would be that of a simple harmonic oscillator if $R$ was a constant (no friction term, constant sound speed) and in absence of gravitational source terms. Then, the solution would be of the form 
\begin{equation}
\Theta_0=\Theta_{\mathrm{ini}}\, \cos(k c_\mathrm{s} \eta+\varphi)~,
\end{equation} 
with two constants of integration $(\Theta_{\mathrm{ini}}, \varphi)$. We know that for adiabatic initial conditions and in the Newtonian gauge, photon density/temperature fluctuations should be constant in the super-Hubble limit, $k \eta \ll 1$: this fixes the phase to $\varphi=0$. In the opposite limit, this solution corresponds to the propagation of acoustic oscillations. Actually, the limit between the constant and oscillatory regime is not set by the value of $k\eta$, but by that of $k c_\mathrm{s} \eta$. In fact, the condition $k c_\mathrm{s} \eta \ll 1$ is equivalent to $\lambda \gg d_\mathrm{s}$, where $\lambda$ is a physical wavelength ($\lambda = 2\pi a/k$), and $d_\mathrm{s}$ is the physical sound horizon, given in the case of a constant sound speed by:
\begin{equation}
d_\mathrm{s}  = a \int^{\eta}_{\eta_\mathrm{ini}} c_\mathrm{s} d\eta
\simeq a c_\mathrm{s} \eta 
\end{equation}
(assuming $\eta \gg\eta_\mathrm{ini}$). Hence, the phase $k c_\mathrm{s} \eta$ of the cosine stands for the ratio $2 \pi d_\mathrm{s}/\lambda$. Modes start oscillating when their wavelength becomes smaller than the sound horizon, and later on, the number of oscillations is given by the ratio between these two scales.

\begin{figure}
\begin{center}
\includegraphics[width=10cm]{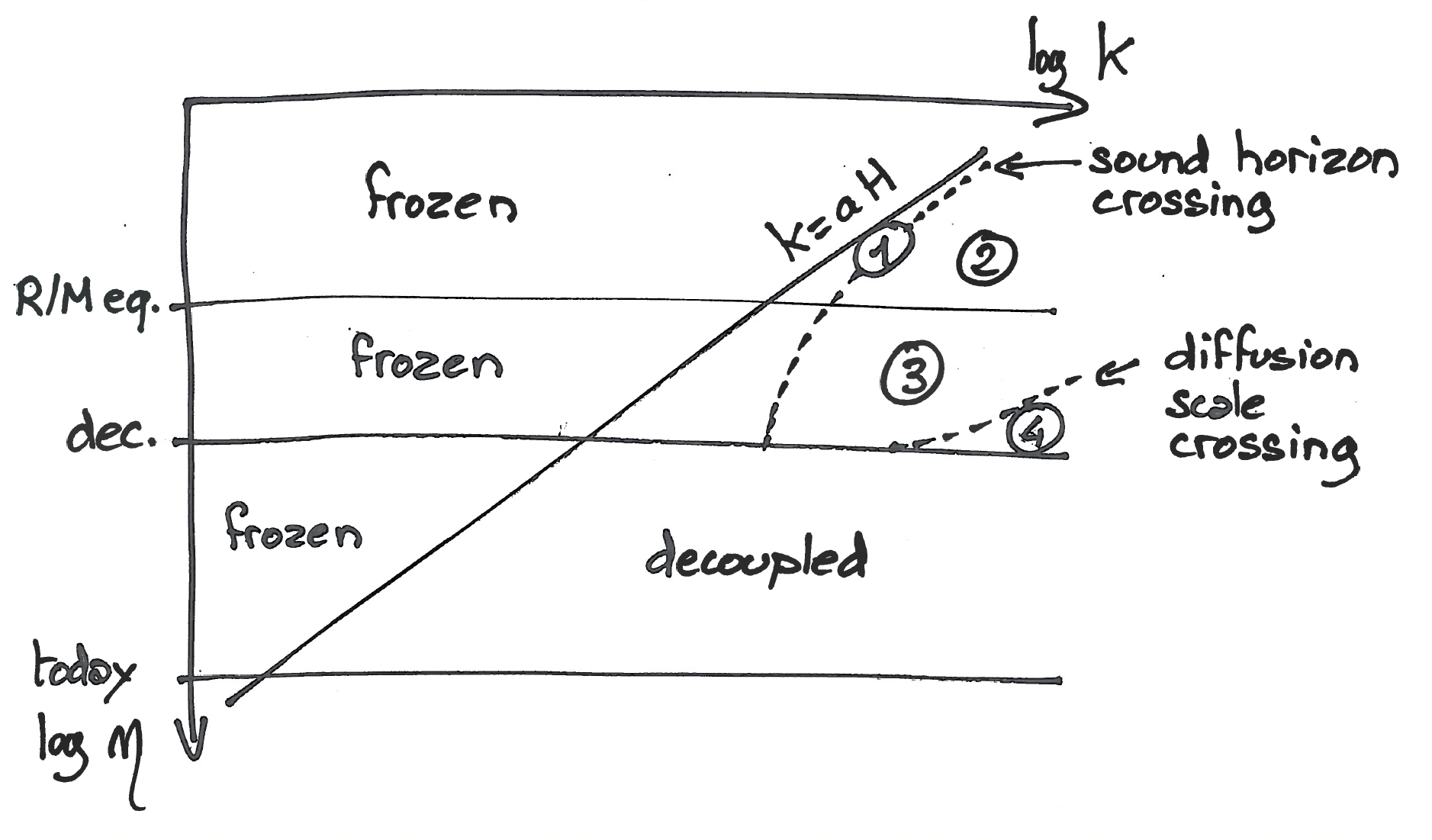}
\end{center}
\caption{Different regions in $(k,\eta)$ space, corresponding to qualitatively different behaviors for photon (and baryon) perturbations.}
\label{jl:fig_25a}
\end{figure}
In reality, $R$ grows with time (and crosses one roughly around the time of decoupling). In addition, the gravitational source terms in Eq.~(\ref{jl:osc_eq}) can play a role in some regimes. Let us describe qualitatively the evolution of $\Theta_0$ in the different regions in $(k,\eta)$ space shown in Fig.~\ref{jl:fig_25a}. In this figure, the horizontal axis corresponds to wavenumbers $k$ (large wavelengths are on the left), and the vertical axis to conformal time, flowing from top to bottom. The super-Hubble and sub-Hubble regions are separated by the solid diagonal line corresponding to $k = aH$ (equivalent to $k\eta=1$ during radiation domination).  From top to bottom, the horizontal lines correspond to the time of equality between radiation and matter, to the time of photon decoupling, and to the time today. The upper dashed line separates wavelengths bigger/smaller than the sound horizon in the baryon-photon fluid before decoupling (after decoupling, this notion does not make sense anymore). As we have just seen, this limit corresponds to $\lambda = d_\mathrm{s}$, or (up to a factor $2\pi$) to $k = (a/d_\mathrm{s})$. At early times, $c_\mathrm{s}=1/\sqrt{3}$, and this condition reads $k \eta = \sqrt{3}$. Just before decoupling, $R$ becomes large, $c_\mathrm{s}$ goes to zero, and the comoving sound horizon $(d_\mathrm{s}/a)$ becomes asymptotically constant, explaining the shape of the upper dashed line. Finally, the lower dashed line separates wavelengths bigger/smaller than the diffusion length defined in section~\ref{jl:sec21}: this line corresponds to $k=1/r_\mathrm{d}$, where $r_\mathrm{d}$ is given in first approximation by Eq.~(\ref{jl:diffusion_scale}).

The evolution of $\Theta_0$ in the super-Hubble region is trivial: as long as $k \eta \ll1$ --- and {\it a fortiori} $k c_\mathrm{s} \eta \ll1$ --- the fluctuation  $\Theta_0$ is frozen, and remains approximately equal to its initial value. 

The region marked with a \textcircled{\footnotesize 1} in the figure corresponds to modes that are crossing the sound horizon before decoupling. This is precisely the region in which gravitational source terms are important. They shift the zero point of oscillations, and boost their amplitude (due to gravitational forces and dilution effects). This happens during a limited amount of time, because the metric fluctuations quickly decay inside the sound horizon during radiation domination, making the gravitational source terms negligible. An approximation for the zero-point of oscillations can be found by setting $\Theta_0''$ and $\Theta_0'$ to zero in equation~(\ref{jl:osc_eq}), and by keeping only the first gravitational term:
\begin{equation}
\Theta_0^{equilibrium}=-\frac{1}{3 c_\mathrm{s}^2} \psi = - (1+R) \psi~.
\end{equation}
Since the gravitational potential is non-zero on super-Hubble scales (we have seen that for adiabatic initial conditions $-2 \psi= \delta_\mathrm{tot}$), the equilibrium point is shifted away from zero on those scales. It reaches asymptotically zero on sub-sound-horizon scales.

Region \textcircled{\footnotesize 2}  corresponds to wavelengths smaller than the sound horizon during radiation domination. In this regime, the metric fluctuations have decayed, so the source term in Eq.~(\ref{jl:osc_eq}) can be neglected. The friction term can also be neglected, because during radiation domination, $R\ll1$.  Finally the effective mass $k^2 c_\mathrm{s}^2$ is constant in time because $R\ll1$ implies $c_\mathrm{s}^2=1/3$. Hence we are in the simple case discussed before, and the solution is proportional to $\cos(k c_\mathrm{s} \eta)$, corresponding to stationary oscillations, symmetric around $\Theta_0=0$.

Region \textcircled{\footnotesize 3} refers to wavelengths smaller than the sound horizon during the intermediate stage between the time of equality and that of photon decoupling. In this region, the metric perturbations have decayed, but $R$ cannot be neglected (baryons and photons contribute to the total energy density with the same order of magnitude). Hence  the oscillator equation has a non-negligible friction term (increasing with time), and a time-varying effective mass (decreasing with time). The solution of the equation corresponds to damped oscillations. Physically, this damping is caused by the increasing inertia and decreasing pressure of the baryon-photon fluid when the energy density of non-relativistic baryons takes over.

Region \textcircled{\footnotesize 4} refers to modes with smaller wavelength than the diffusion scale in the photon-baryon fluid. We have defined this scale in section~\ref{jl:sec21}. At early times, in the tightly-coupled limit, the mean free path of particles in the fluid is negligible, and cosmologically interesting scales are all well above the diffusion length. At the approach of decoupling, the diffusion length suddenly increases, and encompasses most of sub-sound-horizon wavelengths. In this regime, the oscillator equaion~(\ref{jl:osc_eq}) does not apply anymore, because we cannot describe baryons and photons in terms of a perfect fluid. Perturbations are then strongly damped, since diffusion tends to average out any small-scale fluctuation.

\begin{figure}
\begin{center}
\includegraphics[width=10cm]{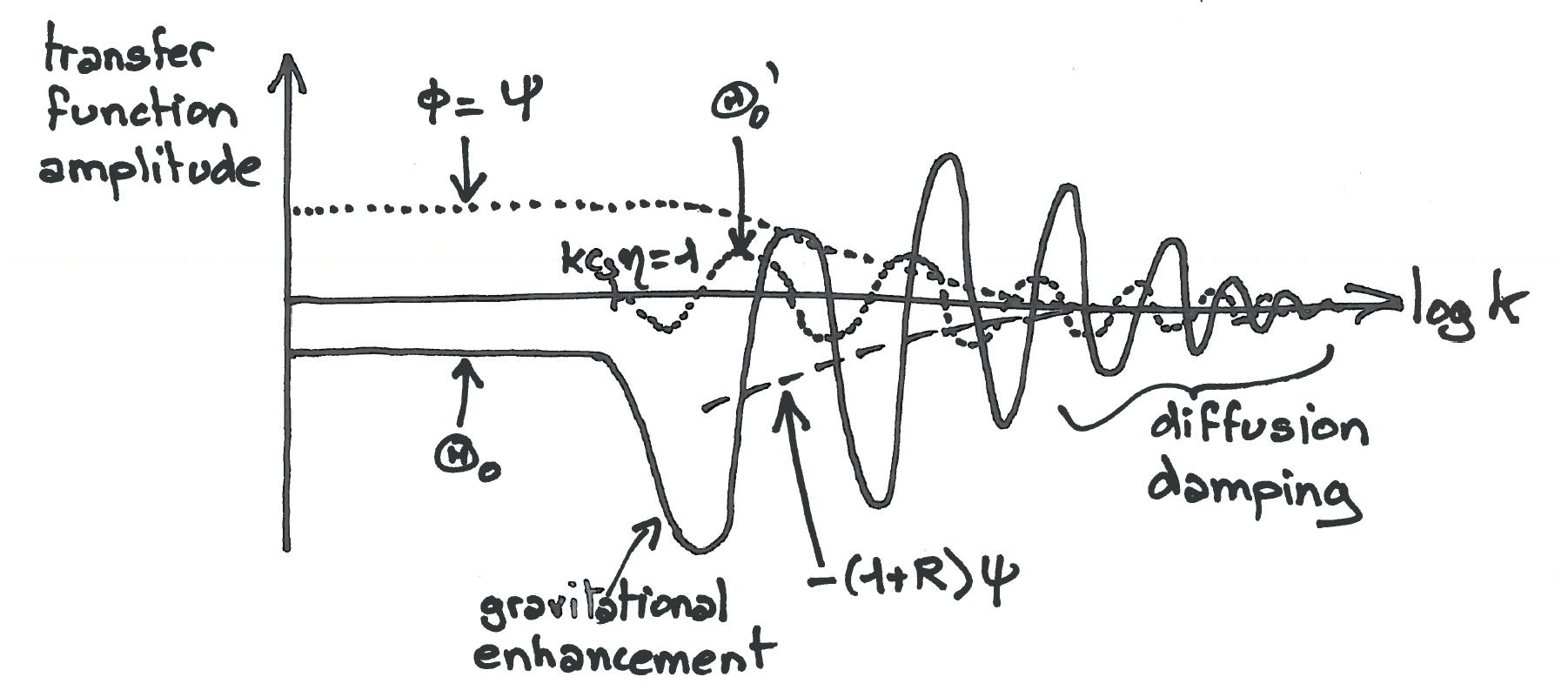}
\end{center}
\caption{Transfer functions at the time of decoupling.}
\label{jl:fig_25b}
\end{figure}
After describing these different regions, we are ready for understanding the qualitative behavior of the various relevant transfer functions, evaluated at the time of decoupling. Figure~\ref{jl:fig_25b} shows the transfer functions $\Theta_0(\eta_\mathrm{dec},k)$ (solid line), $\psi(\eta_\mathrm{dec},k)$ (upper dotted line) and $\Theta_0'(\eta_\mathrm{dec},k)$ (middle dotted line) as a function of $\log k$. For simplicity, we neglect the role of the anisotropic stress generated by neutrinos (and to a lesser extent by photons near decoupling time). Hence we can assume $\phi=\psi$ at all times. For scales above the sound horizon ($k c_\mathrm{s} \eta \ll 1$), the transfer functions are constant: indeed we know that, on those scales and in the Newtonian gauge, density and metric fluctuations are frozen. Adiabatic initial conditions impose $-2 \psi = \delta_\mathrm{tot} \simeq \delta_\mathrm{b} \simeq \frac{3}{4} \delta_\gamma \simeq 3 \Theta_0$. The opposite sign of $\Theta_0$ and $\psi$ reflects the fact that an over-density corresponds to a temperature excess and a gravitational potential well (and vice-versa). We remember that transfer functions are all normalized to ${\cal R}(\vec{k})=1$ (see section~\ref{jl:sec15}): this corresponds to a negative $\delta_\gamma$ (and $\Theta_0$) and to a positive $\psi$, like in the figure.

Because of the decay of metric fluctuations inside the Hubble radius, the $\psi$ curve smoothly decreases and tends towards zero in the small wavelength limit. The behavior of $\Theta_0$ is more complicated.  Modes which are just crossing the sound horizon near $\eta=\eta_\mathrm{dec}$ are experiencing the boost caused by gravitational source terms in Eq.~(\ref{jl:osc_eq}): this explains the first bump in the solid line. For smaller wavelengths, we  see oscillatory patterns corresponding to acoustic oscillations.  Smaller wavelengths crossed the sound horizon earlier, and had more time for oscillating before decoupling. The maxima observed at the time of decoupling correspond to modes that could experience 0.5, 1, 1.5, 2, ..., periods of oscillations before that time. The zero point of oscillations follows $-(1+R)\psi$, represented on the figure with a dashed line: this zero point reaches zero well inside the sound horizon. The amplitude of the oscillations is maximal for the first oscillatory pattern, i.e. for modes that crossed the sound horizon very recently. The second oscillatory pattern is reduced by the fact that those modes stayed for a longer time inside the sound horizon during the matter dominated regime, and experienced more damping due to baryons. The third and higher oscillatory patterns are reduced even more by diffusion damping just before photon decoupling. Temperature fluctuations on very small wavelengths are completely suppressed by photon diffusion.

Figure~\ref{jl:fig_25b} also shows qualitatively the behavior of the time derivative $\Theta_0'(\eta_\mathrm{dec},k)$, which exhibits oscillations that are out of phase with respect to those of $\Theta_0(\eta_\mathrm{dec},k)$. This will be important in a few paragraphs, when discussing the Doppler effect. 

Now that we understand qualitatively the behavior of the metric and photon transfer functions, we can go back to the decomposition of the CMB temperature spectrum $C_l$ in three terms (Sachs-Wolfe, Doppler and integrated Sachs-Wolfe) discussed in section~\ref{jl:sec25}.

\vspace{0.2cm}

{\bf Sachs-Wolfe contribution.}~We have seen that the Sachs-Wolfe contribution to $C_l$ is approximately given by the power spectrum of the combination $(\Theta_0+\psi)$ at $\eta=\eta_\mathrm{dec}$, with a correspondence between $k$ and $l$ given by Eq.~(\ref{jl:cor_kl}). We know that this power spectrum is given by the product of the primordial spectrum ${\cal P}_{\cal R}(k)$ (that we can choose to be scale-invariant in first approximation) by the square of the transfer function $(\Theta_0+\psi)^2$. The qualitative behavior of the latter has no more secrets for us. We can pick up the solid and upper dotted lines in figure~\ref{jl:fig_25b}, add them up, and square the result.
\begin{figure}
\begin{center}
\includegraphics[width=10cm]{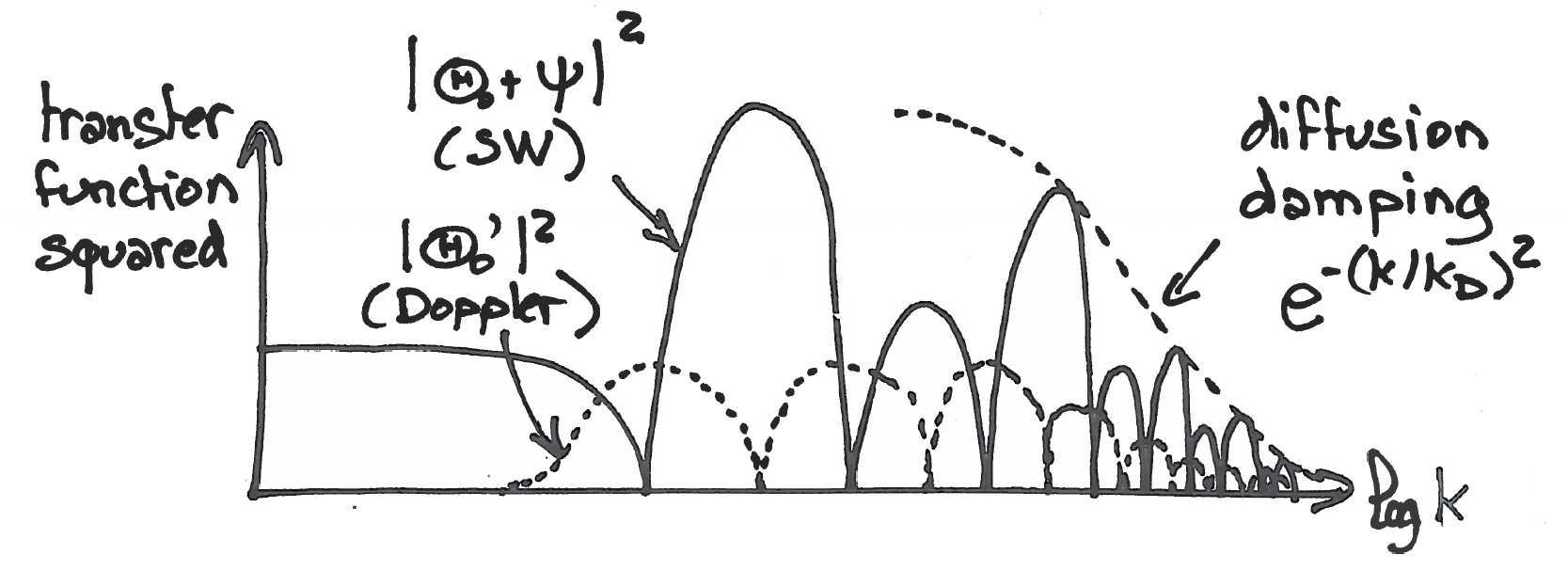}
\end{center}
\caption{Squared transfer functions at the time of decoupling.}
\label{jl:fig_25c}
\end{figure}
The result is shown in figure~\ref{jl:fig_25c} as a function of $\log k$. For modes with $k c_\mathrm{s}^2 \eta\ll1$,  we see a flat plateau: there, our previous calculation of the Sachs-Wolfe effect would apply (this part of the curve is actually called the Sachs-Wolfe plateau). We then observe a series of peaks. Due to the shift of the zero-point of oscillations given by $-(1+R)\psi$ for $\Theta_0$, and hence by $-R\psi$ for the Sachs-Wolfe term $(\Theta_0+\psi)$, there is an asymmetry between the first few odd and even peaks, with odd peaks being enhanced. Moreover the overall amplitude of the peaks is suppressed in the large $k$ limit by diffusion damping. A bit of algebra would show us that the envelope of the peaks is given in first approximation by the function $\exp[-(k/k_\mathrm{d})^2]$, where $k_\mathrm{d}$ is the diffusion wavenumber, related to the diffusion comoving scale of Eq.~(\ref{jl:diffusion_scale}) by $k_\mathrm{d} r_\mathrm{d}=1$.

\vspace{0.2cm}

{\bf Doppler contribution.}~Next, we know that the $C_l$'s receive a second contribution from the Doppler effect, related to the power spectrum of $\theta_\mathrm{b}$, which is equal to $\theta_\gamma$ until baryon and photons decouple from each other. It turns out that the photon velocity divergence  $\theta_\gamma$ is itself related to the time derivative of the temperature fluctuation $\Theta_0$. At a very qualitative level, we can infer the Doppler contribution from the shape of the transfer function $\Theta_0'(\eta_\mathrm{dec},k)$. This contribution is null for scales above the sound horizon, since in this regime there are no oscillations and no significant dynamics in the fluid. On smaller scales, the Doppler contribution has oscillatory patterns, that are out of phase with respect to those of the Sachs-Wolfe term. The Doppler contribution is represented schematically as a dotted line in figure~\ref{jl:fig_25c}.

We can now sum up the Sachs-Wolfe and Doppler contributions. Also, we can transpose our results for power spectra as a function of $k$ in terms of $C_l$'s as a function of $l$. We have already seen in section~\ref{jl:sec25} that there is a mapping betwwen the two, at least in the small-angle and instantaneous decoupling approximation, with a correspondence between values of $k$ and $l$ given by Eq.~(\ref{jl:cor_kl}).
\begin{figure}
\begin{center}
\includegraphics[width=10cm]{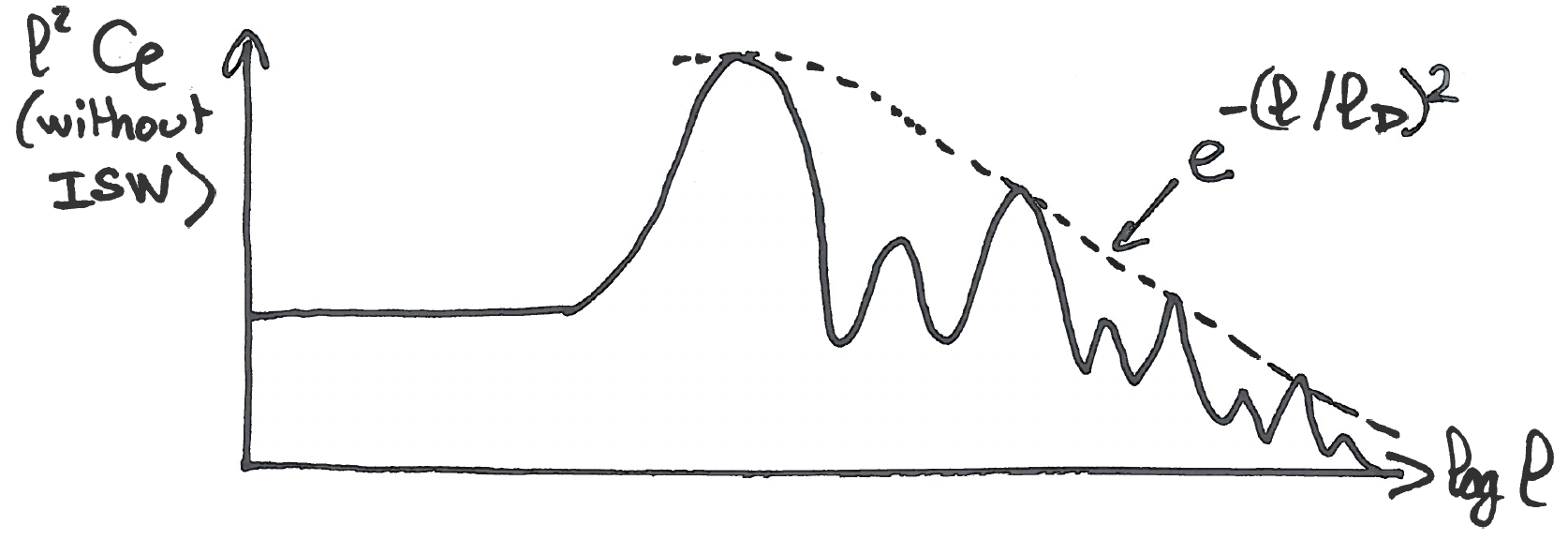}
\end{center}
\caption{Contribution to the $C_l$'s from the SW and Doppler terms.}
\label{jl:fig_25d}
\end{figure}
The result is shows in figure~\ref{jl:fig_25d}. Note that the vertical axis stands for $l^2 C_l$, for the following reason. If the primordial spectrum was scale invariant (${\cal P}_{\cal R}(k)=\mathrm{constant}$) and the transfer functions were flat (as it is the case for large wavelengths/small $l$'s), the quantity $l(l+1) C_l$ would also be flat and independent of $l$. This would follow from Eq.~(\ref{jl:clsw}) if we had been more carefull in keeping all the factors. It is convenient to plot $l^2 C_l$ or $l(l+1) C_l$ instead of $C_l$, in order to display a roughly constant curve, just modulated by acoustic oscillations and diffusion damping.

\vspace{0.2cm}

{\bf Diffusion damping effect.}~In Fig.~\ref{jl:fig_25d}, we can identify all the features mentioned before: the flat Sachs-Wolfe plateau, the series of oscillations with enhanced odd peaks, and the exponentially decaying envelope of the peaks for large $l$. The envelope is now given by $\exp[-(l/l_\mathrm{d})^2]$, where $l_\mathrm{d}$ is the diffusion multipole, related to the diffusion angle by $l_\mathrm{d}=\pi/ \theta_\mathrm{d}$. The diffusion angle is related to the diffusion scale $r_\mathrm{d}$ by the usual angular diameter distance relation,
\begin{equation}
\theta_\mathrm{d} \, d_\mathrm{a}(\eta_\mathrm{dec}) = \lambda_\mathrm{d} = a(\eta_\mathrm{dec}) r_\mathrm{d}(\eta_\mathrm{dec}).
\end{equation}

At this point, we are almost done with the qualitative description of the CMB temperature spectrum $C_l$. We only missed the integrated Sachs-Wolfe contribution, and the effect of reionization.

\vspace{0.2cm}

\begin{figure}
\begin{center}
\includegraphics[width=10cm]{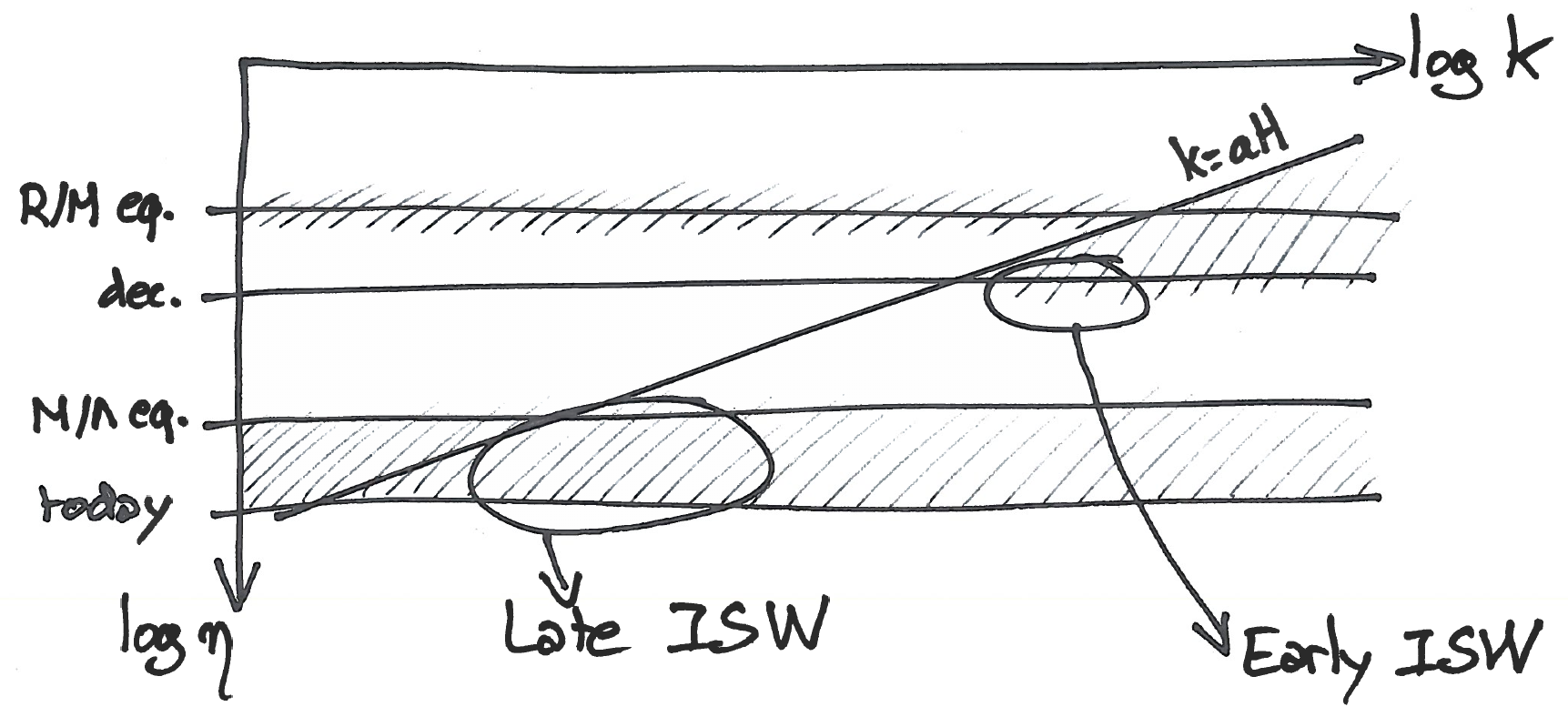}
\end{center}
\caption{Regions in $(k, \eta)$ space where metric fluctuations are expected to vary with time (giving rise potentially to an ISW effect after decoupling).}
\label{jl:fig_25e}
\end{figure}

{\bf Integrated Sachs-Wolfe contribution.}~We have seen that the integrated Sachs-Wolfe effect is given by an integral over $(\psi'+\phi')$ between photon decoupling and today. In fact, $(\psi'+\phi')$ remains vanishingly small in a large part of the space $(k,\eta)$. In Figure~\ref{jl:fig_25e}, we hatched all the regions in $(k,\eta)$ space where metric fluctuations are expected to vary with time. Let us discuss these different regions.

Before photon decoupling, we know that metric fluctuations decay inside the sound horizon. Instead, in the Newtonian gauge, they remain frozen outside the Hubble radius, excepted near times at which the equation of state of the universe changes: namely, at the time of equality between radiation and matter. 

Let us now discuss the variation of metric perturbations after photon decoupling (this is the relevant epoch for the ISW effect). Deep inside the matter dominated regime, one can show that metric fluctuations are static, even inside the Hubble radius (at least within linear perturbation theory). We will justify this result in section~\ref{jl:sec32}. Hence in figure~\ref{jl:fig_25e} there are no hatches during matter domination on whatever scales. Note however that at the beginning of matter domination, it takes some time for sub-sound-horizon metric fluctuations to freeze around a constant value: hence the hatches continue below the line corresponding to the time of equality, and extend till the line corresponding to photon decoupling. 

Later on, during the $\Lambda$ (or dark energy) dominated regime, the equation of state of the universe changes again, so metric fluctuations vary on all scales, like at the time of equality. A simple calculation based on Einstein equations would show that metric fluctuations are damped during this stage.

\begin{figure}
\begin{center}
\includegraphics[width=9cm]{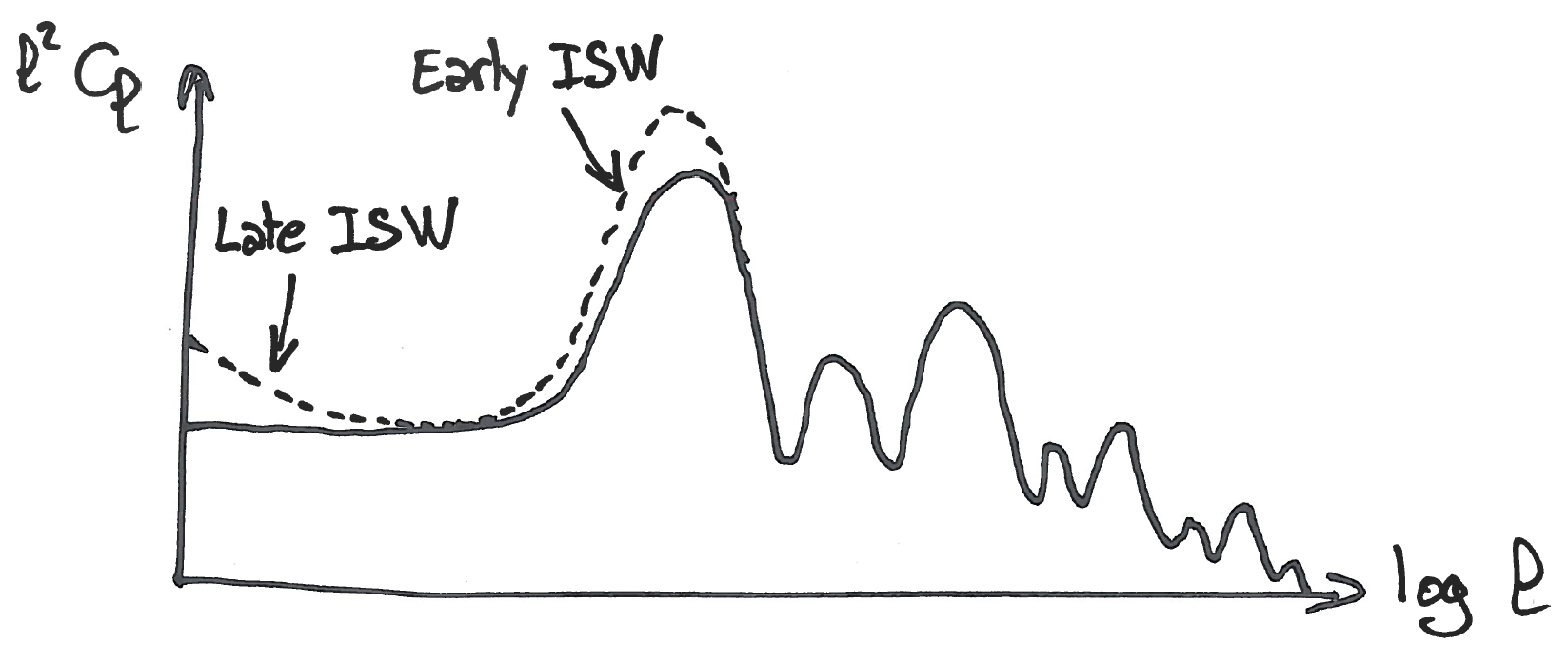}
\end{center}
\caption{ISW contribution to the temperature spectrum.}
\label{jl:fig_25f}
\end{figure}

In summary, contribution to the integrated Sachs-Wolfe effect can only come from two regions: 
\begin{itemize}
\item
just after photon decoupling, on sub-sound-horizon scales, and
\item during $\Lambda$ domination, on all scales.
\end{itemize} 
These two distinct contributions are usually called the Early Integrated Sachs-Wolfe (EISW) and Late Integrated Sachs-Wolfe (LISW) effects. One can show that both effects decrease with wavelength, for geometrical reasons. Hence the EISW effect is maximal for scales crossing the sound horizon just at the time of photon decoupling, while the LISW effect is maximal for the largest observable scales today. In multipole space, this means that the EISW effect contributes mainly to the scale of the first peak, i.e. to $l \sim 200$, while the LISW contributes mainly to the smallest multipoles $l=2,3,4,$ etc. The two ISW contributions are drawn on figure~\ref{jl:fig_25f}. The EISW effect enhances the first peak, while the LISW effect tilts the Sachs-Wolfe plateau even if the primordial spectrum is exactly scale invariant.

\vspace{0.2cm}

\begin{figure}
\begin{center}
\includegraphics[width=9cm]{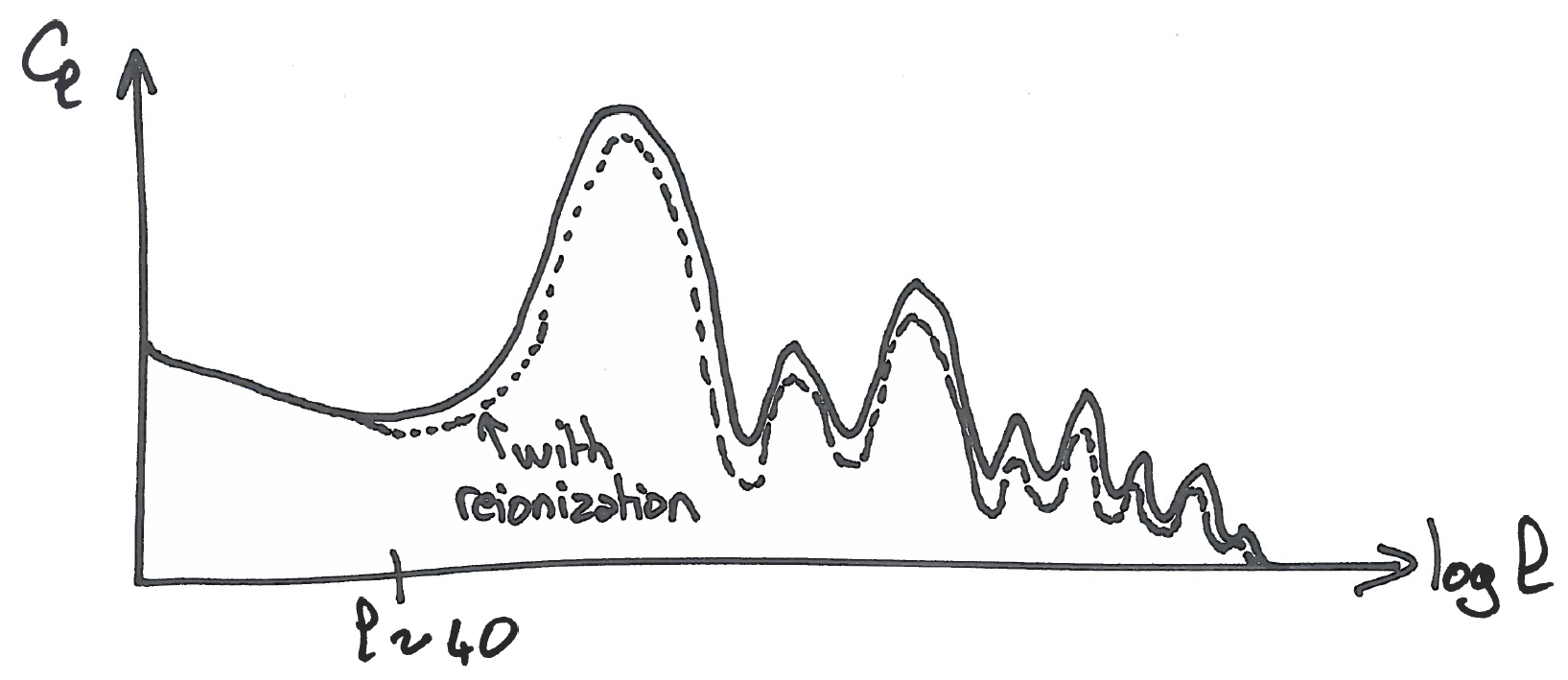}
\end{center}
\caption{Impact of reionization on the temperature spectrum.}
\label{jl:fig_25g}
\end{figure}

{\bf Reionization effect.}~The last effect that we omitted to describe is that of reionization. We have seen in section~\ref{jl:sec21} that at small redshift ($z \simeq 10$), the reionization of the universe produces a small secondary bump in the visibility function, corresponding physically to a small probability for CMB photons to rescatter at late times. This rescattering will tend to smooth out any temperature anisotropy pattern. Hence, reionization lowers the overall amplitude of the $C_l'$, but only a small amount (by approximately 15\%). Note that the suppression of power is not uniform over the whole multipole range: smoothing effects cannot reach the largest observable scales (corresponding to the smallest values of $l$). Hence the effect of reionization is step-like shaped, and saturates for $l$ of the order of 40 or so (as illustrated on figure~\ref{jl:fig_25g}).

\subsection{Parameter dependence of the temperature spectrum}\label{jl:sec26}

We summarized in the last section the various physical effect contributing to the shape of the CMB temperature spectrum $C_l$. We will now recapitulate the effect of the various cosmological parameters on the $C_l$'s, within the framework of the minimal $\Lambda$CDM model. 

This model assumes zero spatial curvature, and a power-law primordial spectrum of scalar perturbations:
\begin{equation}
{\cal P}_{\cal R}(k) = A_\mathrm{s} (k/k_*)^{n_\mathrm{s}-1}~,
\end{equation}
where $k_*$ is an arbitrary fixed pivot scale, $A_\mathrm{s}$ is the spectrum amplitude at this scale, and $n_\mathrm{s}$ is called the scalar tilt (the exponent is chosen to be $n_\mathrm{s}-1$ rather than just $n_\mathrm{s}$ for historical reasons; with such notations, a scale-invariant spectrum corresponds to $n_\mathrm{s}=1$).

We recall that the Hubble parameter today, $H_0$, can be expressed in terms of a dimensionless reduced Hubble parameter $h$:
\begin{equation}
H_0 \equiv 100\, h\, \mathrm{km}.\mathrm{s}^{-1}.\mathrm{Mpc}^{-1}
\end{equation}
The physical energy density of a given component $x$ today can be expressed in terms of a dimensionless parameter $\omega_x \equiv \Omega_x h^2$:
\begin{equation}
\rho_x^0 = \Omega_x \rho_\mathrm{crit}^0 = \Omega_x \frac{3 H_0^2}{8 \pi G} = \beta \, \omega_x
\end{equation}
where $\beta= \frac{3 (H_0/h)^2}{8 \pi G} $ is a fixed number, with the dimension of an energy per volume.

The six free parameters of the minimal $\Lambda$CDM model can be chosen to be
\begin{equation}
\{A_\mathrm{s}, n_\mathrm{s}, \omega_\mathrm{b}, \omega_\mathrm{m}, \Omega_\Lambda, \tau_\mathrm{reio} \}, 
\label{jl:lcdm_basis}
\end{equation}
where $\tau_\mathrm{reio}$ is the optical depth to recombination, which is non-zero because of reionization in the recent universe. At first order, this is the only parameter that one needs to introduce for describing reionization. For a very accurate description, one should introduce other parameters specifying the full reionization history, but these extra parameters are very difficult to probe experimentally. Hence, they are usually not specified.

The above parameter basis specifies the baryon density $\omega_\mathrm{b}$, the total non-relativistic (baryons + CDM) matter density $\omega_\mathrm{m}$, and the fractional density of cosmological constant $\Omega_\Lambda$. The photon density is implicitly assumed to match the measured value of the CMB temperature ($T=2.725$~K implies $\omega_\gamma \sim 2.10^{-5}$). Neutrinos are assumed to be still relativistic today, with an abundance relative to photons given by the prediction of the standard neutrino decoupling model\footnote{In this course, for simplicity, we do not discuss the effect of neutrinos; this effect is far from being negligible, but since we consider the abundance of neutrinos as fixed, and their mass as irrelevant, we can explain the effect of other free parameters without taking neutrinos into account. Neutrino effects are described in Ref.~\refcite{CUP}.}. Finally, we did not include the parameter $H_0$ (or $h$) in the parameter basis  (\ref{jl:lcdm_basis}). Given that we are assuming a flat universe, $h$ can be inferred from other parameters in (\ref{jl:lcdm_basis}): $h=\sqrt{\omega_\mathrm{m}/(1-\Omega_\Lambda)}$.

The parameters basis (\ref{jl:lcdm_basis}) is just one particular choice. Many other bases would be valid, for instance, $\{A_\mathrm{s}, n_\mathrm{s}, \omega_\mathrm{b}, \omega_\mathrm{cdm}, H_0, \tau_\mathrm{reio} \}$). The choice of (\ref{jl:lcdm_basis}) is dictated by purely pedagogical considerations: we will show that the shape of the CMB spectrum can easily be related to these parameters.

\begin{figure}
\begin{center}
\includegraphics[width=10cm]{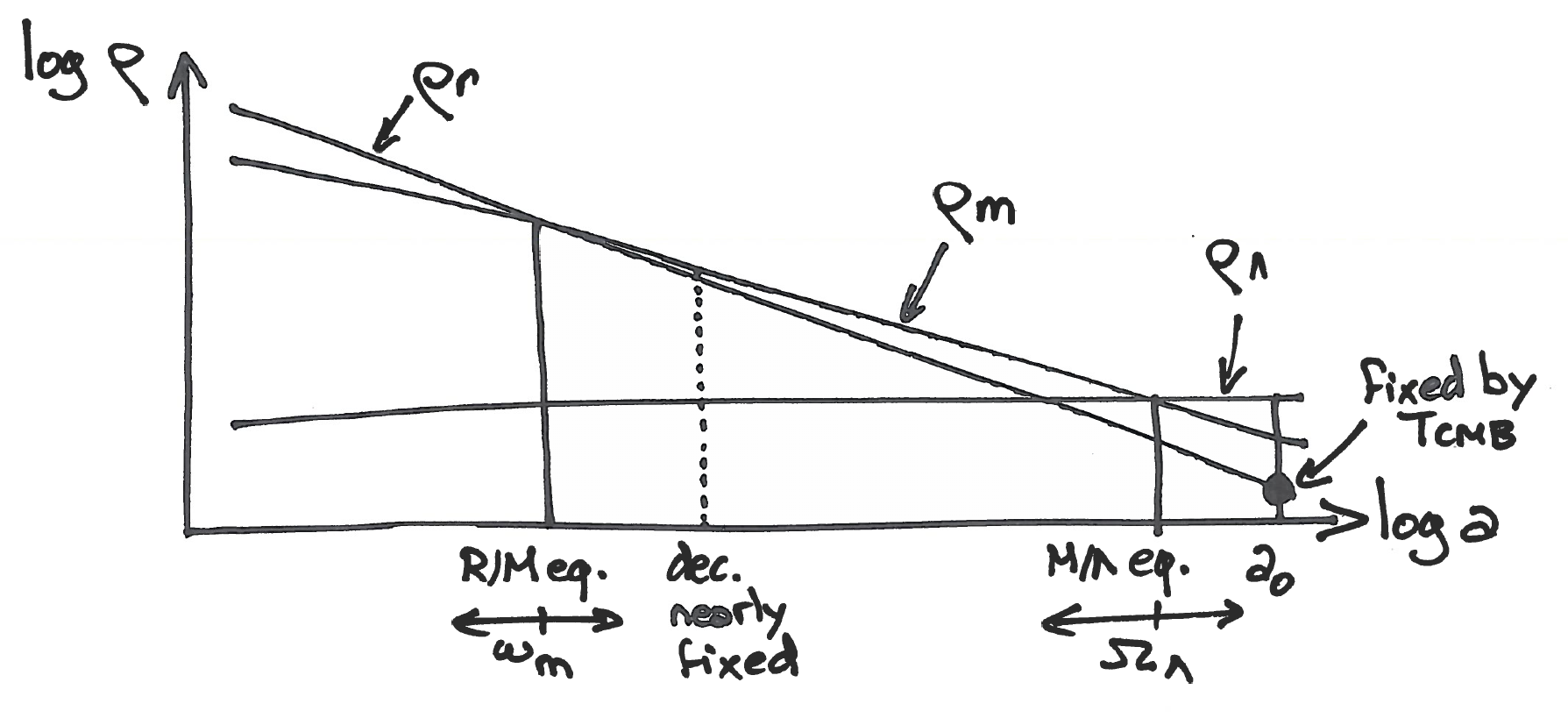}
\end{center}
\caption{In the minimal $\Lambda$CDM model, evolution of the background density of radiation, matter and cosmological constant as a function of the scale factor.}
\label{jl:fig_26}
\end{figure}
We summarize in Fig.~\ref{jl:fig_26} the evolution of background densities in the minimal $\Lambda$CDM model. The normalization of the radiation density $\rho_r$ is fixed by the CMB temperature and by the standard neutrino decoupling model. The normalization of the matter density $\rho_\mathrm{m}$ and of the cosmological constant density $\rho_\Lambda$ is given respectively by the parameters $\omega_\mathrm{m}$ and $\Omega_\Lambda$. The scale factor at radiation/matter equality is given by
\begin{equation}
\rho_r = \rho_\mathrm{m} \quad \Longrightarrow \quad 
\omega_r \left(\frac{a_0}{a_\mathrm{eq}}\right)^4 = \omega_\mathrm{m} \left(\frac{a_0}{a_\mathrm{eq}}\right)^3 \quad  \Longrightarrow \quad 
\frac{a_\mathrm{eq}}{a_0}=\frac{\omega_r}{\omega_\mathrm{m}}~.
\end{equation}
Since $\omega_r$ is considered as fixed, the redshift of equality is controlled by $\omega_\mathrm{m}$ only. 
The scale factor at matter/$\Lambda$ equality is given by
\begin{equation}
\rho_\mathrm{m} = \rho_\Lambda \quad \Longrightarrow  \quad 
\omega_\mathrm{m} \left(\frac{a_0}{a_\Lambda}\right)^3 = \Omega_\Lambda h^2
\quad  \Longrightarrow \quad 
\frac{a_\Lambda}{a_0}=\left(\frac{1-\Omega_\Lambda}{\Omega_\Lambda}\right)^{1/3}~,
\end{equation}
so it is controlled by $\Omega_\Lambda$ only. The scale factor at photon decoupling has a small (logarithmic) dependence on $\omega_\mathrm{b}$, that we can neglect --- we will consider that decoupling takes place at a fixed temperature, and hence a fixed scale factor and redshift.

Among the various physical effects described in the previous section, we can identify eight independent leading effect. We summarize them in Table~\ref{jl:Ceffects}, and show by which parameters they are governed.
\begin{table}[h!]
\tbl{Independent leading effects controlling the shape of the CMB temperature power spectrum $C_l$  in the minimal $\Lambda$CDM model.}
{\tablefont
\begin{tabular}{lllc}
\toprule  & Effect & Relevant quantity &  Parameter\\
\colrule
(C1) & Peak scale & $\theta_\mathrm{peak} = \displaystyle{\frac{\pi}{l_\mathrm{peak}}} \sim \displaystyle{\frac{d_\mathrm{s} |_\mathrm{dec}}{d_\mathrm{a} |_\mathrm{dec}}}$ & \begin{tabular}{l}$\leftarrow \omega_\mathrm{m}, \omega_\mathrm{b}$\\$\leftarrow \Omega_\Lambda, \omega_\mathrm{m}$\end{tabular} \\
&&&\\
(C2) & Odd/even peak amplitude ratio & $R|_\mathrm{dec}$ & $\omega_\mathrm{b}$ \\
&&&\\
(C3) & Overall peak amplitude & $\frac{a_\mathrm{dec}}{a_0}$ & $\omega_\mathrm{m}$ \\
&&&\\
(C4) & Damping enveloppe & $\theta_\mathrm{d} = \displaystyle{\frac{\pi}{l_\mathrm{d}}} = \displaystyle{\frac{a_\mathrm{dec} r_\mathrm{d}|_\mathrm{dec}}{d_\mathrm{a}|_\mathrm{dec}}}$ &  \begin{tabular}{l}$\leftarrow \omega_\mathrm{m}, \omega_\mathrm{b}$\\$\leftarrow \Omega_\Lambda, \omega_\mathrm{m}$\end{tabular} \\
&&&\\
(C5) & Global amplitude & ${\cal P}_{\cal R}(k_*)$ & $A_\mathrm{s}$ \\
&&&\\
(C6) & Global tilt & $\frac{d \log {\cal P}_{\cal R}}{d \log k}$ & $n_\mathrm{s}$ \\
&&&\\
(C7) & Additional plateau tilting (LISW) & $\frac{a_\Lambda}{a_0} $ & $\Omega_\Lambda$ \\
&&&\\
(C8) & Amplitude for $l \geq 40$ only & $\tau_\mathrm{reio}$ & $\tau_\mathrm{reio}$\\
\botrule
\end{tabular}
}
\label{jl:Ceffects}
\end{table}
Below, we give more details on these effects.
\begin{description}
\item[\tablefont(C1)] Since all peaks in multipole space correspond to the harmonics of a single correlation length in real space, the scale of the peak is controlled (in good approximation) by a single number $l_\mathrm{peak}$. It depends on the ratio of the sound horizon at decoupling by the angular diameter distance to decoupling. The first quantity depends on the evolution prior to decoupling, and in particular on the expansion history and sound speed. Hence it depends on $\omega_\mathrm{m}$ (governing the time of equality) and $\omega_\mathrm{b}$ (governing $c_\mathrm{s}^2$ as a function of $a$). The second quantity depends on the expansion and geometry of the universe after decoupling, i.e. on $\Omega_\Lambda$ and $H_0$, or in our parameter basis $\Omega_\Lambda$ and $\omega_\mathrm{m}$.
\item[\tablefont(C2)] When studying the SW contribution to the $C_l$'s, we have seen that the asymmetry between the amplitude of odd and even peaks depends on the shift of the zero-point of acoustic oscillations by a term $-R\psi$. The value of the ratio $R$ at decoupling is governed in our parameter basis by $\omega_\mathrm{b}$.
\item[\tablefont(C3)] A shift in the time of radiation/matter equality affects the amplitude of all peaks for two reasons: it controls the duration of the intermediate stage between equality and decoupling, during which acoustic oscillations are damped by baryonic effects, and the EISW. Both effects go in the same direction. If equality takes place earlier, there is less time for damping (hence all peaks are increased). The metric fluctuations are also less stabilized at decoupling, and the EISW is larger (hence the first peak is increased even more). 
\item[\tablefont(C4)] Diffusion damping near the time of recombination controls the envelope of the peaks (the function $\exp[-(l/l_\mathrm{d})^2]$ suppresses them, starting essentially from the third one). It depends on the ratio of the damping scale at decoupling by  the angular diameter distance to decoupling. The first quantity depends on the Thomson scattering rate prior to decoupling, and in particular on $\omega_\mathrm{m}$ (governing the value of conformal time at equality) and $\omega_\mathrm{b}$ (governing the ionization fraction as a function of $a$). The second quantity depends on the expansion and geometry of the universe after decoupling, i.e. on $\Omega_\Lambda$ and $H_0$, or in our parameter basis $\Omega_\Lambda$ and $\omega_\mathrm{m}$. The parameter dependence of effects (C1) and (C4) could be thought to be similar: in fact, it is not, because the sound horizon and the diffusion scale depend on very different combinations of $\omega_\mathrm{m}$ and $\omega_\mathrm{b}$.
\item[\tablefont(C5)] The global amplitude of the $C_l$'s depends trivially on that of the primordial spectrum, fixed by $A_\mathrm{s}$.
\item[\tablefont(C6)] The global slope of the $C_l$'s depends trivially on the tilt of the primordial spectrum, fixed by $n_\mathrm{s}$.
\item[\tablefont(C7)] The LISW effect tilts the Sachs-Wolfe plateau (on top of the effect of $n_\mathrm{s}$). The plateau is more lifted at small $l$'s if $\Lambda$ domination is longer, i.e. if metric fluctuations decay during a larger amount of time. Hence this effect is enhanced by large values of $\Omega_\Lambda$.
\item[\tablefont(C8)] The $C_l$ amplitude is suppressed if reionization takes place early, i.e. if $\tau_\mathrm{reio}$ is large, but without affecting the largest scales (small $l$'s).  
\end{description}

We see that in the framework of the minimal $\Lambda$CDM model, six parameters control eight distinct physical effects with different impacts on the $C_l$'s. This suggests that an accurate enough measurement of the temperature spectrum is sufficient for fixing the six parameters of the cosmological model describing our universe, at least if the data is consistent with $\Lambda$CDM. This conclusion is roughly correct, but must be refined with some words of caution. Indeed, we remember that for small $l$'s, cosmic variance is large, so that the average $C_l$'s of the ``true model'' describing our universe cannot be measured precisely, even in the case of an ideal experiment. However, two of the previous effects  {\tablefont(C1)} - {\tablefont(C8)} can only affect the smallest multipoles:
\begin{itemize}
\item effect  {\tablefont(C7)} affects only the Sachs-Wolfe plateau,
\item a combination of effects  {\tablefont(C5)} and {\tablefont(C8)}, corresponding to the product $e^{-2\tau_\mathrm{reio}}A_\mathrm{s}$, controls the global amplitude for $l \gg 40$, but a variation of both $\tau_\mathrm{reio}$ and $A_\mathrm{s}$ with the previous product being kept fixed would only affect the smallest multipoles. 
\end{itemize}
Hence, the measurement of the CMB temperature spectrum is sufficient for constraining the six parameters of the $\Lambda$CDM model, but with a relatively large error bar for $\Omega_\Lambda$ and for a particular combination of $\tau_\mathrm{reio}$ and $A_\mathrm{s}$. In the next section, we will say a few words on the measurement of CMB polarization, which allows to better constrain reionization: polarization data allow to remove the degeneracy between $\tau_\mathrm{reio}$ and $A_\mathrm{s}$. Instead, the error bar on $\Omega_\Lambda$ can only be improved by combining CMB data with other cosmological probes (e.g. supernovae luminosity or large scale structure data).

\subsection{A quick word on polarization}\label{jl:sec27}

The Bolztmann equation presented in Eq.~(\ref{jl:boltz1}) was a bit over-simplified. We did as if the only degree of freedom describing photons with a blackbody spectrum was their temperature. In fact, photons are described by more degrees of freedom, called the Stokes parameters, involving also their polarization. In Eq.~(\ref{jl:boltz1}), the Thomson scattering rate was integrated over polarization parameters, but in reality some polarized correction terms are present. 

Well before decoupling, photons remain unpolarized on average. Indeed, we have seen in section~\ref{jl:sec22} that, throughout the tight-coupling regime and in the frame comoving with the fluid (i.e. such that $\theta_\mathrm{b}=\theta_\gamma=0$), the photon temperature is isotropic is every point. This isotropy (resulting from frequent interactions) implies that photons acquire no net polarization patterns when they scatter over electrons.

Instead, at the approach of decoupling, when Thomson scattering becomes inefficient, the photon temperature is no longer isotropic in the frame comoving with the electrons. This means that a given electron will scatter simultaneously some hotter photons coming from one direction, and some colder photons coming from another direction. What is important for polarization is the quadrupolar component of the the temperature distribution in each point. This component starts from zero and grows at the approach of decoupling. When it becomes significant, photon scattering leads to a net linear polarization. Hence, today, CMB photons have a different polarization amplitude and orientation in each direction of the sky. 

The map of CMB temperature anisotropies is a scalar map: it can be represented with a one-dimensional color code. The map of CMB polarization can be represented with sticks of different size and orientation in different points of the map. Roughly speaking, the size and orientation of the sticks can be related to the magnitude and orientation of the quadrupole anisotropy in each point of the last scattering surface. Since these quadrupolar patterns reflect variations of temperature in the region of the last scattering surface, it is clear that there is a non-zero correlation between temperature and polarization maps. Still, both maps contain some independent information.

In general, a vector field can be decomposed into a gradient and a curl component, like the electric and magnetic fields. Similarly, CMB polarization maps can be expanded in two scalar maps called, by analogy, E-type and B-type polarization maps. 

One can show that primordial scalar perturbations can produce both temperature anisotropies and E-type polarization anisotropies. If the early universe only features Gaussian scalar perturbations on cosmological scales, all the information contained in CMB maps is encoded in the temperature power spectrum $C_l^{TT}$, the E-type polarization power spectrum $C_l^{EE}$, and the cross-correlation power spectrum $C_l^{TE}$. The later really contains additional independent information, because polarization patterns are only partially correlated with temperature patterns (in mathematical terms, $[C_l^{TE}]^2 \leq C_l^{TT} C_l^{EE}$).

We will mention in the next section the possibility that a significant amount of tensor perturbations are produced in the primordial universe. Such perturbations can also generate B-type polarization. If this is the case, all the information is encoded in $C_l^{TT}$, $C_l^{EE}$, $C_l^{TE}$ and $C_l^{BB}$. We did not include in this list the cross-correlation spectra $C_l^{TB}$ and $C_l^{EB}$, because the parity symmetry imposes that they should vanish (at the level of primary anisotropies).

Mapping the CMB polarization is interesting because it contains additional information with respect to temperature anisotropies.

\begin{figure}
\begin{center}
\includegraphics[width=10cm]{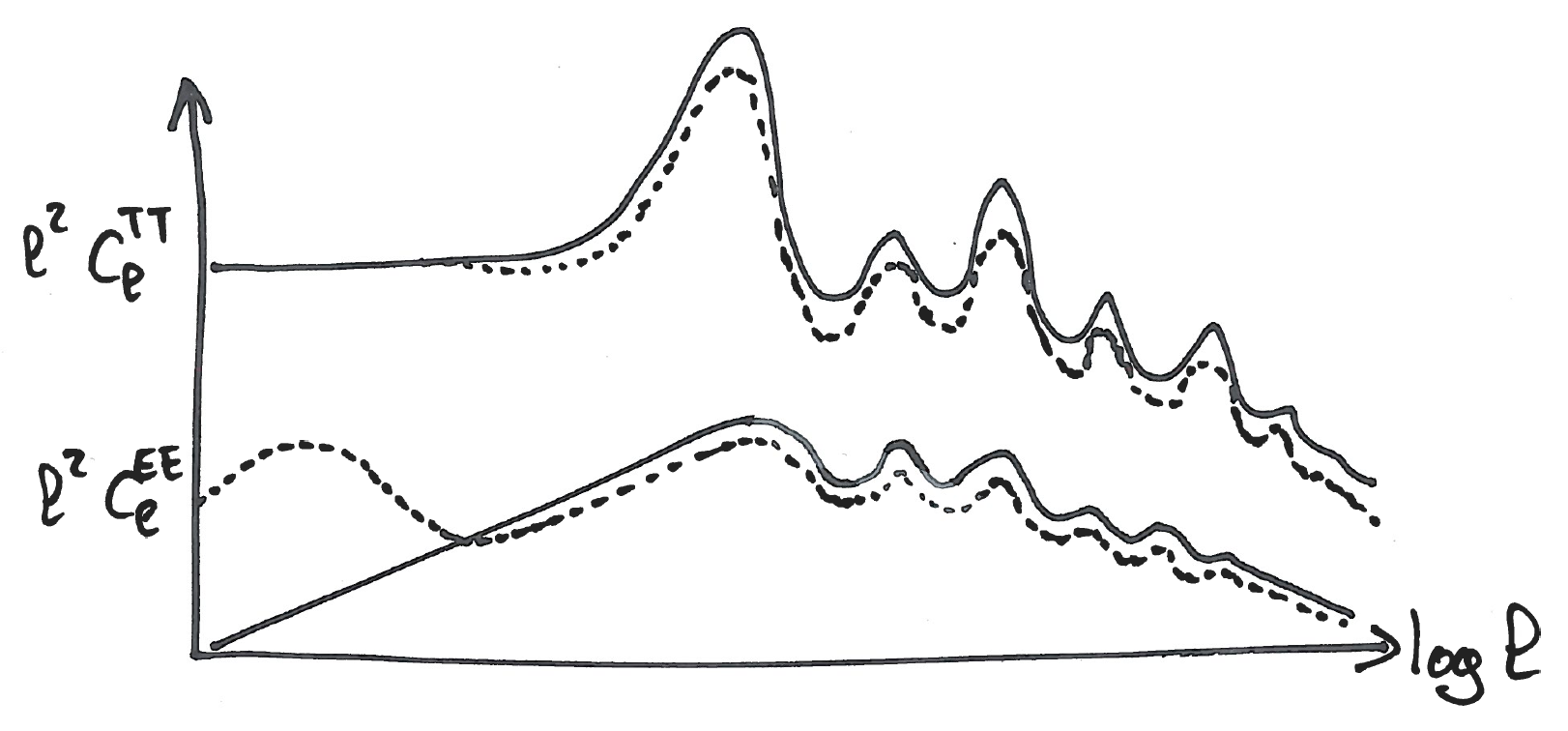}
\end{center}
\caption{Typical shape of the temperature and E-type polarization power spectra in a $\Lambda$CDM model without (solid lines) or with (dotted lines) reionization of the universe at small redshift.}
\label{jl:fig_27}
\end{figure}
First, reionization has a very distinct effect on the polarization spectrum. The small fraction of photons rescattered around the time of reionization acquire additional polarization that will show up in the form of a peak in the small-$l$ branch of $C_l^{EE}$ (see figure~\ref{jl:fig_27}) and $C_l^{TE}$. We have seen in the previous section that with temperature data only, cosmic variance limits our ability to measure $\tau_\mathrm{reio}$, and {\it a fortiori} to constrain additional parameters describing the reionization history. The effect of reionization in the small-$l$ branch of $C_l^{EE}$ and $C_l^{TE}$ is so big that despite of cosmic variance, $\tau_\mathrm{reio}$  can be well measured using polarization data.

So the effect of reionization, described before as {\tablefont(C8)}, is very different for temperature and polarization. Instead, the other effects {\tablefont(C1)} - {\tablefont(C7)} have a qualitatively similar impact on $C_l^{EE}$ and $C_l^{TE}$. Still, measuring polarization is interesting, because temperature anisotropies feature a particular combination of the  SW, Doppler and ISW effects, while polarization provides information on the quadrupole $\Theta_2$ on the last scattering surface, which is correlated with the SW and Doppler terms, but probing a different combination of them. Hence the role of polarization measurements is to remove some of the parameter degeneracies appearing in the analysis of temperature data.

Note that in the full Boltzmann equation, the equation of motion of temperature and polarization degrees of freedom are coupled. In the last sections, we neglected such a coupling. In fact, the impact of polarization on the evolution of temperature anisotropies is very small. Hence our qualitative discussion of the various effects affecting CMB temperature remains valid.

\subsection{A quick word on tensors}\label{jl:sec28}

We have seen in section~\ref{jl:sec11} that tensor modes are related to the traceless divergenceless components of the metric and stress-energy tensor, obeying to 
\begin{equation}
\sum_i \, \delta T_{ii}=0\qquad \mathrm{and}\qquad \forall j, \quad \sum_i \, \partial_i \delta T_{ij}=0~.
\end{equation}
Since these relations impose four constraints on the six degrees of freedom of the $3\times3$ symmetric matrix $\delta T_{ij}$ (or $\delta g_{ij}$), there are two tensor degrees of freedom in $g_{\mu \nu}$ and $T_{\mu \nu}$.

The two degrees of freedom in $\delta g_{ij}$ are the two degrees of polarization of gravitational waves, that can propagate even in the vacuum. In the presence of matter with a non-diagonal stress tensor $ \delta T_{ij}$, gravitational waves can be seeded by the tensor components $ \delta T_{ij}$.

If tensor perturbations are sufficiently large at recombination or later, they can generate CMB temperature and polarization anisotropies: one can show that they interact with CMB photons, and produce effects similar to the SW and ISW ones. As mentioned in the previous section, gravitational waves can even seed B-type polarization.

The stress-energy tensor is diagonal for perfect fluids, and negligible for pressureless components like CDM. Hence, during radiation and matter domination, tensor perturbations can only be seeded by decoupled neutrinos and/or decoupled photons. However, neutrino and photon tensor modes are far too small for generating a detectable amount of CMB anisotropies. However, in the early universe, other mechanisms may produce gravitational waves on scales sufficiently large to be observable in the CMB. The most famous one is related to inflation. During inflation, quantum fluctuations of the metric can excite primordial tensor perturbations at a level that will produce a detectable signature in CMB maps. The amplitude of this signal is directly proportional to the energy scale of inflation. If this scale is large enough, tensors can contribute to $C_l^{TT}$, but only on small $l$'s, because gravitational waves decay quickly inside the Hubble radius. Hence, one way to detect tensors would be through a small distortion of the $C_l^{TT}$ spectrum shape for $l\leq 100$ (i.e. for scales that are equal or larger than the Hubble radius at the time of decoupling). 

However, if the tensor signal is small with respect to the error bars associated to cosmic variance, it will remain undetectable. In that case, further sensitivity could be obtained by measuring $C_l^{BB}$: in absence of tensors,  $C_l^{BB}$ would vanish, so even if the tensor amplitude is very small it can still dominate the $C_l^{BB}$ signal. Unfortunately, this is true only up to some extent, because secondary anisotropies (in particular, those generated by weak lensing) produce a non-zero $C_l^{BB}$ that could mask the primary tensor anisotropy spectrum. 

It is very challenging for CMB experiments to reach the sensitivity level required for detecting a $C_l^{BB}$ signal, even for that generated by weak lensing.  Current limits on the tensor primordial spectrum (and on the energy scale of inflation) mainly come from the observation of $C_l^{TT}$ at small $l$ by WMAP. The sensitivity of Planck to tensors will also mainly come from temperature. Future experiments dedicated to CMB polarization will improve the sensitivity to B-type polarization and obtain more precise bounds, until they reach the theoretical limit set by the lensing contamination.

\section{Matter power spectrum}\label{jl:sec3}

\subsection{Definition}\label{jl:sec31}

The total energy perturbation in the universe can be expanded as
\begin{equation}
\delta \rho_\mathrm{tot} = \delta \rho_\gamma + \delta \rho_\mathrm{b} + \delta_\mathrm{cdm} + \delta \rho_\nu \, (+ \, \delta \rho_\mathrm{de} + ...)
\end{equation}
where $\delta \rho_\mathrm{de}$ refers to possible Dark Energy (DE) perturbations, and the three dots for extra relics. In the minimal $\Lambda$CDM model, only the first four components are present.

Many Large-Scale Structure (LSS) observables are related to the power spectrum of $\delta \rho_\mathrm{tot}$, at different wavenumbers and redshifts. This is the case of the galaxy and of the halo correlation function, of the cluster mass function, of CMB lensing, of the cosmic shear spectrum, etc.

All these observations probe the power spectrum during matter or $\Lambda$/DE domination, when photons are subdominant and $\delta \rho_\gamma$ can be neglected. If neutrinos are still ultra-relativistic today, $\delta \rho_\nu$ can also be neglected (in this course, we do not have time to discuss the impact of small neutrino masses, for which we refer the reader to Ref.~\refcite{CUP}). If the acceleration of the universe is caused by a cosmological constant, there is no term $\delta \rho_{de}$. More generally, most Dark Energy models would predict a negligible amount of DE perturbations. In summary, in a wide category of cosmological scenarii including $\Lambda$CDM, we can use $\delta \rho_\mathrm{tot} \simeq \delta_\mathrm{m} \equiv \delta \rho_\mathrm{b} + \delta_\mathrm{cdm} $. Hence, in the context of LSS observations, it is customary to refer to the power spectrum of the non-relativistic matter fluctuation $\delta_\mathrm{m}$, defined as
\begin{equation}
\delta_\mathrm{m} = \frac{\delta \rho_\mathrm{m}}{\bar{\rho}_\mathrm{m}} =  \frac{\delta \rho_\mathrm{b} + \delta \rho_\mathrm{cdm}}{\bar{\rho}_\mathrm{b} + \bar{\rho}_\mathrm{cdm}}~,
\end{equation}
which is indistinguishable from the ratio $\delta \rho_\mathrm{tot}/\bar{\rho}_\mathrm{m}$. Only in models with large dark energy perturbations or modifications of Einstein gravity, the two quantities might be different, and special care about the definition of the matter power spectrum is needed.

The matter power spectrum $P(z,k)$ of $\delta_\mathrm{m}$ is defined like in section~\ref{jl:sec15}:
\begin{equation}
\langle \delta_\mathrm{m}(z,\vec{k}) \delta_\mathrm{m}^*(z,\vec{k}') \rangle = \delta_D(\vec{k}-\vec{k}') \,\, P(z,k)~.
\end{equation} 
Here we used the redshift as a time variable, but we could have indifferently used proper or conformal time. We have seen in section~\ref{jl:sec15} that for Gaussian initial conditions and as long as perturbations are linear, the power spectrum at a given time can be written as the product of the primordial spectrum by the square of the relevant transfer function, in our case $\delta_\mathrm{m}(z,k)$. Sticking to the same conventions as in section~\ref{jl:sec15} and assuming a power-law primordial spectrum like in section~\ref{jl:sec26}, this gives
\begin{equation}
P(z,k) = \frac{2 \pi^2}{k^3} A_\mathrm{s} \left(\frac{k}{k_*}\right)^{n_\mathrm{s}-1}  \delta_\mathrm{m}^2(z,k)~.
\label{jl:pk_as}
\end{equation}
Hence, by studying qualitatively the evolution of the transfer function $\delta_\mathrm{m}(z,k)$, we will get some insight on the cosmological information encoded in the matter power spectrum.

\subsection{Transfer function evolution}\label{jl:sec32}

{\bf CDM dominated universe.}~ In order to simplify the presentation, let us first assume that we leave in a $\Lambda$CDM universe with a negligible amount of baryons: $\Omega_\mathrm{b} \ll \Omega_\mathrm{cdm}$ and $\delta_\mathrm{m} \simeq \delta_\mathrm{cdm}$. In section~\ref{jl:sec25}, we wrote the master equation governing the evolution of photon perturbations during the tightly-coupled regime. Similarly, by combining the continuity and Euler equation of CDM perturbations, we can write here a master equation for $\delta_\mathrm{cdm}$, actually valid in all regimes:
\begin{equation}
\delta_\mathrm{cdm}'' + \frac{a'}{a} \delta_\mathrm{cdm}' = - k^2 \psi + 3 \phi'' + 3 \frac{a'}{a} \phi'~.
\label{jl:cdm_clus}
\end{equation}
In an expanding universe, the clustering rate depends on the expansion rate: expansion increases distances, weakens gravitational forces, and slows down clustering processes. In the above equation, this is accounted by the second term, often called the Hubble friction term. On the right-hand side, the first term represents gravitational forces, and the last two terms account for dilation effects.

On super-Hubble scales, we have seen that in the Newtonian gauge, adiabatic ICs  predict constant density fluctuations $\delta_\mathrm{cdm}$. To be precise, $\phi$ and $\delta_\mathrm{cdm}$ vary on super-Hubble scales only when the total equation of state of the universe changes (i.e. around the time of radiation/matter equality, and during $\Lambda$ domination). Instead, they remain constant on those scale during the radiation and matter dominated regime.

Inside the Hubble radius, we can neglect dilation terms, and replace the gravitational potential term:
\begin{equation}
\delta_\mathrm{cdm}'' + \frac{a'}{a} \delta_\mathrm{cdm}' - \frac{3}{2} \left(\frac{a'}{a}\right)^2 \Omega_\mathrm{cdm}(a) \,\, \delta_\mathrm{cdm} =0~,
\label{jl:meszaros}
\end{equation}
where $\Omega_\mathrm{cdm}(a)$ is the fraction of the critical density coming from CDM at a given value of time (or of the scale factor). This equation is often called the M\'esz\'aros equation. It can be obtained by combining Eq.~(\ref{jl:cdm_clus}) with the (00) component of the Einstein equation (or its Poisson limit) and the Friedmann equation. The careful reader might have noticed something suspicious: the gravitational force term $k^2 \psi$ has been eliminated in favor of $\delta_\mathrm{cdm}$, while the first Einstein equation (or its Poisson limit) involves the total density fluctuation. Hence, shouldn't Eq.~(\ref{jl:meszaros}) feature also $\delta_\gamma$, $\delta_\mathrm{b}$ and $\delta_\nu$? Actually, it turns out that the above equation is a really good approximation for the CDM equation of evolution in all regimes, under our assumption $\Omega_\mathrm{b} \ll \Omega_\mathrm{cdm}$. It applies even during radiation domination, when photon fluctuations are potentially large. In this course, we  do not have enough time for justifying this point, but we refer the reader to Weinberg\cite{Weinberg:2002kg,Weinberg:1102255} for a detailed explanation.

During radiation domination, the Friedmann equation gives $a \propto \eta$, and $\Omega_\mathrm{cdm}$ is much smaller than one. Hence we can neglect the last term in the M\'esz\'aros equation, and find that the two solutions are $\delta_\mathrm{cdm}=\mathrm{constant}$ and $\delta_\mathrm{cdm}\propto \log \eta$. Hence CDM fluctuations grow logarithmically. During matter domination, $a \propto \eta^2$ and $\Omega_\mathrm{cdm} \simeq 1$, so the solutions are $\delta_\mathrm{cdm}\propto \eta^2$ and $\delta_\mathrm{cdm}\propto \eta^{-3}$. Then, CDM fluctuations grow quadratically with $\eta$. (These are only asymptotic solutions, but the M\'esz\'aros equation can actually be solved analytically at all times). During $\Lambda$ domination, the function $a(\eta)$ is more complicated, and $\Omega_\mathrm{cdm}$ decreases. With a bit of work, one can show that $\delta_\mathrm{cdm}$ grows at a smaller rate than during matter domination (i.e. slowlier than $\eta^2$), and that the reduction of the growth rate does not depend on $k$.

\begin{figure}
\begin{center}
\includegraphics[width=9cm]{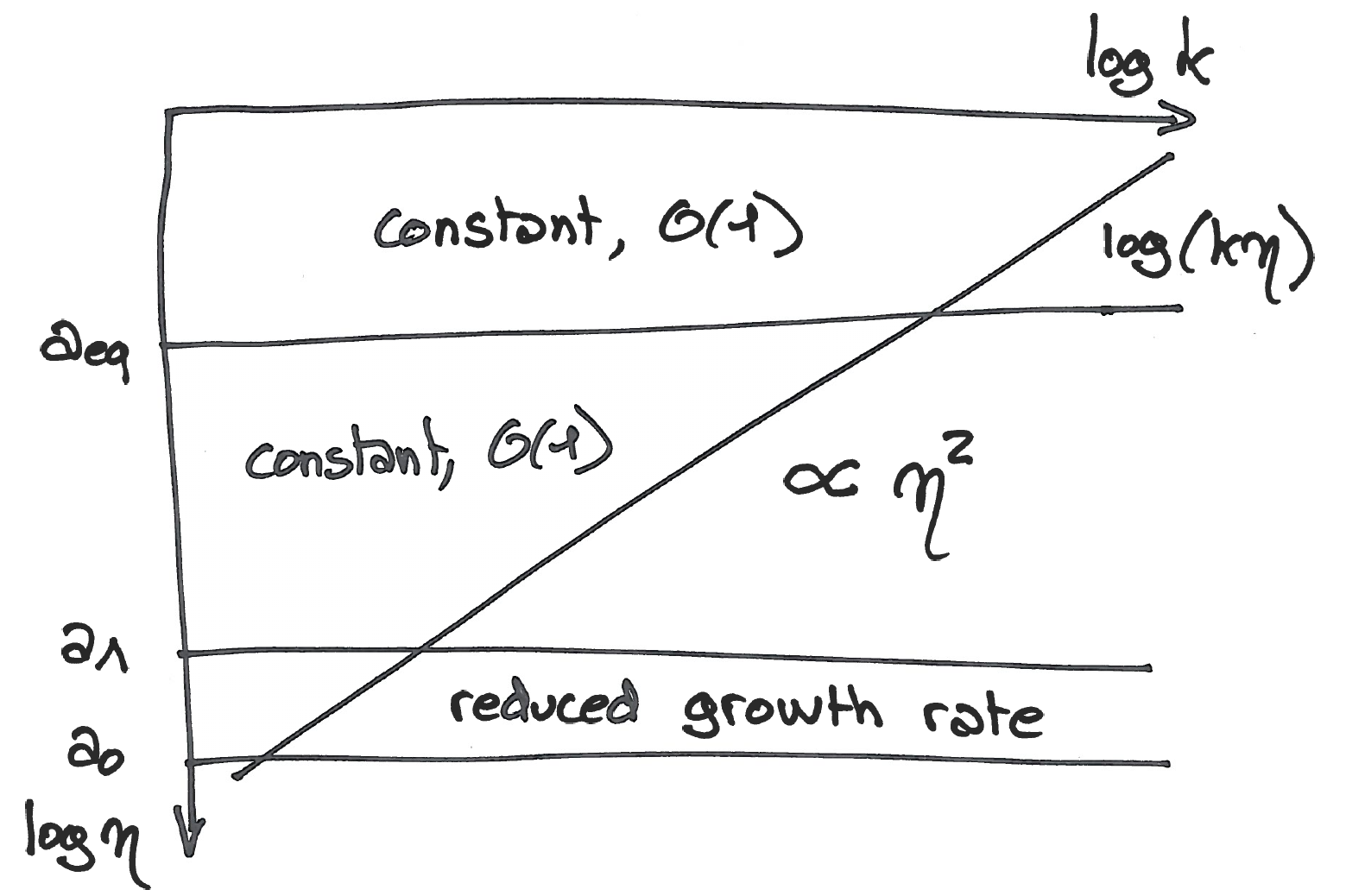}
\end{center}
\caption{Qualitative evolution of the transfer function $\delta_\mathrm{cdm}(\eta,k)$ (normalized as usual to ${\cal R}(k)=1$ at initial time) in different regimes: during radiation domination, matter domination, $\Lambda$ domination, and on super/sub-Hubble scales.}
\label{jl:fig32a}
\end{figure}

In summary, during radiation domination, $\delta_\mathrm{cdm}(\eta,k)$ is constant on super-Hubble scales and grows logarithmically on sub-Hubble scales. A more precise calculation would show that up to a numerical factor of order one, $\delta_\mathrm{cdm}$ is given on sub-Hubble scales by $\delta_\mathrm{cdm}(\eta,k)=\log(k\eta)$. During matter domination, 
$\delta_\mathrm{cdm}(\eta,k)$ is still constant on super-Hubble scales, and grows like $\eta^2$ on sub-Hubble scales. Finally, during $\Lambda$ domination, it grows more slowly. These different behaviors are reported in Fig.~\ref{jl:fig32a}.

\begin{figure}
\begin{center}
\includegraphics[width=10cm]{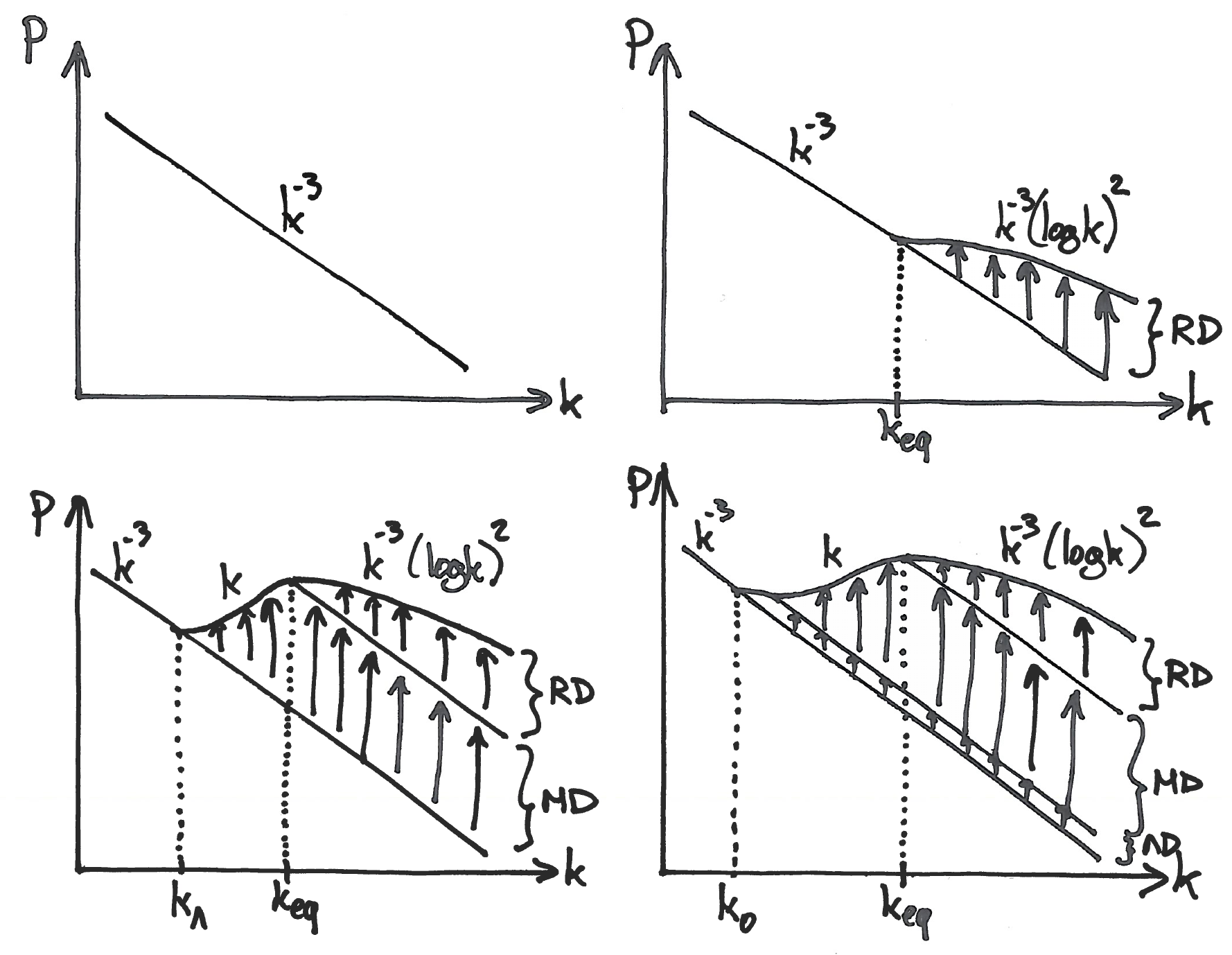}
\end{center}
\caption{Shape of the matter power spectrum $P(k)$ (log-log scale) at four different times: {\it (upper left)} when initial conditions are imposed (and all wavenumbers are super-Hubble); {\it (upper right)} at radiation/matter equality (arrows show the logarithmic growth during radiation domination); {\it (lower left)} at matter/$\Lambda$ equality (lower set of arrows show the growth during matter domination); {\it (lower right)} today (lower set of arrows show the growth during $\Lambda$ domination).}
\label{jl:fig32b}
\end{figure}

This simple discussion is sufficient for understanding the shape of the matter power spectrum at different time. Like in a cartoon, Fig.~\ref{jl:fig32b} shows this shape at four different times: at some initial time when all relevant modes are super-Hubble; at radiation/matter equality; at matter/$\Lambda$ equality; and today. Let us comment these plots. We first need to define the comoving wavenumbers corresponding to wavelengths crossing the Hubble radius at the time of radiation/matter equality, of matter/$\Lambda$ equality, and today:
\begin{equation}
k_\mathrm{eq} = a_\mathrm{eq} H_\mathrm{eq}~, \qquad
k_{\Lambda} = a_\Lambda H_\Lambda~, \qquad
k_0 = a_0 H_0~.
\end{equation}
We can now review the evolution of $P(k)$ with respect to time, following the same steps as in Fig.~\ref{jl:fig32b}.
\begin{enumerate}
\item
At initial time, if we assume a scale-invariant spectrum with $n_\mathrm{s}=1$, we see from Eq.~(\ref{jl:pk_as}) that $P(k) \propto k^{-3}$, with an amplitude given by $A_\mathrm{s}$.

\vspace{0.3cm}

\item
During radiation domination, modes grow logarithmically inside the Hubble radius, like $\log(k\eta)$. At equality, super-Hubble modes ($k_\mathrm{eq} \eta \ll 1$) are still shaped like at initial time, while sub-Hubble modes ($k_\mathrm{eq} \eta \gg 1$) have been enhanced by a factor $[\delta_\mathrm{cdm}(\eta_\mathrm{eq},k)/\delta_\mathrm{cdm}(\eta_\mathrm{ini},k)]^2 \simeq [\log(k \eta_\mathrm{eq})]^2$. The two asymptots of $P(k)$ are then given by $k^{-3}$ for $k \ll k_\mathrm{eq}$ and $k^{-3}[\log(k)]^2$ for $k \gg k_\mathrm{eq}$.

\vspace{0.3cm}

\item
At the end of matter domination, when $a=a_\Lambda$, modes still outside the Hubble radius keep being shaped like at initial time. This concerns all modes with $k \ll k_\Lambda$. Modes $k \gg k_\mathrm{eq}$ have been amplified during matter domination by a factor $[\delta_\mathrm{cdm}(\eta_\Lambda,k)/\delta_\mathrm{cdm}(\eta_\mathrm{eq},k)]^2 \simeq (\eta_\Lambda/\eta_\mathrm{eq})^4$. This factor does not depend on $k$ and preserves the shape of the power spectrum on those scales. Finally, intermediate modes entering the Hubble scale during matter domination have been amplified by $(\eta_\Lambda/\eta_*)^4$, where $\eta_*$ is their time of Hubble crossing, given approximately by $\eta_*=1/k$. Hence they have been amplified by $(k\eta_\Lambda)^4$. Putting all these informations together, we see that the spectrum has three branches, scaling respectively like:\\

\begin{tabular}{ll}
$\bullet \,\, P(k) \propto k^{-3}$ & for $k < k_\Lambda$,\\
$\bullet \,\, P(k) \propto k^{-3} k^4 = k$ & for $k_\Lambda < k < k_\mathrm{eq}$,\\
$\bullet \,\, P(k) \propto k^{-3} (\log k)^2$ & for $k>k_\mathrm{eq}$.\\
\end{tabular}

\vspace{0.3cm}

\item
During $\Lambda$ domination, $\delta_\mathrm{cdm}(k,\eta)$ grows more slowly than $\eta^2$, but it still grows at the same rate for all sub-Hubble modes. So the shape of the power spectrum today is unaltered by this stage, and given by:\\

\begin{tabular}{ll}
$\bullet \,\, P(k) \propto k^{-3}$ & for $k < k_0$,\\
$\bullet \,\, P(k) \propto k^{-3} k^4 = k$ & for $k_\Lambda < k < k_\mathrm{eq}$,\\
$\bullet \,\, P(k) \propto k^{-3} (\log k)^2$ & for $k>k_\mathrm{eq}$.\\
\end{tabular}\\

We do not enter into details for the small range of modes obeying $k_0<k<k_\Lambda$.
\end{enumerate}

All this discussion was carried under the assumption of a scale-invariant primordial spectrum. If $n_\mathrm{s}\neq1$, the above shape should simply be rescaled by $k^{n_\mathrm{s}-1}$, and the three branches of the power spectrum are given by:\\
\begin{tabular}{ll}
$\bullet \,\, P(k) \propto k^{n_\mathrm{s}-4}$ & for $k < k_0$,\\
$\bullet \,\, P(k) \propto k^{n_\mathrm{s}}$ & for $k_\Lambda < k < k_\mathrm{eq}$,\\
$\bullet \,\, P(k) \propto k^{n_\mathrm{s}-4} (\log k)^2$ & for $k>k_\mathrm{eq}$.
\end{tabular}

\vspace{0.3cm}

This closes the presentation of the shape of $P(z,k)$ in the limit $\Omega_\mathrm{b} \ll \Omega_\mathrm{cdm}$. What remains to be seen is the impact of a non-negligible baryon fraction on the power spectrum. 

\vspace{0.5cm}

{\bf Baryon corrections.}~ We must define a new important time in the evolution of the universe: the baryon drag time. The photon decoupling time was defined through the maximum of the photon visibility function $g(\eta)$. But there are many more photons than baryons in the universe. Hence, for some amount of time after photon decoupling, baryons keep tracking photon perturbations: in other words, the baryon population decouples later than the photon population (seen as a whole). 

Until the baryon drag time, we know that $\delta_\mathrm{b} = \frac{3}{4} \delta_\gamma$. After that time,  baryons do not experience significant Thomson scattering anymore. They only feel gravity, and collapse in gravitational potential wells. 

We know qualitatively the behavior of the baryon transfer function $\delta_\mathrm{b}(\eta,k)$ just before baryon drag, since in the CMB section we have studied the behavior of the photon transfer function. We know that $\delta_\mathrm{b} = \frac{3}{4} \delta_\gamma$ is constant on super-sound-horizon scales, experiences stationary oscillations on sub-sound-horizon scales during radiation domination, and finally damped oscillations on sub-sound-horizon scales during matter domination. Moreover, we expect that at any time --- and particularly at the approach of decoupling --- the relation $\delta_\mathrm{b} = \frac{3}{4} \delta_\gamma$ breaks on very small wavelengths comparable to the mean free path of baryons in the imperfect baryon-photon fluid, but here we will not discuss such small scales.

The behavior of the CDM transfer function $\delta_\mathrm{cdm}(\eta,k)$ before baryon drag is still given by the M\'esz\'aros equation: deep inside the sound horizon, baryons are much less clustered than CDM, so that CDM is almost self-gravitating.

\begin{figure}
\begin{center}
\includegraphics[width=10cm]{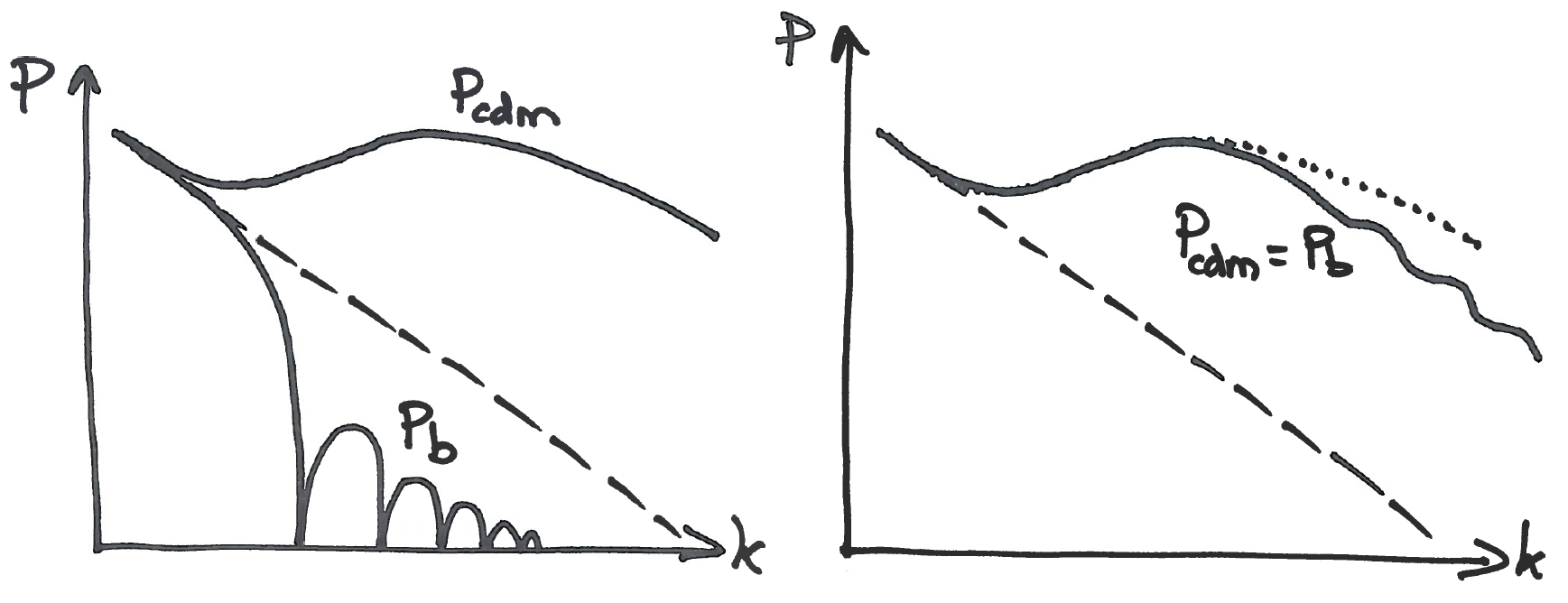}
\end{center}
\caption{{\it (Left)} Shape of the baryon and CDM power spectra just before the baryon drag time, compared to the shape of the primordial spectrum (dashed line). {\it (Right)} Shape of the matter power spectrum well after the baryon drag time, when baryons and CDM perturbations have reached gravitational equilibrium, compared to the shape of the primordial spectrum (dashed lines) and to the spectrum of a CDM-dominated universe (dotted line).}
\label{jl:fig32c}
\end{figure}

In the left plot of Fig.~\ref{jl:fig32c}, we sketch the qualitative behavior of the individual power spectrum of baryons and CDM just before baryon drag ($P_\mathrm{b}(k)$ and $P_\mathrm{cdm}(k)$ are defined as in Eq.~(\ref{jl:pk_as}), with $\delta_\mathrm{m}$ replaced either by $\delta_\mathrm{b}$ or $\delta_\mathrm{cdm}$).

After baryon drag, baryons collapse in potential gravitational wells. Since $\delta_\mathrm{b}$ grows, CDM will start to feel the gravity of baryons. Finally, because CDM and baryons are two collisionless species feeling the same gravitational forces, they will equilibrate with $\delta_\mathrm{b}=\delta_\mathrm{cdm}$. The gravitational potential is then related to this common value by the (00) Einstein equation (or its Poisson sub-Hubble limit). This is not true however on very small scales, for which the baryon pressure cannot be neglected, but here we do not discuss such small scales.

Hence, after a quick relaxation period, the two power spectra of baryons and CDM are equal to each other, $P_\mathrm{b}(k)=P_\mathrm{cdm}(k)=P(k)$. In order to relate this common power spectrum to the individual power spectrum of baryons and CDM before baryon drag, one must follow a matching process described in details in Einsteinstein and Hu~\cite{Eisenstein:1997ik}. Intuitively, the matching depends on the relative weight of baryons and CDM, i.e. on the ratio $\Omega_\mathrm{b}/\Omega_\mathrm{cdm}$. In the limit $\Omega_\mathrm{b} \ll \Omega_\mathrm{cdm}$, what we wrote at the beginning of this section applies. In the limit $\Omega_\mathrm{b} \gg \Omega_\mathrm{cdm}$, the power spectrum is very suppressed with respect to the CDM-dominated case, with a much more negative slope on average, and some large oscillations corresponding to photon-baryon acoustic waves before decoupling. Finally, if $\Omega_\mathrm{b}$ is a bit smaller than $\Omega_\mathrm{cdm}$ but not negligible (as it is the case in our universe), the power spectrum departs slightly from a pure CDM one, with a smooth step-like suppression plus small oscillatory patterns, which are the smoking gun in the recent universe of baryon-photon acoustic oscillations happening before photon decoupling. This is illustrated on the right plot of Fig.~\ref{jl:fig32c}. These oscillations are called Baryon Acoustic Oscillations (BAO).

\vspace{0.2cm}

LSS observations can only probe the power spectrum on scales much smaller than the radius of the observable universe, i.e. than the current Hubble radius. For this reason, the first of the three branches described above is unobservable\footnote{The first unobservable branch is actually gauge-dependent: it behaves like $k^{n_\mathrm{s}-4}$ in the Newtonian gauge, but not in other gauges. Note that truly observable quantities are always gauge-invariant. We have been a bit uncareful when saying that LSS data probes the power spectrum of $\delta_\mathrm{m}$. In fact, different LSS observations probe different quantities, each of them being gauge-invariant. However, well inside the Hubble region, all these quantities coincide with each other and with the spectrum of $\delta_\mathrm{m}$ computed in an arbitrary gauge (up to small corrections that may become important in the future, but not with current experimental sensitivities).}. We can only measure the second and third branches, behaving respectively like $P(k\ll k_\mathrm{eq}) \propto k^{n_\mathrm{s}}$ and, in first approximation, $P(k\gg k_\mathrm{eq}) \propto[k^{n_\mathrm{s}-4} (\log k)^2]$, plus the titling and superimposed oscillations coming from baryons. 

\subsection{Parameter dependence}\label{jl:sec33}

The discussion presented in the previous section allows us to understand which effects and which parameters impact the shape of the matter power spectrum in the minimal (flat) $\Lambda$CDM model, parametrized by
\begin{equation}
\{A_\mathrm{s}, n_\mathrm{s}, \omega_\mathrm{b}, \omega_\mathrm{m}, \Omega_\Lambda, \tau_\mathrm{reio} \}
\label{jl:lcdm_basis2}
\end{equation}
(see Sec.~\ref{jl:sec26} for details on this parametrization). We can already notice that the reionization optical depth is relevant for the CMB spectrum but not for $P(k)$, since it only impacts the scattering rate of photons in the recent universe. Other effects are described in Table~\ref{jl:Peffects}.
\begin{table}
\tbl{Independent leading effects controlling the shape of the matter power spectrum $P(k)$  in the minimal $\Lambda$CDM model.}
{\tablefont
\begin{tabular}{lllc}
\toprule  & Effect & Relevant quantity &  Parameter\\
\colrule
(P1) & scale of the maximum & $k_\mathrm{eq}$ & $\omega_\mathrm{m}, \Omega_\Lambda$ \\
&&&\\
(P2) & \hspace{-0.2cm}\begin{tabular}{l}slope for $k \gg k_\mathrm{eq}$\\ and BAO amplitude\end{tabular} & $\Omega_\mathrm{b}/\Omega_\mathrm{cdm}$ & $\omega_\mathrm{b}, \omega_\mathrm{m}$ \\
&&&\\
(P3) & BAO scale & $r_\mathrm{s}(\eta_\mathrm{drag}$) & $\omega_\mathrm{b}, \omega_\mathrm{m}$ \\
&&&\\
(P4) & Global amplitude & \hspace{-0.2cm}\begin{tabular}{l}amplitude of primordial spectrum\\and duration of $\Lambda$D\end{tabular} &  $A_\mathrm{s}, \Omega_\Lambda$ \\
&&&\\
(P5) & Global tilt & tilt of primordial spectrum & $n_\mathrm{s}$ \\
&&&\\
\botrule
\end{tabular}
}
\label{jl:Peffects}
\end{table}
Below, we give more details on these effects.
\begin{description}
\item[\tablefont(P1)] The time of equality determines the scale $k_\mathrm{eq}$ of the power spectrum peak. More precisely, if this scale is expressed in units of $h\,$Mpc$^{-1}$, which is the usual convention, then one can show that the scale of the maximum depends on both $z_\mathrm{eq}$ and $\Omega_\mathrm{m}$ (i.e. on our parameter basis on $\omega_\mathrm{m}$ and $\Omega_\Lambda=1-\Omega_\mathrm{m}$).
\item[\tablefont(P2)] The baryon abundance (relative to CDM) is crucial at the matching time: a high baryon abundance leads to more suppression of the power spectrum for $k \geq k_\mathrm{eq}$, and to more pronounced BAOs.
\item[\tablefont(P3)] We have seen that the scale of acoustic oscillations is set by the sound horizon at a given time. Since photons decouple at $\eta_\mathrm{dec}$, the scale of oscillations on the last scattering surface is set by $d_s(\eta_\mathrm{dec})$, corresponding to the comoving scale $r_s(\eta_\mathrm{dec})=d_s(\eta_\mathrm{dec})/a(\eta_\mathrm{dec})$. Similarly, since baryons decoupled at $\eta_\mathrm{drag}$, the scale of BAOs depends on $d_s(\eta_\mathrm{drag})$ at baryon drag, corresponding to the comoving scale $r_s(\eta_\mathrm{drag})=d_s(\eta_\mathrm{drag})/a(\eta_\mathrm{drag})$. As explained in Sec.~\ref{jl:sec26}, the sound horizon at a given time depends on $\omega_\mathrm{b}$ and $\omega_\mathrm{m}$. In the case of $d_s(\eta_\mathrm{drag})$ the dependence on $\omega_\mathrm{b}$ is even stronger because the time of baryon drag itself depends strongly on the baryon abundance.
\item[\tablefont(P4)] The global amplitude depends of course on the primordial spectrum amplitude, i.e. on $A_\mathrm{s}$. Also, we have seen that during $\Lambda$ domination, the growth rate of fluctuations is reduced with respect to matter domination, but is independent of $k$ for all sub-Hubble scales. Hence, $\Omega_\Lambda$ also affects the global amplitude of the power spectrum.
\item[\tablefont(P5)] The global tilt depends of course on the primordial spectrum tilt, i.e. on $n_\mathrm{s}$. 
\end{description}
This discussion shows that in principle, a precise measurement of the matter power spectrum today (or at a given redhsift) would allow to measure independently $\omega_\mathrm{b}$, $\omega_\mathrm{m}$, $n_\mathrm{s}$, $A_\mathrm{s}$ and $\Omega_\Lambda$ (assuming a flat $\Lambda$CDM universe). In practice, given the limited precision of current data sets, some of the effects described above are degenerate with each other, and matter power spectrum observations are mainly useful in combination with CMB observations, since they bring independent and complementary information. 

Future experiments will use all the discriminating power of the matter power spectrum. They will perform accurate measurements of $P(k,z)$ at different redshifts. This is crucial for at least two reasons:
\begin{itemize}
\item
First, the effect of $\Omega_\Lambda$ in {\tablefont(P5)} is redshift dependent: at different redshifts, fluctuations have spent more or less time during the $\Lambda$ dominated regime, and the power spectrum amplitude has been more or less affected. Hence, the comparison of the amplitude at different redshifts allows to measure $\Omega_\Lambda$, or more generally to test the compatibility of the data with a cosmological constant (rather than some dynamical Dark Energy model).
\item
Second, BAOs appear at a fixed comoving wavenumber $k_\mathrm{BAO}$ in Fourier space (related to $r_s(\eta_\mathrm{drag})$), but if we measure it in different LSS datasets at different redshifts (corresponding to different shells in real space), this scale will be seen under different angles\footnote{Here we are assuming that the BAO scale is measured in each redshift shell transversally, i.e. in the direction orthogonal to the line of sight. Real experiments probe the BAO scale both transversally and longitudinally, so the situation is a bit more subtle than in this simplified discussion.}. By comparing the BAO angle at different redshifts, one can reconstruct the angular diameter distance, and therefore the expansion history at different redshifts. This is another way of measuring $\Omega_\Lambda$ or testing the $\Lambda$ model versus DE models.
\end{itemize}

\section{Effects of dark matter on CMB and matter spectrum}\label{jl:sec4}

The existence of Dark Matter (DM) was first postulated by astronomers in order to account for the velocity of stars in the Milky Way (Oort, 1932), and that of galaxies in the Coma cluster (Zwicky, 1933). Later on, it has been firmly established that if general relativity applies to galactic scales and above, some form of DM halo must be present in most galaxies and clusters. 

\vspace{0.2cm}

\noindent {\it \bf Most of the dark matter is cold.}
If DM is made of collisionless and stable particles, the only DM property that can be probed by cosmological perturbations is its velocity distribution.

At the level of homogeneous cosmology, the DM phase-space distribution $f$ only depends on momentum $p$. The number of particles in a given momentum bin is given by $dp\,p^2f(p)$. If  $p^2f(p)$ is very peaked near $p=0$ or the DM mass is very large, DM particles have a negligible velocity dispersion. In this case, they can be represented effectively as a pressureless fluid, as explained in Sec.~\ref{jl:sec13}. We recall that they are not forming a real fluid, due to the absence of interactions, but since their velocity dispersion is negligible, all particles have approximately the same bulk velocity in each point (described by the divergence $\theta_\mathrm{cdm}$). In that case the DM component is called Cold Dark Matter (CDM).

If $p^2f(p)$ is not very peaked near $p=0$ or the DM mass is not very large, a significant fraction of DM particles may have a velocity $p/m$ comparable to the speed of light, or at least to that of astrophysical objects. Typically, if the velocity dispersion is of the order of  $10^{-3}$ (in units of $c$) or larger, it cannot be neglected when studying the CMB or LSS spectra. In that case, the velocity of individual particles results from two physical mechanisms: the dynamics imposed by gravitational forces, plus an intrinsic velocity distributed stochastically, according to the phase-space distribution. The notion of ``velocity dispersion'' refers to the second component, since the first effect tends to impose the same bulk velocity to all particles in a given point. DM with a large velocity dispersion cannot be described by fluid equations like CDM. A famous example of DM particles with a large velocity dispersion is neutrinos. The velocity dispersion is then fixed by the neutrino mass, and by their Fermi-Dirac distribution. Such DM is called Hot Dark Matter. Possible DM species with a smaller (but still significant) velocity dispersion are called Warm Dark Matter\footnote{The limit between HDM and WDM is usually placed in the following way: both types of particles have a sizable velocity dispersion, but most HDM particles become non-relativistic after radiation/matter equality, while most WDM particles become non-relativistic before that time.}.

Since CDM is a very generic limit, and neutrinos are know to exist since the 1950's, the two most popular DM models until the 1990's were CDM and HDM. CDM can be motivated by many possible extensions of the standard model of particle physics (supersymmetry, Peccei-Quinn symmetry, etc.), predicting some early decoupled stable particles, with a very small velocity dispersion. HDM was motivated at that time by massive neutrinos with masses of the order of $\sim 10$~eV.

These two models lead to very different predictions for the matter power spectrum. We have discussed before the case of CDM. A study of perturbations in a HDM (or $\Lambda$HDM) model would show that for $k\geq k_\mathrm{eq}$, the power spectrum $P(k)$ would decrease with a much steeper slope. The physical explanation is that the velocity dispersion of HDM particles prevents them from clustering on small scales: hence the growth rate of $\delta_\mathrm{m}$ inside the Hubble radius is considerably reduced in the HDM model.

In the 90's, the first estimates of the matter power spectrum $P(k)$ (or of some related quantities) on galaxy scales and cluster scales showed clearly that the HDM scenario was ruled out. Today, using more accurate LSS data, it is firmly established that most of the DM must be cold (or eventually, warm, but with such a small velocity dispersion that WDM would be indistinguishable from CDM at least on the largest observable scales). Still, we cannot exclude the presence of a subdominant hot or warm component. In fact, since we know that neutrinos have a small mass (of the order of $\sim 0.1$~eV), we do expect that a few percents of DM is hot. Since the signature of this sub-leading component has not yet been detected, the minimal model is still $\Lambda$CDM.

In the $\Lambda$CDM case, cosmological perturbations can only measure one DM-related parameter: its relic density, parametrized by $\omega_\mathrm{cdm}$. This is better than nothing: the value of $\omega_\mathrm{cdm}$ can be used to constrain some fundamental parameters governing DM production, within the framework of a postulated model.

\vspace{0.2cm}

\noindent {\it \bf Measuring $\omega_\mathrm{cdm}$ with the CMB.}
Going back to Sec.~\ref{jl:sec26}, we see that $\omega_\mathrm{cdm}$ can be measured ``indirectly'' using CMB observations: effect {\tablefont(C2)} measures $\omega_\mathrm{b}$, effect {\tablefont(C3)} measures $\omega_\mathrm{m}$, and $\omega_\mathrm{cdm}$ is given by the difference between these two parameters.

There is no more ``direct'' effect of CDM that could be probed with the CMB. A priori, we could think that photons are affected by gravitational forces generated by CDM. This gravitational interaction could only be relevant before photon decoupling, and around the time of sound horizon crossing for each given wavelength (we have seen in Sec.~\ref{jl:sec24} that well inside the sound horizon, photons do not feel gravitational forces anymore, because pressure forces are much larger). But around sound horizon crossing, photons feel mainly the gravitational forces generated by themselves, by neutrinos, and to a lesser extent by baryons. In fact, in this regime, CDM and other components are effectively decoupled\cite{Weinberg:2002kg}. 

Hence the CMB probes the CDM abundance only indirectly, as a ``missing mass'', exactly like in astrophysics: it measures the total matter density through the time of equality, the baryon density through the ratio between odd and even peaks, and the difference must be attributed to a non-relativistic component not interacting with the photons: Dark Matter. 

In this discussion, we concentrated only on effects {\tablefont(C2)} and {\tablefont(C3)}. Indeed, effects {\tablefont(C1)} and {\tablefont(C4)} are also sensitive to $\omega_\mathrm{b}$ and $\omega_\mathrm{cdm}$, but only in combination with other parameters. So, unlike {\tablefont(C2)}+{\tablefont(C3)}, they do not bring a robust and nearly model-independent evidence for DM.

\vspace{0.2cm}

\noindent {\it \bf Measuring $\omega_\mathrm{cdm}$ with the matter power spectrum.}
The overall shape of $P(k)$ for $k\geq k_\mathrm{eq}$ brings very strong support for CDM (or eventually WDM with a small enough velocity dispersion). Indeed, a universe with only baryonic matter, or baryonic plus hot dark matter, would have a radically different power spectrum (with a very steep slope  for $k\geq k_\mathrm{eq}$, and eventually huge BAOs in the purely baryonic case). More precisely, data on the matter power spectrum allows to measure the ratio $\omega_\mathrm{b}/\omega_\mathrm{cdm}$ through effect  {\tablefont(P2)} of Sec.~\ref{jl:sec33}. Hence the matter power spectrum probes the CDM abundance independently of the CMB. This probe is more direct, since $P(k)$ is really sensitive to the gravitational clustering of CDM inside the Hubble radius.

\vspace{0.2cm}

\noindent {\it \bf Signatures of DM annihilation or decay in the CMB.}
Depending on the underlying particle physics model, DM particles could either annihilate or decay. Of course, the annihilation or decay rate should be very small, since we need a stable dark matter population. Still, a small annihilation or decay rate could play a role in the recent thermal history of the universe, and in particular in the evolution of the ionized fraction of atoms, $x_e(z)$, to which the CMB is sensitive. Indeed, DM annihilation/decay products may include photons that could directly ionize hydrogen atoms, or particles that could reheat the inter-galactic medium. These mechanisms tend to increase $x_e$ at a given time. Their efficiency depends on the mass of DM particles, on their cross-section in the case of annihilation, and on their lifetime in the case of decay.

Before recombination and photon decoupling, the ionization fraction is anyway very close to one. In the minimal cosmological model, $x_e$ drops abruptly after recombination, freezes out when particles leave chemical equilibrium, and increases again to one at the time of reionization, due to ionizing photons emitted at star formation. DM annihilation or decay could potentially modify this picture and play two roles:
\begin{itemize}
\item at high redshift (typically $600<z<1100$), it could slow down the drop of $x_e$ after recombination, and increase its freeze-out value. This would appear in the CMB as a shift in the time of decoupling (impacting effect {\tablefont(C1)} and {\tablefont(C4)} of Sec.~\ref{jl:sec26}), as a broadening of the peak of the visibility function $g(\eta)$ (i.e. the last scattering ``surface'', which is in fact more like a shell, would be thicker, impacting {\tablefont(C4)}), and as an increase of the optical depth to recombination (impacting {\tablefont(C8)}, but not on the same scales as reionization).
\item at small redshift (typically $z<40$), the concentration of DM in halos could enhance its annihilation/decay rate, and annihilation/decay products could reionize atoms and increase $x_e$. This would be equivalent to some extra reionization: it could modify the reionization time or history, and impact {\tablefont(C8)}.
\end{itemize}

The recent literature shows that for realistic models, CDM annihilation could play these two roles: hence, interesting bounds on DM annihilation can be derived from the CMB. These bounds are complementary from those inferred from cosmic ray data, or direct detection experiments. 

DM decay could play the second role, but the bounds on the DM lifetime inferred in this way from the CMB are not as strong as those from cosmic ray data.

These effects had to be mentioned in this section, because they are part of the interplay between DM physics and cosmological perturbation theory. We do not discuss here bounds on DM annihilation/decay from cosmic ray data, since they have little to do with the evolution of linear cosmological perturbations.

\vspace{0.2cm}

\noindent {\it \bf What else could we learn on DM from CMB or LSS data?} In the simplest cosmological scenario, cosmological perturbations only probe the relic abundance of DM particles. If we are lucky, they may provide some hint of DM annihilation. There are other properties of DM that can be tested with cosmological perturbations, even if at the moment, everything is consistent with the minimal $\Lambda$CDM picture:
\begin{itemize}
\item the small mass of neutrinos implies a small fraction of HDM, with distinct effects on the CMB and LSS power spectrum\cite{CUP} (effects on $P(k)$ are larger: a small neutrino mass implies a step-like suppression of the power spectrum on small scales, by at least 4\%  if we assume the smallest possible neutrino masses that are consistent with neutrino oscillations experiments).
\item a non-negligible velocity dispersion of the dominant DM component (like in the WDM model) or of a fraction of it (in the case of several DM species) would also induce a suppression of the small-scale matter power spectrum, at a scale and with a shape depending on the mass and momentum distribution of the warm species. However, such a feature is clearly not observed in the matter power spectrum on linear scales. Such a signature could only be found on smaller scales, affected by non-linear gravitational clustering.
\item instead of being treated as collisionless, DM could experience small interactions with itself or with other species (photons, neutrinos baryons). All these possibilities require a case-by-case analysis, and usually predict observable features similar to WDM. Both the CMB and the matter power spectrum can be used to put very strong bounds on a possible DM interactions.
\end{itemize}

In conclusion of this section, it appears that CMB and LSS observations bring some crucial information concerning the abundance and the nature of Dark Matter. But as long as everything is compatible with the minimal $\Lambda$CDM scenario,  CMB and LSS data are only telling us what DM cannot consist of, rather than what it is.

\section{Effects of dark energy on CMB and matter spectrum}\label{jl:sec5}

In this course, we assumed for simplicity that the (at least apparent) acceleration of the universe is caused by a cosmological constant. The possibility that acceleration would be caused by a dynamical component, called generically Dark Energy (DE), is one of the most investigated questions in cosmology. Of course, the acceleration could also be caused by some alternative theory of gravity, different from General Relativity on very large scales. Even in that case, everything could be reformulated in terms of a dark energy component (by moving new terms from the left to the right hand-side of the Einstein equation). Hence, measuring DE parameters is a rather generic way to look for deviations from a simple cosmological constant\footnote{Note that alternatively, the observed apparent acceleration might be the consequence of a departure from the homogeneous Friemann-Lema\^{\i}tre model, but we will not discuss such a possibility (which is very constrained and maybe already excluded by observations) in this course.}.

\vspace{0.2cm}

\noindent {\it \bf Possible Dark Energy parameters.}
At the background level, the stress-energy tensor of any species compatible with the assumption of a homogeneous and isotropic FL universe must be diagonal, involving only two function $\bar{\rho}_\mathrm{de}(a)$ and $\bar{p}_\mathrm{de}(a)$, related to each other by an equation of conservation of energy. Hence these two functions are not independent of each other, at least as long as DE is not interacting with other species. Even if in general a DE model has no reasons to obey to an equation of state, one can always define the function $w(a)\equiv \bar{p}_\mathrm{de}(a)/\bar{\rho}_\mathrm{de}(a)$.

A cosmological constant corresponds to $w(a)=-1$, i.e. to a constant density $\bar{\rho}_\mathrm{de}=-\bar{p}_\mathrm{de}$, parametrized by the value of $\Omega_\mathrm{\Lambda} = \Omega_\mathrm{de} \equiv \bar{\rho}_\mathrm{de} / \rho_\mathrm{crit}$ today. 

A simple extension of this model is the $w$CDM scenario, in which one assumes a linear and constant equation of state $\bar{p}_\mathrm{de}=w\bar{\rho}_\mathrm{de}$, leading to $\bar{\rho}_\mathrm{de}\propto a^{-3(1+w)}$ by virtue of energy conservation. This model has two independent DE parameters, $\Omega_\mathrm{de}$ and $w$.

However, $w$ could in principle vary with time. In a purely phenomenological approach, after confronting the $w$CDM model to observations, one can test a model with one more free parameter, accounting for the variation of $w$ today. Then, there are three DE parameters, for instance $\Omega_\mathrm{de}$, $w(a_0)$ and $\frac{dw}{da}(a_0)$.

At the level of perturbations, different DE models with a given background evolution could differ through their pressure perturbation, or their anisotropic stress. First, let us stress that DE cannot be a perfect fluid. If this was the case, DE would have an adiabatic sound speed obeying to
\begin{equation}
c_a^2 \equiv \frac{\bar{p}'}{\bar{\rho}'} = \frac{(w \bar{\rho})'}{\bar{\rho}'}
=  w + \frac{w'}{3 \frac{a'}{a} (1+w)}
\end{equation}
where in the last equality we have used energy conservation. Given observational constraints on $w$ and $w'$ today, the squared adiabatic sound speed of dark energy must be negative. If the actual squared sound speed $c_s^2 \equiv\frac{\delta p_\mathrm{de}}{\delta \rho_\mathrm{de}}$ was negative, then DE perturbations would be unstable, and would explode exponentially, in contradiction with observations. This proves that for DE, $c_s^2\neq c_a^2$. The stability of DE perturbations imposes $c_s^2 > 0$. On the other hand, the absence of superluminal propagation in conventional physics imposes an upper bound, $c_s^2\leq 1$.

In phenomenological studies of DE models, the default assumptions concerning DE perturbations is that {\it (i)} pressure perturbations are related to density perturbations through $c_s^2=1$, and that {\it (ii)} anisotropic stress is negligible. This is justified by the fact that DE could be a scalar field (often called quintessence). If this is the case, one can easily show that the sound speed is indeed equal to one\footnote{Actually, the quantity ${\delta p_\mathrm{de}}/{\delta \rho_\mathrm{de}}$ is gauge dependent. Everything becomes clearer if one decides to call ``sound speed'' the ratio ${\delta p_\mathrm{de}}/{\delta \rho_\mathrm{de}}$ calculated {\it in the frame comoving with the DE fluid}, i.e. in a gauge such that the velocity divergence $\theta_\mathrm{de}$ vanishes. If the sound speed of a given component derives from some fundamental principle, it should take a simple form precisely in this frame. Actually, it is only when this condition is imposed that one finds $\delta p_\mathrm{de}=\delta \rho_\mathrm{de}$ for a scalar field.}, while anisotropic stress only appears at second order in perturbations.

A DE model with $c_s^2=1$ cannot lead to sizable DE perturbations. Indeed, $c_s^2=1$ implies that pressure forces in the fluid resist to gravitational collapse. Astrophysicists like to formulate this problem in terms of the Jeans instability. In this language, $c_s^2=1$ means that the Jeans length in the fluid is as large as the Hubble radius. Then DE perturbations remain as small as in the primordial universe (of the level of $10^{-5}$), and can never leave an observable signature, even during DE domination.

Only if $c_s^2$ is assumed to be very small, DE perturbations can grow and impact LSS observables. Indeed, if $c_s^2$ goes to zero, the Jeans length can be arbitrarily small, and perturbations on cosmological scales can be amplified by the Jeans instability. In that case, one needs to postulate whether DE perturbations feature anisotropic stress or not. In a phenomenological approach, one may try to parameterize the anisotropic stress using some effective parameter (like a viscosity parameter). One should stress that a very small sound speed can be assumed, but these models sound artificial: $c_s^2 \ll 1$ is not a generic prediction of ``realistic'' DE models (if any such model exist).

\vspace{0.2cm}

\noindent {\it \bf Probing DE models with CMB and LSS data.}
If the DE sound speed is not very small, DE models can only be probed through their impact on the background evolution, which affects indirectly the evolution of metric and matter perturbations. 

Since the background expansion at late time impacts the angular diameter distance, the CMB is sensitive to DE through effects {\tablefont(C1)} and {\tablefont(C4)} of Sec~\ref{jl:sec26}, both depending on $d_\mathrm{a}(z_\mathrm{dec})$. However, {\tablefont(C1)} and {\tablefont(C4)} are also sensitive to many other quantities, so this probe is only useful in combination with other ones: in the best case it may remove parameter degeneracies.

The LISW effect {\tablefont(C7)} probes the duration of the DE dominated era, and could bring one bit of information on DE models. In practise, it is however impossible to infer precise constraints from {\tablefont(C7)} due to the large cosmic variance affecting the $C_l$'s on the scale of the Sachs-Wolfe plateau. 

We have seen in Sec~\ref{jl:sec33} that LSS data can be useful especially if the matter power spectrum is measured accurately at several distinct redshifts. Then, both the evolution of the power spectrum amplitude (in other words, of the linear growth factor) and of the BAO apparent scale encode information on the expansion rate at several redshifts, and therefore on $w$ or even $w(a)$.  This is why measurements of $P(k,z)$ with future ambitious programs, like for instance the Large Synoptic Survey Telescope or the Euclid satellite, are expected to provide the most accurate tests of DE models. 

In summary, the CMB is not very useful for testing DE models, but can bring a few independent pieces of information; while LSS data at different redshifts is sensitive to the evolution of the background DE density. LSS data should anyway be used in combination with other probes of homogeneous cosmology (like supernovae luminosity) in order to get a maximum amount of information on DE.

The situation that we depicted would change radically if: 
\begin{itemize}
\item
DE would account for a significant fraction of the total energy density even at early times ($z \gg 1$), like in ``early dark energy'' models: then, DE could affect the CMB more strongly than assumed in the previous discussion;
\item
large DE perturbations would develop as a consequence of a very small sound speed $c_s^2 \ll 1$: then the total power spectrum (including DE fluctuations) may depart significantly from the $\Lambda$CDM one.
\end{itemize}
Both possibilities are still open, but no such effects are indicated by current data, and there are no solid theoretical motivations for our universe to fall in one of these two cases.

\section{Conclusions and bibliography}

We hope that this short --- but very dense --- introduction to cosmological perturbations was clear enough, and that the reader will feel like going further. 

Some readers might regret that, after exposing in details the effect of cosmological parameters on the CMB and LSS spectra (in Secs.~\ref{jl:sec26}, \ref{jl:sec33}, \ref{jl:sec4} and \ref{jl:sec5}), we did not show concrete bounds from current experiments. There are three reasons for this. First, the duration of the lectures and the length of these notes would have exploded. Second, such information would be outdated very quickly by new data releases. Third, it is actually a very good exercise (and an efficient way to digest this course) to look at published bounds on cosmological parameters (inferred from CMB data, or LSS data, or from both), and to figure out by oneself whether everything is consistent with these notes: we encourage the reader to check that the best and worst constrained parameter combinations can be related to the physical effects described here.

We promised a bibliography in the introduction. We already mentioned Ref.~\refcite{CUP}, in which one can find a similar but more detailed presentation, plus a discussion of neutrino effects. We have cited in the core of the text three seminal papers (Refs.~\refcite{Seljak:1996is,Weinberg:2002kg,Eisenstein:1997ik}) providing explicit calculations on precise topics: the line-of-sight integral in Fourier space\cite{Seljak:1996is}, the justification of the M\'esz\'aros equation and of the fact that CDM behaves effectively as a self-gravitating species during radiation domination\cite{Weinberg:2002kg}, and the matching between CDM and baryon perturbations around the time of baryon drag, in view of computing analytically baryon effects in the matter power spectrum\cite{Eisenstein:1997ik}.

Specialists of the field refer regularly to a concise paper of Ma \& Bertschinger~\cite{Ma:1995ey}, containing a remarkably clear presentation of gauge issues and of all equations of motion in two gauges (including the Newtonian one).

Many recent textbooks cover most topics in cosmology, including linear perturbations, CMB and large scale structure. We only mention in the reference list those of S.~Dodelson\cite{Dodelson:1282338}, V.~Mukhanov\cite{Mukhanov:991646} and S.~Weinberg\cite{Weinberg:1102255}, all three outstanding, but there exist many excellent alternatives. For a presentation more focused on the CMB, we recommend the thesis of W.~Hu\cite{Hu:1995em} and the book of R.~Durrer\cite{Durrer:1127831}. 

\section*{Acknowledgements}

I warmly thank all TASI 2012 organizers --- and particularly Elena Pierpaoli --- for giving me the opportunity to lecture in such an inspiring atmosphere, and for patiently collecting and processing the lecture notes. I am grateful to the TASI students for their interest and feedback.

\bibliographystyle{ws-procs9x6}
\bibliography{lesgourgues}

\begin{thebibliography}{10}

\bibitem{CUP}
J.~Lesgourgues, G.~Mangano, G.~Miele and S.~Pastor, {\em Neutrino Cosmology}
  (Cambridge Univ. Press, Cambridge, 2013).

\bibitem{Seljak:1996is}
U.~Seljak and M.~Zaldarriaga, {\em Astrophys.J.} {\bf 469}, 437 (1996).

\bibitem{Weinberg:2002kg}
S.~Weinberg, {\em Astrophys.J.} {\bf 581}, 810 (2002).

\bibitem{Weinberg:1102255}
S.~Weinberg, {\em Cosmology} (Oxford Univ. Press, Oxford, 2008).

\bibitem{Eisenstein:1997ik}
D.~J. Eisenstein and W.~Hu, {\em Astrophys.J.} {\bf 496}, p. 605 (1998).

\bibitem{Ma:1995ey}
C.-P. Ma and E.~Bertschinger, {\em Astrophys.J.} {\bf 455}, 7 (1995).

\bibitem{Dodelson:1282338}
S.~Dodelson, {\em Modern cosmology} (Academic Press, San Diego, CA, 2003).

\bibitem{Mukhanov:991646}
V.~Mukhanov, {\em Physical Foundations of Cosmology} (Cambridge Univ. Press,
  Cambridge, 2005).

\bibitem{Hu:1995em}
W.~T. Hu, {\it Wandering in the Background: A CMB Explorer}, astro-ph/9508126,
  (1995).

\bibitem{Durrer:1127831}
R.~Durrer, {\em The Cosmic Microwave Background} (Cambridge Univ. Press,
  Cambridge, 2008).

\end{thebibliography}

%\begin{thebibliography}{9}

\end{document}